\documentclass[10pt,journal,compsoc]{IEEEtran}
\usepackage[sort&compress,numbers]{natbib}
\usepackage{amsmath, bm}
\usepackage{url}
\usepackage{xr-hyper}
\usepackage{xcolor}
\usepackage{comment}
\usepackage{graphicx}
\usepackage{booktabs}
\ifCLASSINFOpdf
\else
\fi
\usepackage{algorithm}
\usepackage{algpseudocode}
\usepackage{subfigure}
\usepackage{array}
\usepackage{hyperref}
\newcolumntype{P}[1]{>{\centering\arraybackslash}p{#1}}

\newcommand{\jhc}{\textcolor{blue}}
\newcommand{\new}{\textcolor{black}}

\usepackage{enumitem}
\setlist[itemize]{leftmargin=*, noitemsep}
\setlist[enumerate]{leftmargin=*, noitemsep}

\begin{document}

\title{Beyond Binary Opinions: A Deep Reinforcement Learning-Based Approach to Uncertainty-Aware Competitive Influence Maximization}

\author{Qi~Zhang,
Dian~Chen,     
Lance M.~Kaplan,~\IEEEmembership{Fellow,~IEEE,}
Audun J{\o}sang, 
Dong Hyun Jeong,~\IEEEmembership{Member,~IEEE,} 
Feng Chen,~\IEEEmembership{Member,~IEEE,} 
Jin-Hee Cho,~\IEEEmembership{Senior Member,~IEEE}

\thanks{Qi Zhang, Dian Chen and Jin-Hee Cho are with Department of Computer Science, Virginia Tech, USA (emails: \{qiz21, dianc, jicho\}@vt.edu).}
\thanks{Lance M.~Kaplan is with U.S. Army DEVCOM Army Research Laboratory, USA (email: lance.m.kaplan.civ@army.mil).}
\thanks{Audun J{\o}sang is with Department of Informatics, University of Oslo, Norway  (email: josang@ifi.uio.no).}
\thanks{Dong Hyun Jeong is with Department of Computer Science, University of the District of Columbia, USA (email: djeong@udc.edu).}
\thanks{Feng Chen is with Department of Computer Science, University of Texas at Dallas, USA (email: feng.chen@utdallas.edu)}
}

\markboth{ IEEE Transactions on Network Science and Engineering}%
{Zhang \MakeLowercase{\textit{et al.}}: IEEE Transactions on Network Science and Engineering}

\IEEEtitleabstractindextext{
\begin{abstract} 
The \textit{Competitive Influence Maximization (CIM)} problem involves multiple entities competing for influence in online social networks (OSNs). While \textit{Deep Reinforcement Learning (DRL)} has shown promise, existing methods often assume users' opinions are binary and ignore their behavior and prior knowledge. We propose DRIM, a multi-dimensional uncertainty-aware DRL-based CIM framework that leverages \textit{Subjective Logic} (SL) to model uncertainty in user opinions, preferences, and DRL decision-making. DRIM introduces an \textit{Uncertainty-based Opinion Model} (UOM) for a more realistic representation of user uncertainty and optimizes seed selection for propagating true information while countering false information. In addition, it quantifies uncertainty in balancing exploration and exploitation. Results show that UOM significantly enhances true information spread and maintains influence against advanced false information strategies. DRIM-based CIM schemes outperform state-of-the-art methods by up to 57\% and 88\% in influence while being up to 48\% and 77\% faster. Sensitivity analysis indicates that higher network observability and greater information propagation boost performance, while high network activity mitigates the effect of users’ initial biases.
\end{abstract}

}

\maketitle

\IEEEdisplaynontitleabstractindextext
\IEEEpeerreviewmaketitle

\IEEEraisesectionheading{\section{Introduction}\label{sec:introduction}}
Online Social Networks (OSNs) have become dominant platforms for information exchange and opinion formation. However, they also facilitate the rapid spread of false information, leading to reputational harm, financial loss, and manipulated public opinion~\cite{guo2020online}. The \textit{Competitive Influence Maximization} (CIM) problem arises in environments where competing entities, such as political parties or corporations, strategically select seed nodes to maximize their influence in OSNs. This study focuses on two competing parties: The {\em true party} (TP), disseminating true information, and the {\em false party} (FP), spreading false information. To mitigate the impact of false information, it is crucial to develop strategies that empower the TP to counteract the FP effectively and enhance the reach of true information in OSNs.

Traditional CIM approaches typically model user opinions as binary (i.e., supporting either the TP or the FP), overlooking the complexity and uncertainty of real-world opinion dynamics. In reality, OSN users often encounter ambiguous, evolving, and multi-dimensional information that shapes their decisions. A binary assumption oversimplifies influence propagation, neglecting belief strength, uncertainty, and adaptability. This limitation highlights the need for an advanced CIM framework that incorporates opinion uncertainty and dynamic belief updates.

\textit{Deep Reinforcement Learning} (DRL)~\cite{sutton2018reinforcement} has emerged as a powerful tool for real-time decision-making and strategy optimization in CIM tasks. By learning policies from interactions, DRL enables adaptive decision-making that evolves with the changing OSN landscape. A key challenge in DRL-based CIM is balancing exploration and exploitation. The agent must discover new influential nodes (exploration) while leveraging known high-impact nodes (exploitation) to maximize long-term influence. Conventional methods such as Epsilon-greedy~\cite{sutton2018reinforcement} and Upper Confidence Bound (UCB)~\cite{Lai85UCB} either fail to adapt the exploration rate dynamically or rely solely on reward or state values, ignoring the DRL model's intrinsic uncertainty.

To address these challenges, we propose an uncertainty-aware DRL framework for CIM that integrates \textit{Subjective Logic} (SL)~\cite{Josang16}, a belief model that explicitly quantifies multiple types of uncertainty, to represent opinion uncertainty. By incorporating uncertainty estimates into the exploration-exploitation trade-off, our approach enables adaptive decision-making that considers belief strength and confidence levels in user opinions. This allows the DRL agent to make more informed decisions under uncertainty, enhancing its ability to counteract false information while maximizing the spread of true information in OSNs.

Our work has made the following {\bf key contributions}: 
\begin{enumerate}
\item {Beyond Binary Opinion Models:} We advance CIM research by moving beyond traditional binary opinion models, which dominate existing studies~\cite{Ali20-CIM, Ali21_DRLCIM, Chung2019-DeepRL, Lin15_LearningBasedCIM}. Instead, we formulate opinion dynamics using {\em Subjective Logic} (SL)~\cite{Josang16}, a belief model that explicitly allows estimating uncertainty and dynamically updating user opinions based on real-world interactions. Unlike static binary models, our approach reflects the evolving nature of user beliefs in OSNs. Furthermore, we introduce and compare three different opinion models to evaluate how user attitudes toward opinion updates influence the effectiveness of false information mitigation.

\item {Dual-Agent DRL for CIM:} We adopt a dual-agent DRL framework, where both the TP and the FP engage in a CIM process. While DRL has been previously applied to CIM~\cite{Chung2019-DeepRL, li2020community, Lin15_LearningBasedCIM}, prior works have only considered DRL to model a single agent's decision-making. By explicitly modeling the interaction between two competing parties, our framework captures the adversarial nature of influence propagation in OSNs, enabling a more realistic and strategic seed selection process.

\item {Uncertainty-Aware Decision-Making:} We introduce uncertainty-aware decision-making at three critical levels: user opinion modeling, balance between exploration and exploitation in DRL, and network environment simulation. Our framework integrates SL-based opinion modeling with uncertainty estimation to better capture the uncertainties in user beliefs. Moreover, we develop multiple uncertainty-aware exploration-exploitation strategies using Evidential Neural Networks (ENNs)~\cite{sensoy2018EDL, zhao2020uncertainty}, bridging SL with Deep Neural Networks (DNNs) for more informed decision-making. Furthermore, we consider partially observable networks, where network edges are only partially visible, reflecting real-world constraints. To the best of our knowledge, no prior CIM work has incorporated such a multi-faceted uncertainty-aware approach across opinions, decision-making, and environmental modeling.

\item {Extensive Empirical Validation:} We conduct extensive simulation experiments on three real-world social network datasets to rigorously evaluate our proposed uncertainty-aware DRL-based CIM framework. Our experiments assess the impact of different opinion models, uncertainty-aware DRL strategies, and sensitivity analyses across various settings, ensuring the robustness and applicability of our findings.
\end{enumerate}

While preliminary results of this research were presented in~\cite{zhang2024uncertainty}, this work significantly extends~\cite{zhang2024uncertainty} by (1) introducing uncertainty-aware DRL with novel exploration-exploitation strategies, (2) leveraging ENNs to estimate multiple uncertainty types for seed node selection, and (3) validating the proposed framework across three real-world datasets with extensive sensitivity analyses to ensure higher applicability and robustness.

We specifically use the term {\em false information} instead of {\em misinformation} or {\em disinformation} to maintain clarity. Misinformation refers to mistakenly shared information without intent to deceive, while disinformation is deliberately spread to mislead others, characterized by malicious purpose. The nuanced differences between misinformation and disinformation and their separate impacts fall outside the purview of our research focus.

\section{Related Work}\label{sec:literature-review}

\new{This section presents an overview of related work in competitive information maximization, opinion and information propagation models, exploitation-exploration balance in DRL, and evidential deep learning. Additionally, we highlight the distinctions of our proposed approach from these works and summarize them in Table~\ref{tab:comparison} and Appendix C (Tables 3–5) in the supplement for easy reference.}

\begin{table*}[t]
\centering
\caption{\sc \new{Comparison of Related Works in CIM Research and Our Proposed Approach}}
\label{tab:comparison}
\vspace{-3mm}
\begin{tabular}{|p{1cm}|p{8cm}|p{8cm}|}
\hline
\textbf{Work} & \multicolumn{1}{c|}{\textbf{Contribution}} & \multicolumn{1}{c|}{\textbf{Difference from Our Work}} \\ 
\hline
\cite{domingos2001IM, kempe2003maximizing} & Introduced and formalized the IM problem. & Focuses on traditional IM, not competitive settings. \\ 
\hline
\cite{li2015getreal} & Applied game theory and Nash Equilibrium to CIM. & Assumes rational competitors and static strategies. \\ 
\hline
\cite{Pham19-CIM, liu2020CIM_UnwanedUser, hong2020efficient} & Developed heuristics like SPBA and CRIE for seed selection. & Lacks opinion modeling and uncertainty-awareness. \\ 
\hline
\cite{tsaras2021collective, galante2024competition} & Used game theory and probabilistic influence propagation. & Does not incorporate uncertainty-aware opinion evolution. \\ 
\hline
\cite{bozorgi2017community, xie2021competitive, liu2024user, bagheri2024community} & Proposed community structure-based CIM. & Relies on predefined communities, limiting adaptability. \\ 
\hline
\cite{Lin15_LearningBasedCIM, Chung2019-DeepRL, Ali20-CIM, Ali21_DRLCIM} & Applied reinforcement learning (RL) for multi-round CIM. & Models only a single agent, lacks adversarial interaction. \\ 
\hline
\cite{he2024multistage} & Integrated budget constraints and multi-stage CIM. & Assumes full network knowledge and static competitors. \\ 
\hline
\cite{li2018dominated, tong2022novel, liang2023targeted, dai2024competitive} & Considered user constraints such as time delays and interests. & Does not model uncertainty-aware opinion updates. \\ 
\hline
\cite{lin2015analyzing, liang2025mecim} & Addressed CIM uncertainty via probabilistic modeling. & Lacks belief-based opinion dynamics and uncertainty modeling. \\ 
\hline
\cite{gao2020fair, zhang2023fairness} & Explored fairness-aware seed selection strategies. & Uses simplistic fairness metrics, lacking opinion evolution. \\ 
\hline
\textbf{Our Work} & Proposes a dual-agent DRL CIM framework with SL for opinion dynamics and DRL-based uncertainty-aware decision-making. & Moves beyond binary opinion models, integrates adversarial RL for CIM, and applies multi-faceted uncertainty modeling.  \\ 
\hline
\end{tabular}
\end{table*}

\subsection{Competitive Information Maximization} \label{subsec:CIM}

\citet{domingos2001IM} introduced Influence Maximization (IM), focusing on selecting optimal seed nodes to maximize influence in Online Social Networks (OSNs). \citet{kempe2003maximizing} formalized IM as a discrete stochastic optimization problem, laying its theoretical foundation. 

\new{IM extends to Competitive IM (CIM) by introducing multiple competing entities. Early works applied game theory, such as \cite{li2015getreal} finding Nash Equilibrium in competitive networks. More recent methods include \cite{Pham19-CIM} with Sandwich Approximation, eliminating Monte Carlo simulations~\cite{liu2020CIM_UnwanedUser}, and developing Competitive Reverse Influence Estimation (CRIE)~\cite{hong2020efficient}. \citet{tsaras2021collective} proposed Awareness-to-Influence model collective IM, \citet{galante2024competition} leveraged game theory for influencer competition.  Community structure has been explored for CIM \cite{bozorgi2017community, bagheri2024community}. \citet{xie2021competitive, liu2024user} improved influence estimation using community borders and user preferences. These methods assume well-defined community structures, which may not always exist.}

\new{Reinforcement Learning (RL) has also been applied. \citet{Lin15_LearningBasedCIM} introduced \texttt{STORM}, a RL framework for multi-round CIM. \citet{Chung2019-DeepRL} extended this with spectral community detection. \citet{Ali20-CIM, Ali21_DRLCIM} refined DRL-based CIM by incorporating network exploration and dynamic graph embeddings, and further integrated Transfer Learning (TL) to reduce training time \cite{ali2022leveraging}. \citet{he2024multistage} incorporated Nash Equilibrium with Q-learning for strategic robustness.  On the other hand, other works address specific constraints in terms of time~\cite{li2018dominated}, user behaviors~\cite{tong2022novel, dai2024competitive}, and specific group~\cite{liang2023targeted}. Uncertainty \cite{lin2015analyzing, liang2025mecim} and fairness in CIM problem \cite{gao2020fair, zhang2023fairness} also have been explored.}

\new{Despite advancements, CIM research faces key limitations. Many rely on greedy algorithms \cite{bozorgi2017community, xie2021competitive, liang2023targeted, gao2020fair, lin2015analyzing, hong2020efficient}, lacking global optimality. Others assume full knowledge on network and competitor \cite{Chung2019-DeepRL, Lin15_LearningBasedCIM, Pham19-CIM, liu2020CIM_UnwanedUser}, or oversimplify user's and competitor's behavior\cite{galante2024competition, tong2022novel, tsaras2021collective, Ali20-CIM, Ali21_DRLCIM, he2024multistage}. Fairness-aware methods use simplistic metrics \cite{gao2020fair, zhang2023fairness}. Our work improves CIM realism by modeling opinion dynamics and incorporating uncertainty in user beliefs and network structure.  Table~\ref{tab:comparison} highlights our contributions by summarizing key differences between our work and existing approaches.}

\subsection{Opinion Models} \label{subsec:rw-opinion-models}
The Voter \cite{liggett2013stochastic}, DeGroot \cite{degroot1974reaching}, \new{and Friedkin-Johnsen (FJ) \cite{friedkin1999social} models focus on opinion dynamics but lack a competitive nature. They adopted a random neighbor's opinion, the DeGroot model averages neighbors’ opinions, and FJ extends it by introducing stubborn users. The Deffuant-Weisbuch (DW) \cite{deffuant2000mixing} and Hegselmann-Krause (HK) \cite{hegselmann2002opinion} models use bounded confidence, emphasizing pairwise interactions and clustering, being less suitable for competitive scenarios. }

\new{The Galam model \cite{galam2008sociophysics} focuses on binary opinions and employs majority rule, oversimplifying CIM. The Sznajd model \cite{sznajd2000opinion} relies on local interactions, making it unsuitable for global influence spread. Noisy models \cite{grauwin2012opinion, su2017noise} add randomness to existing models to simulate real-world phenomena, but randomness complicates CIM modeling. }

\new{Wildly used CIM propagation models include Independent Cascade (IC) and Linear Threshold (LT) \cite{kempe2003maximizing, li2015getreal, liang2023targeted, tong2022novel, bagheri2024community}, with many variants \cite{liu2020CIM_UnwanedUser, Pham19-CIM, lin2015analyzing, hong2020efficient, li2018dominated, ali2022leveraging, Ali20-CIM, Ali21_DRLCIM, tsaras2021collective, liang2025mecim, dai2024competitive, liu2024user, gao2020fair, zhang2023fairness}. IC assumes independent neighbor influence, while LT activates nodes based on cumulative influence thresholds. Despite their effectiveness, their assumption cannot fully reflect the realistic CIM process. Our work enhances these models by incorporating dynamic opinion dynamics and evaluating their resilience to misinformation.}

\subsection{Exploitation-Exploration Balance in DRL} \label{subsec:rw-ee-drl}

A key challenge in reinforcement learning is balancing exploration and exploitation (EE) \cite{sutton2018reinforcement}. One simple method is \textit{Epsilon-greedy} \cite{sutton2018reinforcement}, where the agent explores randomly with probability $\epsilon$ and otherwise exploits the best-known action. $\epsilon$ decays over time to shift towards exploitation.

\new{\textit{Reward shaping} \cite{devidze2022exploration} introduces additional rewards to guide agents to learn, especially in sparse reward environments. Other methods encourage agents to visit new states or choose diverse actions, such as \textit{Entropy Regularization (ER)}~\cite{ziebart2010modeling}, which tries to maximize policy entropy. \textit{Random Network Distillation (RND)} \cite{burda2018exploration} uses prediction errors from another network as intrinsic rewards. \textit{Pseudo-counts} \cite{bellemare2016unifying} and \textit{Never Give Up (NGU)} \cite{badia2020never} reward visiting less frequent or novel states.  Goal-based methods focus exploration on unknown areas. \textit{Go-Explore} \cite{ecoffet2019goexplore} memorizes visited states and selects goals probabilistically for further exploration. \textit{Reverse Curriculum Generation} \cite{florensa2017reverse} starts from goal states and works backward to explore the environment.}

\new{Uncertainty-based methods balance EE using confidence measures. \textit{Upper Confidence Bound (UCB)} \cite{Lai85UCB} selects actions based on reward estimates plus a confidence term, widely used in multi-armed bandits and RL. \textit{Bootstrapped DQN} \cite{osband2016boostrappedDQN, kalweit2017boostrappedDDPG} uses multiple Q-networks to estimate uncertainty and guide exploration, effective in continuous control but computationally expensive.} \new{\textit{Variational Information Maximizing Exploration (VIME)} \cite{houthooft2016vime} uses Bayesian neural networks for information gain by reducing uncertainty. }However, existing EE strategies have yet to fully explore the role of different uncertainty types in decision-making.

\subsection{Evidential Deep Learning} \label{subsec:rw-edl}

\new{Evidential Deep Learning (EDL), leveraging Dempster–Shafer Theory (DST) for uncertainty estimation, was refined by \cite{sensoy2018EDL} to enhance confidence measures via a specialized loss function. This improves performance in noisy environments and out-of-distribution (OOD) detection. Further advancements integrated Wasserstein GANs (WGANs) and Normal Inverse Gamma (NIG) distributions for better OOD detection and stereo matching uncertainty \cite{hu2021muENN, wang2022stereoMatchingUncertainty}.}

\new{\citet{deng2023FisherInfoEDL} introduced Fisher Information-based EDL (IEDL), dynamically reweighting loss via the Fisher Information Matrix (FIM) for improved high-uncertainty estimation. \citet{shao2024dual} proposed Dual-level Deep Evidential Fusion (DDEF), enhancing multimodal learning by linking neural outputs with Dirichlet parameters and fusing evidence across modalities. \citet{ancha2024deep} developed a Dirichlet-based framework for pixel-wise uncertainty estimation in semantic segmentation, aiding robotic navigation. \citet{duan2024evidential} extended classification uncertainty quantification using total covariance, offering richer insights beyond entropy-based approaches.}

\new{EDL has diverse applications. \citet{li2023region} adapted it for brain tumor segmentation, generating reliable uncertainty maps for clinical use. \citet{schreck2024evidential} showed EDL quantifies uncertainty effectively in Earth system science, aiding weather and climate modeling. However, careful tuning of its loss function is required to balance accuracy and uncertainty, posing challenges in parameter optimization.}

\new{We outline the key differences between our approach and existing works in Sections~\ref{subsec:rw-opinion-models}--\ref{subsec:rw-edl} of Appendix C (Tables 3–5) in the supplement.}

\section{Background: Subjective Logic (SL)} \label{sec:SL-background}

We employ \textit{Subjective Logic} (SL) to model user uncertainty in Section~\ref{sec:uncertainty-opinion-models} and develop an uncertainty-aware DRL-based CIM framework in Section~\ref{sec:our-algorithm}. The reason SL was chosen over Bayesian Networks and Fuzzy Logic is because it explicitly models opinion uncertainty using belief, disbelief, and uncertainty mass. Additionally, SL incorporates trust propagation and dynamic opinion updates, making it well-suited for capturing evolving user beliefs in competitive influence maximization. This section provides the necessary background on SL to facilitate understanding of our proposed opinion model and CIM framework.

\subsection{Binomial Opinion in SL} \label{subsec:sl-op-form}
In binary logic, a user's opinion toward something is either 100\% believed in or not, which barely happens. To better capture real-world opinion dynamics, we model user beliefs using Subjective Logic (SL)~\cite{Josang16}, which quantifies degrees of uncertainty in beliefs. For propositions classified as true or false, we use SL's binomial opinion model to represent user opinions in OSNs, defined as:  
\begin{eqnarray} 
\omega = (b, d, u, a), 
\end{eqnarray}  
where belief $b$ represents agreement with true information, disbelief $d$ denotes agreement with false information (or disagreement with true information), and uncertainty $u$ accounts for insufficient evidence. For simplicity, we omit the user ID (e.g., $i$) in this opinion representation.  These components lie within the range $[0,1]$ and satisfy the additivity constraint $b + d + u = 1$.  These components are:
\begin{eqnarray} 
\label{eq:sl-mapping}
b = \frac{r}{r+s+W}~,
d = \frac{s}{r+s+W}~,
u = \frac{W}{r+s+W}~,
\end{eqnarray}
where $r$ and $s$ are the numbers of evidence to support $b$ and $d$, respectively, and $W$ refers to the number of uncertain evidence that cannot be judged as true or false, supporting neither $b$ nor $d$.  The base rate $a$ represents the prior belief favoring true information, with $1-a$ corresponding to false information. It is a real number within the range $[0,1]$. While this paper focuses on two opinions using SL's binomial opinion model, the approach can be readily extended to multiple opinions involving additional parties.

When making opinion-based decisions (e.g., selecting a product), users account for uncertainty by leveraging the {\em projected probability} in SL, represented by the projected belief ($P(b)$) and projected disbelief ($P(d)$).They are defined as:  
\begin{eqnarray} 
\label{eq:expected-opinion}
P(b) = b + a \cdot u, \;
P(d) = d + (1-a) \cdot u,
\end{eqnarray}
where $P(b) + P(d) = 1$. Since users choose between $b$ and $d$, they interpret uncertainty based on their prior belief, $\bm{a} = \{a, 1-a\}$. Specifically, user $i$ is considered part of TP if $P(b_i) > 0.5$ and is deemed to believe false information if $P(d_i) > 0.5$. When a user $i$ has $P(b_i) = P(d_i)$ under a uniform prior belief (i.e., base rate $\bm{a}$), they are not counted toward TP or FP. However, such cases are extremely rare, so the sum of users in TP and FP is nearly equal to the total number of users in a given network.

\subsection{Dissonance in SL} \label{subsec:diss-sl}

We consider two types of uncertainty in the user's opinion formulated by SL~\cite{Josang18-fusion}: {\em vacuity} and {\em dissonance}. {\em Vacuity} refers to uncertainty caused by a lack of evidence, while {\em dissonance} indicates uncertainty due to conflicting evidence. 

For a multinomial opinion in SL, the dissonance in the opinion is calculated as~\cite{Josang18-fusion}:
\begin{gather}
\label{eq:multinomial-dissonance}
    b^\mathrm{diss} = \sum_{i \in K} \frac{{\beta}_i \sum_{j \in K/i} {\beta}_j \text{Bal}({\beta}_j, {\beta}_i)}{\sum_{j \in K/i} {\beta}_j}
\end{gather}
where 
\begin{gather}
    \text{Bal}({\beta}_j, {\beta}_i) = 1 - \frac{|{\beta}_j - {\beta}_i)|}{{\beta}_j + {\beta}_i}. \nonumber
\end{gather}
Here, $\text{Bal}({\beta}_j, {\beta}_i)$ measures the degree to which belief masses are evenly distributed, indicating uniformity among them. A higher balance level suggests greater inconclusiveness in the opinion, making decision-making more challenging.

A binomial opinion is a special case of a multinomial opinion, relying on the Beta distribution~\cite{Josang16}. For clarity, the dissonance of a binomial opinion is computed as follows:
\begin{gather}
b^\mathrm{diss} = {(b + d) \cdot \mbox{Bal}(b,d)}, 
\label{eq:binary-dissonance}
\end{gather}
where $\mbox{Bal}(b,d)$ represents the relative balance between belief masses $b$ and $d$ and is defined as:
\begin{equation}
\label{eq:belief-balance}
\mbox{Bal}(b,d) = 1 - \frac{|b-d|}{b+d}.
\end{equation}
We incorporate vacuity (i.e., $u$ in an opinion) and dissonance uncertainty measures in developing an uncertainty-aware opinion model (UOM) in Section~\ref{sec:uncertainty-opinion-models} and in designing uncertainty-aware exploration-exploitation strategies for DRL, as described in Section~\ref{sec:our-algorithm}.

\subsection{Uncertainty Maximization} \label{subsec:uncertainty-maximization}

Uncertainty maximization (UM) in SL~\cite{Josang16} is computed by adjusting the uncertainty mass ($\ddot{u}_i$) to its highest possible value while preserving the projected probabilities. To formally put, $\ddot{u}_i$ is estimated by: 
\begin{gather}
\label{eq:vacuity-maximization}
\ddot{u}_i = \min \Big[\frac{P(b_i)}{a_i}, \frac{P(d_i)}{1-a_i} \Big], \\ \ddot{b}_i = P(b_i) - a_i \cdot \ddot{u}_i, \; \;
\ddot{d}_i = P(d_i) - (1-a_i) \cdot \ddot{u}_i,
\nonumber
\end{gather}
The UM $\ddot{u}_i$ is determined as the minimum of the scaled projected belief and disbelief, ensuring consistency with the prior base rate ($a_i$). The belief ($\ddot{b}_i$) and disbelief ($\ddot{d}_i$) are then recalibrated by subtracting their respective contributions to uncertainty, maintaining the overall probability distribution. We incorporate this UM in designing uncertainty-aware exploration-exploitation strategies in Section~\ref{subsec:ee-strategies}. 

\subsection{Discounting Operator in SL} \label{subsec:discounting-operator-sl}

In SL, a {\em trust filter}, $c_i^j$, is used to discount $j$'s opinion because user $i$ accepts user $j$'s opinion to the extent that user $i$ trusts user $j$.  Each component of $\omega_{i \otimes j} = (b_{i \otimes j}, d_{i \otimes j}, u_{i \otimes j}, a_{i \otimes j})$ is estimated by: 
\begin{gather} \label{eq:discounting}
b_{i \otimes j}  =  c_i^j b_j,~  
d_{i \otimes j}  =  c_i^j d_j,~ \\
u_{i \otimes j}  =  1-c_i^j(1-u_j),~
a_{i \otimes j}  = a_j. \nonumber 
\end{gather}

\subsection{Consensus Operator in SL} \label{subsec:consensus-operator-sl}

The consensus operator $\oplus$ updates user $i$'s opinion by combining its belief weighted by $j$'s uncertainty and vice versa. Disbelief is updated using the same weighting mechanism. User $i$'s uncertainty decreases as the product of both users' uncertainty levels, and opinion updates continue unless either user's uncertainty reaches zero, indicating complete confidence.  User $i$'s updated opinion, incorporating trust in user $j$, is obtained by applying the consensus operator $\oplus$ to its original opinion and the discounted opinion of $j$. The result is given by $\omega_i \oplus \omega_{i \otimes j}$: 
\begin{gather} \label{eq:consensus_op}
b_i \oplus b_{i \otimes j} =  \frac{b_i  (1-c_i^j(1-u_j)) + c_i^j b_j u_i}{\beta}~, \\ \nonumber
d_i \oplus d_{i \otimes j} =  \frac{d_i (1-c_i^j(1-u_j)) + c_i^j d_j u_i}{\beta}~, \\ \nonumber
u_i \oplus u_{i \otimes j} =  \frac{u_i(1 -c_i^j(1 - u_j))}{\beta}~, \\ \nonumber
a_i \oplus a_{i \otimes j} = \frac{(a_i - (a_i + a_j)u_i)(1 - c_i^j(1-u_j)) + a_j u_i}{\beta - u_i(1 - c_i^j(1-u_j))}~, \\ \nonumber
\beta = 1-c_i^j(1-u_i)(1-u_j)\neq 0.
\end{gather}

Therefore, user $i$'s updated opinion, $\omega_i'$, is given by:
\begin{gather}
\label{eq:opinion-update}
    \omega_i' = \omega_i \oplus \omega_{i \otimes j} = (b_i', d_i', u_i', a_i)\\ =
    (b_i \oplus b_{i \otimes j}, d_i \oplus d_{i \otimes j},
 u_i \oplus u_{i \otimes j}, a_i \oplus a_{i \otimes j}),\nonumber 
\end{gather}
where $\omega_i$ is user $i$'s prior opinion, and $\omega_{i \otimes j}$ is $j$'s discounted opinion based on $i$'s acceptance level. Using SL's discounting and consensus operators, we propose uncertainty-aware opinion models in Section~\ref{sec:uncertainty-opinion-models}.

\section{Uncertain Opinion Models} \label{sec:uncertainty-opinion-models}
This section outlines the user types and uncertainty-aware opinion models, detailing how uncertain opinions are updated, shared, and interpreted in this work.

\subsection{User Types} \label{subsec:user-types}
We categorize users in OSNs into three types:
\begin{itemize}
\item {\em True information propagators} (TIPs) are seed nodes selected by the {\em true party} (TP) with an initial opinion $\omega = (b, d, u, a) = (b \rightarrow 1, d \rightarrow 0, u \rightarrow 0, a=1)$. This reflects a strong belief in true information ($b \approx 1$) and no belief in false information ($d \approx 0$).
\item {\em False information propagators} (FIPs) are seed nodes selected by the {\em false party} (FP), with an initial opinion of $\omega = (b, d, u, a) = (b \rightarrow 0, d \rightarrow 1, u \rightarrow 0, a=0)$. This indicates FIPs strongly believe in false information ($d \approx 1$) while lacking belief in true information ($b \approx 0$).
\item {\em Legitimate users} in OSNs are regular users with highly uncertain opinions, initialized as $\omega = (b, d, u, a) = (b \rightarrow 0, d \rightarrow 0, u \rightarrow 1, a=0.5)$.
\end{itemize}
Since TIPs and FIPs hold strong beliefs, they do not change their opinions, while {\em legitimate users} update their opinions by adjusting ($b, d, u, a)$ based on Eq.~\eqref{eq:opinion-update} and the consensus operation in Eq.~\eqref{eq:consensus_op}, describing how opinions shift by supporting evidence during interactions.

\begin{figure}
    \centering
    \includegraphics[width= 0.48\textwidth, height = 0.33 \textwidth]{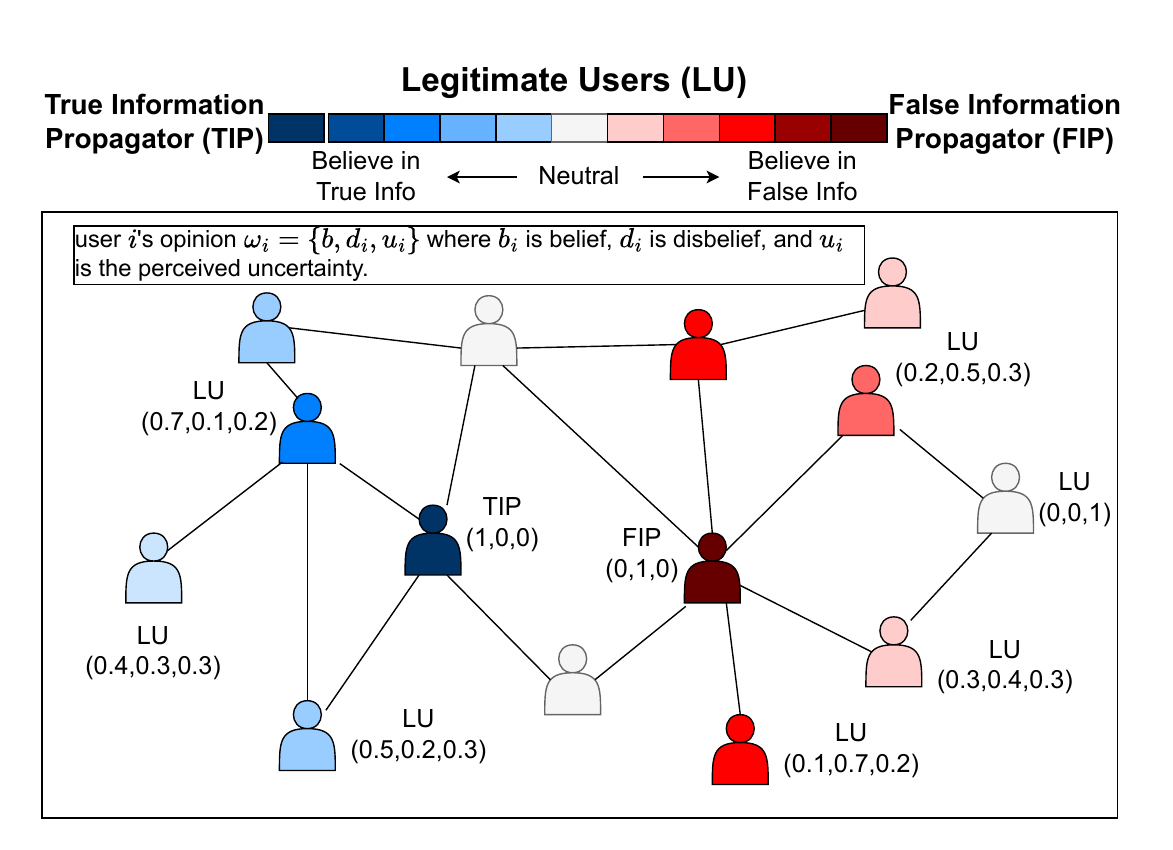}
    \caption{Overview of the CIM problem with Subjective Logic. An \textit{SL-based competitive network} visualizes belief strength via a color gradient where lighter shades indicate higher uncertainty, with white for neutrals. Blue represents true information supporters (darkest: TIP), and red denotes false information supporters (darkest: FIP). Each user $i$'s opinion is represented by $\omega_i = \{b_i, d_i, u_i\}$, we ignore prior belief $a_i$ because it is predefined as a constant for all users.
    }
    \label{fig: SL-based CIM}
\end{figure}

To illustrate SL-based opinion propagation in a competitive OSN with FP and TP, Fig.~\ref{fig: SL-based CIM} presents an overview of the CIM problem with Subjective Logic. The color gradient represents the level of belief, where lighter shades indicate lower certainty, and white denotes neutral users who are totally uncertain. Blue signifies belief in true information, with the darkest blue representing the TIP chosen by the True party (TP). Red represents belief in false information, with the darkest red indicating the FIP selected by the False party (FP).

\subsection{Updating, Sharing, and Reading Opinions} \label{subsec:opinion-models}
User $i$'s opinion is represented by $\omega_i = (b_i, d_i, u_i, \bm{a}_i)$. The user's behavior, $\bm{uc_i}$, comprises opinion updating, sharing, and reading, expressed as $\bm{uc}_i = \{\mathrm{updating}_i, \mathrm{reading}_i, \mathrm{sharing}_i\}$, as detailed below.

\subsubsection{Opinion Updating} \label{susbsubsec:opinion-updating} 
Recall that SL is chosen for its capability to formulate subjective opinions while handling multiple types of uncertainty causes, namely {\em vacuity} and {\em dissonance}. When user $i$ encounters user $j$ and reads their information (i.e., $\omega_j$), user $i$ updates their opinion accordingly. The {\em consensus operator} in SL~\cite{Josang16} is utilized to compute user $i$'s updated opinion, as detailed in Section~\ref{subsec:consensus-operator-sl}.

This study presents three opinion models (OMs): uncertainty-based trust (UOM), homophily-based trust (HOM), and no-trust (NOM), each with a distinct {\em trust filter} definition (see Section~\ref{subsec:discounting-operator-sl}):
\begin{eqnarray} \label{eq:dc}
c_i^j = \begin{cases} uc_i^j, &\text{if UOM} \\ hc_i^j, &\text{if HOM} \\ nc_i^j, &\text{if NOM} \end{cases}
\end{eqnarray}
where $c_i^j$ refers to the trust filter used in Eq.~\eqref{eq:discounting}.

\begin{itemize}
\item {Uncertainty-based OM (UOM):} This model represents users who seek new information when lacking sufficient evidence to decide~\cite{cho2019uncertainty}. UOM applies uncertainty-based trust, where the trust filter $uc_i^j$ is defined as:  
\begin{gather}  
\label{eq:uom-discounting-factor}  
uc_i^j = (1-u_i)(1-u_j),  
\end{gather}  
where $u_j \neq 1$ and $u_i \neq 1$, as the initial uncertainty of a legitimate user is close to 1 but not exactly 1 (see Sections~\ref{subsec:user-types} and \ref{subsec:parameterization}).  To prevent zero $\beta$ in Eq.~\eqref{eq:consensus_op}, we apply uncertainty maximization (UM)(Section~\ref{subsec:uncertainty-maximization}) when a user's uncertainty is low, but conflicting evidence prevents a decision. Given thresholds $T_v$ and $T_d$, if $u_i < T_v$ (low vacuity) and $b_i^{\mathrm{diss}} > T_d$ (high dissonance), user $i$'s opinion is updated as $\ddot{\omega}_i = (\ddot{b}_i, \ddot{d}_i, \ddot{u}_i, \bm{a}_i)$, where $\ddot{b}_i$, $\ddot{d}_i$, and $\ddot{u}_i$ follow Eq.~\eqref{eq:vacuity-maximization}.  The updated uncertainty $\ddot{u}_i$ enables user $i$ to accept new information and refine its opinion. 

\item {\bf Homophily-based OM (HOM)}: Homophily, the tendency of like-minded individuals to associate, influences opinion updates~\cite{Li11}. The homophily {\em trust filter} $hc_i^j$ is computed using {\em cosine similarity}, measuring alignment between users' beliefs and disbeliefs:
\begin{eqnarray} \label{eq:hc}
hc_i^j & = & \frac{b_i b_j + d_i d_j }{\sqrt{b_i^2 + d_i^2} \sqrt{b_j^2 + d_j^2}}.
\end{eqnarray}
This captures opinion similarity through belief ($b_i, b_j$) and disbelief ($d_i, d_j$), indirectly incorporating uncertainty since $b + d + u = 1$.

\item {\bf No-Trust-based OM (NOM)}: Without a trust filter, the No-Trust-based OM sets $nc_i^j = 1$, allowing user $i$ to fully accept user $j$'s information without any discount in opinion updates.
\end{itemize}

\subsubsection{Reading \& Sharing Opinions}  
We model a user's reading and sharing behavior following \cite{cho2019uncertainty}, which derived these behaviors from a survey on social media usage patterns. A user's reading probability, $P_r$, indicates how frequently they read information from neighbors, with values randomly assigned as 1 (multiple times per day), 0.5 (daily), 0.25 (weekly/monthly), or 0.1 (rarely) to reflect diverse reading habits. Similarly, a user's sharing probability, $P_s$, determines how often they share opinions $\omega$, with $P_s$ randomly assigned as 1 (always/mostly), 0.5 (half the time), 0.25 (sometimes), or 0.1 (rarely).

Since reading encourages opinion sharing~\cite{karnowski2018users}, users share only after reading. When user $i$ shares, it transmits its updated opinion $\omega_i$, not the original `True' or `False' information. Only seed nodes, {\em True} and {\em False information propagators} (TIPs and FIPs), share the original information.

\section{Uncertainty-Aware DRL-based CIM}
\label{sec:our-algorithm}

\subsection{Seed Set Selection Process} \label{subsec:seed-set-selection}

We use DRL to optimize strategy selection for maximizing influence spread in a multi-round process. In each round, the FP selects a seed node (turns it into an FIP), and then the FIP starts sharing its opinion with its neighbors and propagates further until all nodes are reached. The TP next selects a TIP to propagate true information. Information spreads through the network via {\em Breadth-First Search}, with nodes deciding to update and propagate based on their reading and sharing frequency. An episode consists of $T$ rounds, matching the number of seed nodes.

\subsection{Formulation of DRL-based Best Seed Selection} \label{subsec:drl-seed-selection}  
In this section, we describe the formulation of a DRL agent's states, actions, and rewards in this work.

{States.} At each round $t$, the state $s_t$ is defined as:
\begin{equation}
s_t = \Bigg \{\sum_{i,j \in \mathcal{U}} e_{i,j}, \max_{i \in \mathcal{U}} \mathrm{deg}_i \Bigg \}.    
\end{equation}
Here, {\em free nodes}, defined as $\{j \in \mathcal{U} | u_j \geq 0.5\}$, represent users with high uncertainty who have not aligned with either party, where $u_j$ denotes $j$'s uncertainty (i.e., vacuity). The term $\sum_{i,j \in \mathcal{U}} e_{i,j}$ counts edges among free nodes, while $\max_{i \in \mathcal{U}} \mathrm{deg}_i$ gives the highest degree among them.

{\bf Actions.} The action space at round $t$ is ${\bf a}_t = \{a_t^{AF}, a_t^{BF}, a_t^{SGF}, a_t^{CF}\}$, focusing on user behavior and centrality in the OSN, and each action means as follows:
\begin{itemize}
\item \textit{Active First (AF, $a_t^{AF}$)} prioritizes the most active user, determined by the highest $P_r \times P_s$, where $P_r$ and $P_s$ represent the user's reading and sharing probabilities.
\item \textit{Blocking First (BF, $a_t^{BF}$)} targets neighbors of the opponent’s party with the highest {\em free degree} (connected to {\em free nodes}. A user belongs to the TP if $P(b_i) > 0.5$, or the FP if $P(d_i) > 0.5$ where $P(b)$ refers to an expected probability for a given belief $b$, as discussed in Eq.~\eqref{eq:expected-opinion}.
\item \textit{SubGreedy First (SGF, $a_t^{SGF}$)} selects the node with the most neighbors within $k$-hops~\cite{Lin15_LearningBasedCIM, Chung2019-DeepRL} (e.g., $k=2$ to balance efficiency and effectiveness).
\item \textit{Centrality First (CF, $a_t^{CF}$)} selects the user with the highest degree centrality. We use degree as a centrality metric, considering its efficiency. Investigating the performance of using a different centrality metric is beyond the scope of our work.

\end{itemize}

{\bf Rewards.} We use immediate rewards to enhance the learning process. At each round $t$, the rewards for each party, TP or FP, are the sum of users' beliefs aligned with each party by:
\begin{equation}
\label{eq: instant reward}
R^{TP}_t = \sum_{i \in TP} b_t^i, \; \; R^{FP}_t = \sum_{i \in FP} d_t^i,
\end{equation}
where $b_t^i$ and $d_t^i$ refer to user $i$'s belief mass at round $t$. 

The respective accumulated rewards over an episode from $t=0$ to $t=T$ are:
\begin{equation}
\label{eq: accumulated reward}
R_T^{TP} = \sum_{t=1}^{T}\gamma^{T-(t-1)}R_t^{TP}, \; \; R_T^{FP} = \sum_{t=1}^{T}\gamma^{T-(t-1)}R_t^{FP},
\end{equation}
where $\gamma \in [0,1]$ is the discount factor.
We integrate model prediction uncertainty to enhance decision-making in the EE trade-off.  After obtaining the action distribution from the policy network's final layer, we interpret it as belief mass $\beta_i$ for each action, ensuring $\sum_{i \in K} \beta_i = 1$ with initial uncertainty mass $\nu = 0$. To introduce positive uncertainty mass $\ddot{\nu}$, we apply the uncertainty maximization (UM) technique~\cite{Josang16}, detailed in Section~\ref{subsec:uncertainty-maximization}. We assume no prior knowledge for belief propositions, assigning an equal base rate probability. 

\begin{figure*}[t]
    \centering
    \includegraphics[width= \textwidth, height = 0.36 \textwidth]{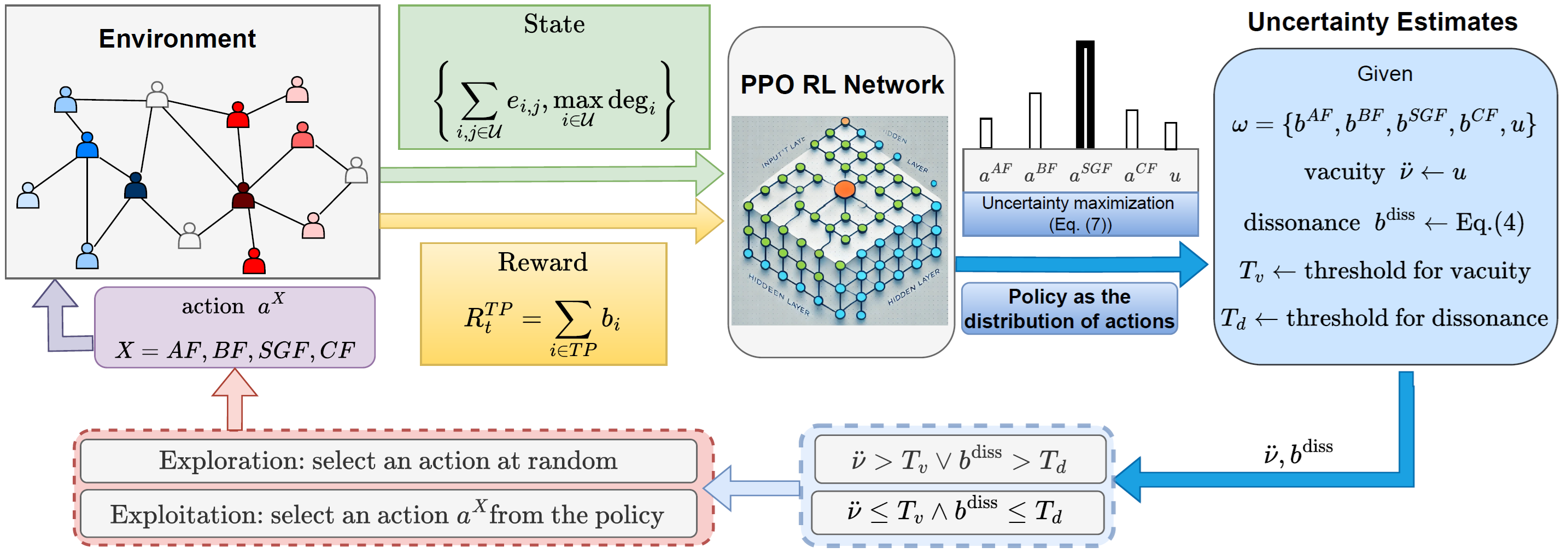}
    \caption{{\bf Overview of the Proposed Uncertainty-Aware DRL-based Node Selection}: Each agent (i.e., DRL agents for TP and FP, respectively) processes OSN states through a policy network to determine the action distribution. Vacuity and/or dissonance guide the exploration-exploitation decision-making: \textit{low uncertainty} favors exploitation (selecting the highest-probability action for maximum influence), while \textit{high uncertainty} promotes exploration (random action selection). The \textit{policy network} is iteratively updated based on rewards.
    }
\label{fig:framework_overview}
    \vspace{-5mm}
\end{figure*}

In addition to vacuity (uncertainty mass $\ddot{\nu}$) for assessing evidence sufficiency, we compute dissonance to detect conflicting decisions, where a flatter action distribution indicates lower confidence. For example, with belief masses $0.24, 0.25, 0.24, 0.27$, the system’s highest probability ($0.27$) does not imply confidence in selecting the optimal action due to the distribution's uniformity. Dissonance (see Eq.~\eqref{eq:multinomial-dissonance}) quantifies this distinction, implying that the flatter the distribution, the lower the dissonance. These uncertainty estimates guide EE decisions, as discussed below. Fig.~\ref{fig:framework_overview} illustrates the proposed Uncertainty-Aware DRL-based node selection process, where each party's DRL agent processes OSN states through a policy network to determine action distributions. 
 
\subsection{Exploration-Exploitation (EE) Strategies}  \label{subsec:ee-strategies}

A DRL agent looks for the best action by randomly choosing an action called \textit{exploration} and relying on their experiences, called \textit{exploitation}.  In this work, we evaluate the following EE strategies: the first three are our proposed uncertainty-aware approaches, and the rest are state-of-the-art methods. Their performance will be compared to highlight the strengths of each strategy. The considered EE strategies are:  
\begin{itemize}
\item {EE Using Vacuity (VAC-EE)}: This strategy balances EE based on vacuity ($\ddot{\nu}$), which quantifies uncertainty from insufficient evidence. If vacuity exceeds a threshold ($T_{v}$), the agent explores due to limited knowledge. Otherwise, it exploits by selecting the best-known action. This approach explores under high uncertainty from a lack of information and exploits otherwise.

\item {EE Using Dissonance (DIS-EE)}: This strategy balances EE based on dissonance ($b^{\mathrm{diss}}$), which quantifies conflicting evidence in decision-making. If dissonance exceeds a threshold ($T_{d}$), the agent explores to resolve inconsistencies. Otherwise, it exploits the best-known action. This approach adapts decisions by exploring under conflict and exploiting otherwise.

\item {EE Using Vacuity and Dissonance (VD-EE)}: This strategy dynamically balances exploration and exploitation using vacuity ($\ddot{\nu}$) and dissonance ($b^{\mathrm{diss}}$). If vacuity exceeds a threshold ($T_{v}$), the agent explores due to insufficient knowledge. Otherwise, it evaluates dissonance; if above ($T_{d}$), exploration occurs to address conflicting information. When both measures are low, the agent exploits the best-known action. This adaptive approach optimizes decision-making by exploring under high uncertainty and exploiting when confidence is sufficient.

\item {EE Using Entropy (ENT-EE)}: This strategy measures the randomness in the agent's action selection using the entropy of the action probability distribution \cite{namdari2019review}. High entropy indicates greater exploration, while low entropy reflects a more deterministic policy, favoring exploitation.

\item {EE Using Epsilon Greedy (EPS-EE)}~\cite{sutton2018reinforcement}: A widely used exploration strategy in reinforcement learning, epsilon-greedy balances exploration and exploitation through random action selection. With probability $\epsilon$, the agent explores by selecting a random action, while with probability $1-\epsilon$, it exploits by choosing the action it currently estimates to be the best based on gathered information.

\item {EE Using Entropy Regulation (ER-EE)}~\cite{ziebart2010modeling}: This strategy incorporates entropy into the reward function by adding an entropy term scaled by a coefficient (which determines the importance of entropy or exploration), the modified objective encourages the agent not only to maximize the expected reward but also to maintain a level of randomness in its action selection.

\item {Upper Confidence Bound (UCB-EE)}~\cite{Lai85UCB}: UCB balances exploration and exploitation by adjusting exploration based on action uncertainty. The agent estimates expected rewards from past observations and computes an uncertainty measure, which decreases as more data is gathered. Less-explored actions are prioritized for exploration, increasing the likelihood of discovering optimal choices.
\end{itemize}
Algorithm~\ref{algo:uncertainty-ee-algorithms} details the first three proposed uncertainty-aware EE strategies.

\begin{algorithm}
\footnotesize
\caption{Uncertainty-Aware Exploration-Exploitation Strategies (VAC-EE, DIS-EE, and VD-EE)} \label{algo:uncertainty-ee-algorithms}
\begin{algorithmic}[1]
\State Given an opinion formulated based on the actions taken by a DRL agent for seed node selection,
\State $\ddot{\nu} \leftarrow$ vacuity,  $b^{\mathrm{diss}} \leftarrow$ dissonance
\State $T_v \leftarrow$ vacuity threshold, $T_d \leftarrow$ dissonance threshold
\Procedure{Vacuity-Based EE (VAC-EE)}{}
    \If{$\ddot{\nu} > T_v$}
        \State {explore}
    \Else
        \State {exploit}
    \EndIf
\EndProcedure
\vspace{1mm}
\Procedure{Dissonance-Based EE (DIS-EE)}{}
    \If{$b^{\mathrm{diss}} > T_d$}
        \State {explore}
    \Else
        \State {exploit}
    \EndIf
\EndProcedure
\vspace{1mm}
\Procedure{Vacuity and Dissonance-Based EE (VD-EE)}{}
    \If{\textit{$\ddot{\nu}$} $>$ \textit{$T_v$}}
        \State {explore}
    \Else
        \If{\textit{$b^{\mathrm{diss}}$} $>$ \textit{$T_d$}}
            \State {explore}
        \Else
            \State {exploit}
        \EndIf
    \EndIf
\EndProcedure
\end{algorithmic}
\end{algorithm}

\subsection{Partially Observable Network} \label{subsec:uncertain-network-modeling}

Following~\cite{nasim2016PartialObservable}, a partially observable network is defined as an undirected graph $G = (V, E)$ with an observable subset $G' = (V, E')$, where $E' \subset E$. Only a subset of edges is visible to the DRL agent when observing the state in OSNs. 

In an OSN, two parties compete to maximize user belief in their opinion, starting with all users neutral. In each round, FP selects a seed node based on the policy network's action distribution to maximize influence. The chosen node (FIP) propagates opinions as per Section~\ref{subsec:opinion-models}. TP then selects a seed node to counter FP's influence. This process repeats until the predefined seed limit is reached or no further selections are possible. TP's influence is evaluated by the number of nodes adopting its opinion.

\section{Experiment Setup} \label{sec:exp-setup}

\subsection{Algorithm Settings}

We employ Proximal Policy Optimization (PPO)~\cite{schulman2017ppo} in our DRL framework for optimal seed node selection, as PPO enables multiple updates with the same data batch, reducing computational complexity given CIM's long data collection time. TP utilizes a trained DRL model for node selection, while FP can choose from six strategies, such as DRL, Active First (AF), Blocking First (BF), SubGreedy First (SGF), and Centrality First (CF), as described in Section~\ref{subsec:drl-seed-selection}.  We train TP's DRL agent against each FP strategy and evaluate their performance in Section~\ref{sec:numerical-analysis-results}.

\subsection{Parameterization} \label{subsec:parameterization}

We initialize SL-based opinions for legitimate users using Eq.~\eqref{eq:sl-mapping} with $(r, s, W) = (1, 1, 101)$, indicating high uncertainty. TIPs start with $(100, 1, 2)$, reflecting strong confidence in true information, while FIPs begin with $(1, 100, 2)$, showing confidence in false information. For uncertainty-aware EE strategies, we set vacuity and dissonance thresholds at $T_v=0.01$ and $T_d=0.6$ to optimize influence. In SubGreedy First, $k=2$ balances efficiency and effectiveness, while DRL employs a discount factor $\gamma=0.99$ for long-term influence.

FP and TP alternately select 50 seeds for propagation, with FP initiating the process, followed by TP. This models real-world false information mitigation, where true information counters ongoing false information. TP propagates twice after each seed selection, while FP propagates once. Given the critical role of human judgment and public awareness, true information is assumed to propagate more effectively in OSNs. All results are averaged over 50 runs.

Experiments ran on an HPE Apollo 6500 with AMD EPYC 7742 processors (2.25 GHz base, 3.4 GHz boost). Key design parameters, definitions, and defaults are in Table~1, Appendix A of the supplementary document. Unless stated otherwise, performance evaluations assume perfect network observability, except when analyzing its variations.

\subsection{Network Datasets}  \label{subsec:datasets}
We validate our framework using three datasets: the URV Email Network~\cite{rossi2015network_dataset_c}, an undirected graph of email communications at {\em Universitat Rovira i Virgili}, Spain, a Facebook social circles dataset~\cite{leskovec2012facebook} (FBN, Facebook network); and a Facebook page network (FBPN)~\cite{rozemberczki2021facebookpage}. These networks effectively model CIM, capturing real user interactions. The URV Email and Facebook datasets reflect how individuals share ideas, opinions, and information, while the Facebook page-page network represents content spread by entities like brands or political campaigns via influential hubs. The datasets’ diverse structures and sizes enable a comprehensive evaluation of our framework’s generalizability on complex OSNs. Table~\ref{tab:datasets} provides detailed statistics. For centrality measures, we report unnormalized medians instead of means, as social networks are scale-free, where most nodes have low degrees and a few have very high degrees, making the median more representative.
   
\begin{table}[t]
\centering
\caption{\sc \centering Characteristics of Network Datasets}
\label{tab:datasets}
\vspace{-3mm}
\begin{tabular}{|P{2.6cm}|P{0.7cm}|P{0.7cm}|P{0.8cm}|P{1.5cm}|}
\hline
Dataset & Nodes & Edges & Degree & Betweenness \\ \hline
URV Email~\cite{rossi2015network_dataset_c} & 1133 & 5452 & 7 & 703.94 \\ \hline
Facebook~\cite{leskovec2012facebook} & 4039 & 88234 & 25 & 47.57 \\ \hline
FB Page~\cite{rozemberczki2021facebookpage} & 22470 &	171002 & 7 & 5496.91 \\ \hline
\end{tabular}
\label{tab:simple_table}
\vspace{-5mm}
\end{table}

\subsection{Metrics} \label{subsec:metrics}  
We evaluate the proposed approaches using the following metrics: (1) {Percentage (\%) of nodes in TP}, measured by the number of users aligned with TP (Eq.~\eqref{eq:expected-opinion}). Users with $P_i(b_i) > 0.5$ are classified under TP, indicating its influence, denoted as $n^{TP}$;  
(2) {Algorithmic efficiency}, assessed by the simulation's runtime per round (Section~\ref{subsec:complexity-analysis}, Table~\ref{tab:running_time});  
(3) {Uncertainty estimates}, evaluated by the measures of vacuity, dissonance, and entropy.  

\subsection{Comparing Schemes} \label{subsec:comparing-schemes}

To evaluate our framework, we compare its performance and runtime against existing DRL-based CIM frameworks. Unlike~\cite{Chung2019-DeepRL, Lin15_LearningBasedCIM, Ali20-CIM, Ali21_DRLCIM}, our approach employs an SL-based dynamic opinion model, enabling opinion evolution through interactions. For a fair comparison, all models operate under identical conditions, including the same opinion models (HOM, UOM, or NOM), seed selection settings (FP moves first, equal propagation times, $T=50$ seed nodes), and opinion update mechanisms (Section~\ref{sec:our-algorithm}). We use ER-EE as the general EE trade-off strategy across all schemes. While adapting these frameworks may affect their original performance, this standardization ensures a realistic evaluation.  

We evaluate the following CIM algorithms:  
(1) {\bf DRIM-A}: Our proposed framework DRIM, where `A' denotes the AF strategy (Section~\ref{subsec:seed-set-selection}), distinguishing it from DRIM-NA.  
(2) {\bf DRIM-NA}: A variant excluding the AF strategy (No-AF) from the action space to assess AF's impact in DRIM.  
(3) {\bf STORM}~\cite{Lin15_LearningBasedCIM}: Originally designed with binary opinions for node occupation, we adapt it by defining {\em free nodes} ($\{j| u_j \geq 0.5\}$) as unoccupied and merging max-weight and max-degree actions, as our datasets are unweighted.  
(4) {\bf C-STORM}~\cite{Chung2019-DeepRL}: An extension of STORM incorporating a preliminary community detection step for seed selection. We adapt it using our opinion model and the same definition of a free node as used in STORM.

\section{Experimental Results \& Analyses}
\label{sec:numerical-analysis-results}
This section analyzes the performance of the considered approaches using results from Tables~\ref{tab:OM-vary-table}--\ref{tab:running_time} and Figs.~\ref{fig:T-drl-F-vary-schemes}--\ref{fig:uncertainty-T-drl-F-vary-EE-strategy}.  

\subsection{Performance Analyses of Opinion Models}
We first analyze the effect of opinion models by comparing performance across three models using four entropy regulation schemes and our uncertainty-aware approaches, with TP facing various FP strategies. Table~\ref{tab:OM-vary-table} presents results on the URV Email Network, where TP's influence ($n^{TP}$) is measured across five FP node selection strategies. 

As shown in Table~\ref{tab:OM-vary-table}, UOM consistently achieves the highest impact for true information, as users with high uncertainty ($u$) are more receptive to new evidence, enabling continuous updates and broader true information spread. This aligns with human intuition -- when uncertain, people seek more information before forming conclusions. Thus, even with FP's first-mover advantage, true information can still propagate effectively in the OSN.  In contrast, HOM and NOM yield poor TP performance, as users exposed to false information first tend to form rigid beliefs that true information struggles to correct. This underscores the importance of early intervention and swift responses to counteract misinformation.  Under HOM and NOM, STORM and C-STORM outperform DRIM-A-based schemes against BF and DRL strategies. BF is not a primary concern in false information mitigation, as false parties rarely use it due to their first-mover advantage. DRL dominance occurs when both parties employ identical DRL designs, favoring FP as the initial actor.  However, against other strategies, DRIM-A performs comparably to STORM and C-STORM under HOM and NOM, demonstrating competitive efficacy.

{For all subsequent experiments, we adopt UOM as the opinion model}, as it facilitates true information spread despite FP’s first-mover advantage. In contrast, HOM and NOM inherently block true information, leaving minimal room for improvement.

\begin{table}
\centering
\caption{\sc \centering True Party (TP)'s Influence in $n^{TP}$ Under Various CIM Algorithms with Different OMs}
\label{tab:OM-vary-table}
\vspace{-2mm}
\scriptsize
\begin{tabular}{|c|c|c|c|c|c|}
\hline
{Scheme / OM}  & {AF} & {BF} & {SGF} & {CF} & {DRL} \\ \hline
DRIM-A/UOM & {1077.74}  & {463.79} & {827.05} & {374.1} & {691.37} \\
DRIM-A/HOM & 15.7  & 263.38 & 4.18 & 4.39 & 15.74 \\ 
DRIM-A/NOM & 27.59 & 354.64 & 5.85  & 4.86  & 17.56 \\  \hline
DRIM-NA/UOM & {1078.26} & {529.52} & {959.63} & {525.7} & {576.47} \\ 
DRIM-NA/HOM & 38.24  & 274.32 & 5.18 & 4.58  & 4.15  \\  
DRIM-NA/NOM & 60.64  & 331.34 & 6.26  & 4.78  & 6.51 \\ \hline
C-STORM/UOM & {1070.75} & {1082.53} & {524.39} & {229.15} & {551.44} \\ 
C-STORM/HOM & 26.88  & 475.27 & 4.57 & 4.55 & 533.17 \\  
C-STORM/NOM & 15.47  & 509.03 & 6.38 & 4.82 & 442.93 \\ \hline
STORM/UOM & {1072.45} & {1058.80} & {461.19} & {63.61} & {424.09} \\ 
STORM/HOM & 60.61 & 477.76 & 4.96 & 4.92  & 375.17  \\  
STORM/NOM & 60.58  & 487.43 & 6.42 & 4.69 & 377.3 \\ \hline
DRIM-A-VDEE/UOM & {1077.64}  & {646.94} & {960.66} & {581.9}
 & {794.33} \\
DRIM-A-VDEE/HOM & 38.23  & 195.8 & 5.05 & 4.47 & 4.68 \\
DRIM-A-VDEE/NOM & 38.06 & 241.91 & 6.23 & 4.82 & 6.91 \\ \hline
\end{tabular}
\vspace{-5mm}
\end{table}

\subsection{Performance Analyses of CIM Schemes} \label{subsec:cim-PA}

\begin{figure*}[htb]
  \centering
  \subfigure{
    \includegraphics[width=0.75\textwidth, height=0.025\textwidth]{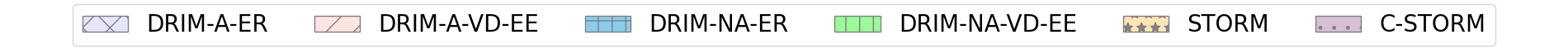}}
    \setcounter{subfigure}{0}
 \vspace{-3mm}    
    
  \subfigure[Email Network]{
    \includegraphics[width=0.3\textwidth, height=0.2\textwidth]{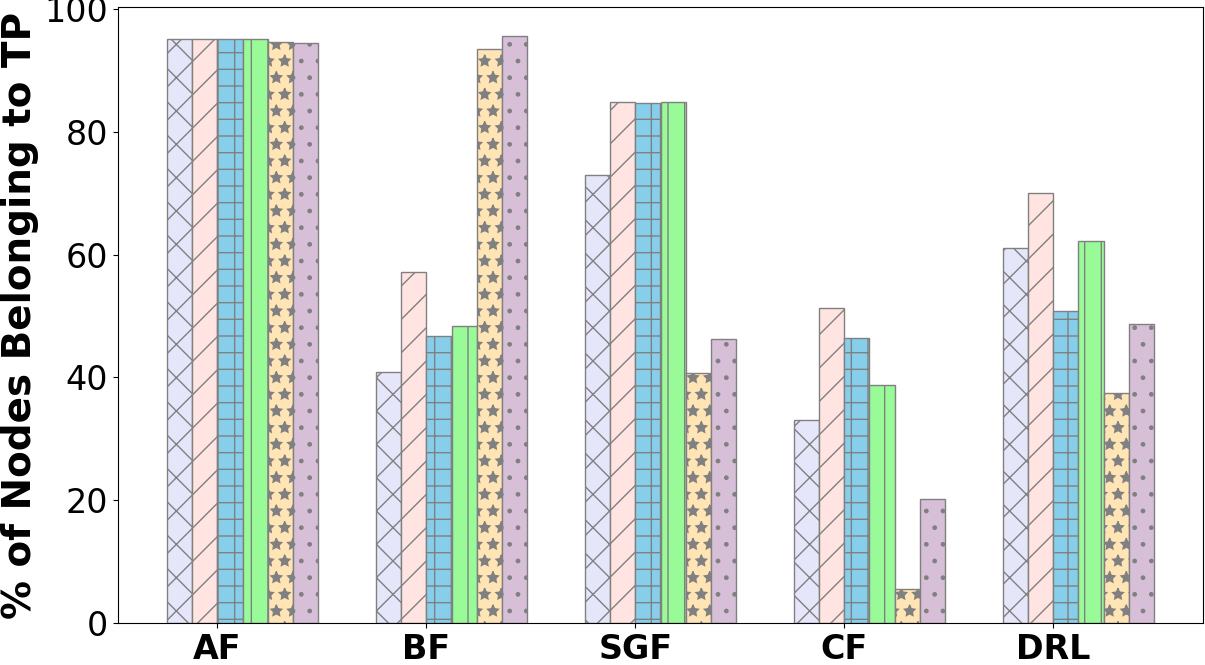}\label{fig:T-drl-F-schemes-vary-strategy_dataset1}}
  \subfigure[Facebook Dataset]{
    \includegraphics[width=0.3\textwidth, height=0.2\textwidth]{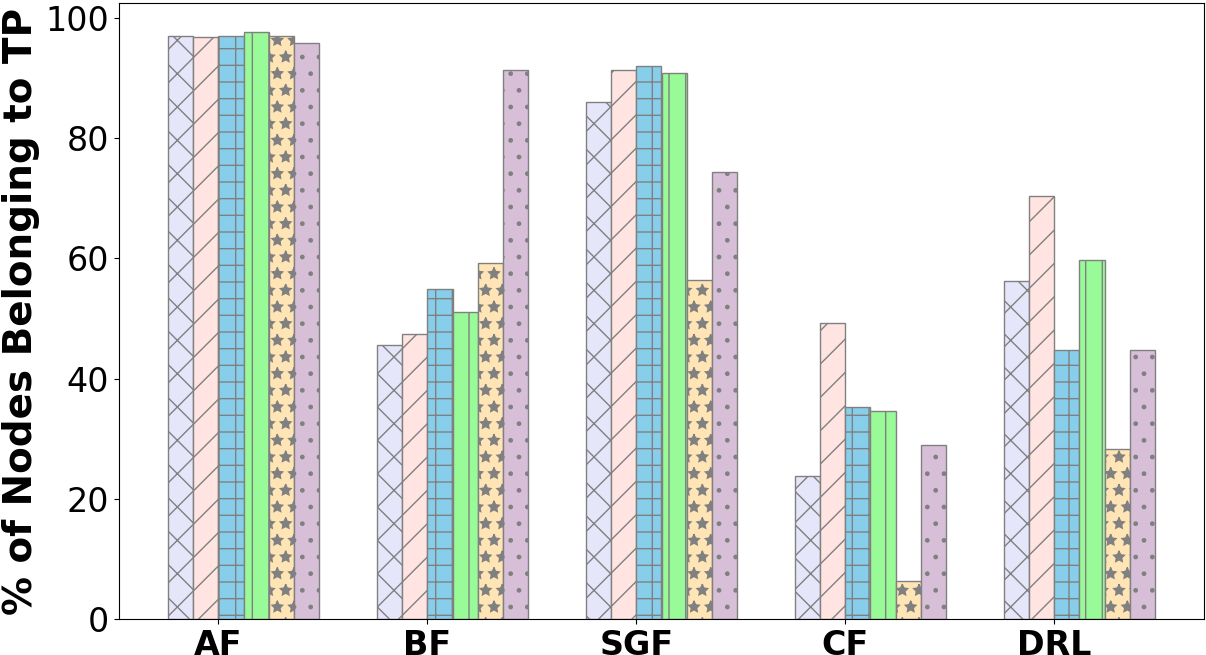} \label{fig:T-drl-F-schemes-vary-strategy_dataset2}}
  \subfigure[Facebook Page Dataset]{
    \includegraphics[width=0.3\textwidth, height=0.2\textwidth]{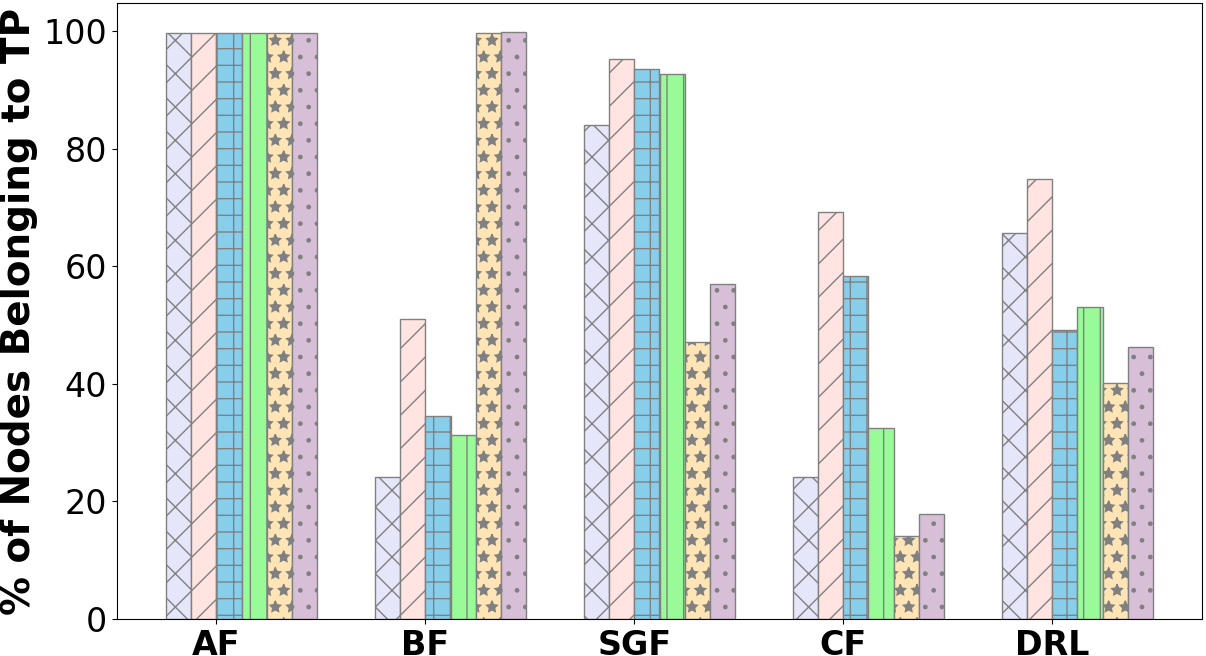} \label{fig:T-drl-F-schemes-vary-strategy_dataset3}}
    \caption{Performance comparison of the considered CIM algorithms with respect to \% of nodes in the true party (TP), representing TP's influence. 
    }
\label{fig:T-drl-F-vary-schemes}
\end{figure*}

{\bf Comparative Analysis under Three Datasets:}  
This section evaluates the performance of existing CIM algorithms and uncertainty-aware exploration-exploitation schemes. All non-uncertainty-aware schemes use entropy regulation to balance exploration and exploitation. Fig.~\ref{fig:T-drl-F-vary-schemes} compares the performance of four ER-DRL-based TP agents under five FP seed selection strategies. The $x$-axis denotes FP's strategy, while the $y$-axis shows the percentage of nodes aligned with TP, normalized across varying network sizes.

Fig.~\ref{fig:T-drl-F-vary-schemes} shows that across all three datasets, our DRIM-based schemes outperform under SGF, CF, and DRL strategies, remain competitive against AF, and underperform against BF. When FP uses SGF or CF, DRIM-based TPs achieve greater influence than STORM and C-STORM, indicating that DRIM-based approaches can surpass state-of-the-art algorithms when FP selects nodes via traditional centrality measures maximizing local reach.

In the dual DRL agent case, the four DRIM-based schemes (i.e., DRIM-A-ER, DRIM-A-VDEE, DRIM-NA-ER, and DRIM-NA-VDEE) consistently outperform STORM and C-STORM. Among them, DRIM-A-VDEE is the most effective, achieving a peak influence of 70.11\% in the Email dataset, while DRIM-NA-ER, though the least effective, still reaches 50.88\%. In comparison, C-STORM and STORM achieve 48.67\% and 37.43\%, respectively. DRIM-A-VDEE surpasses them by 44.05\% and 87.31\%. Similar trends hold for the Facebook and Facebook Page networks, where DRIM-A-VDEE achieves 70.31\% and 74.75\% influence, compared to 44.82\% and 28.23\% for C-STORM and STORM on Facebook, and 46.26\% and 40.12\% on Facebook Page. DRIM-A-VDEE thus outperforms these schemes by at least 56.87\% and 61.59\% across datasets.

When FP employs AF, all schemes achieve nearly 100\% influence. Since AF does not prioritize nodes effectively, it behaves similarly to random selection, allowing all approaches to gain high influence.  When FP uses BF, DRIM-based schemes perform worse due to BF’s defensive nature, which prioritizes blocking over proactive influence. This suggests that DRIM-based approaches are better suited against aggressive strategies rather than defensive ones.

Across datasets, performance variations reveal differences in adaptability. The gap between STORM and C-STORM is notably larger in the Facebook dataset than in the Email and Facebook Page datasets. When FP employs BF or DRL, C-STORM significantly outperforms STORM on Facebook but only slightly on the other two datasets. This is due to C-STORM’s community-based approach, which aligns well with Facebook’s social structure, while the other two datasets exhibit more uniform connectivity. In contrast, DRIM-based schemes maintain stable performance across all datasets, demonstrating their robustness and adaptability to diverse network structures.

\begin{figure*}[htb]
  \centering
  \subfigure{
    \includegraphics[width=0.75\textwidth, height=0.025\textwidth]{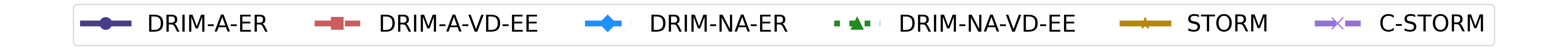}}
    \setcounter{subfigure}{0}
    \vspace{-3mm}    
    
  \subfigure[Varying \# of IP in Email Net.]{
    \includegraphics[width=0.3\textwidth, height=0.2\textwidth]{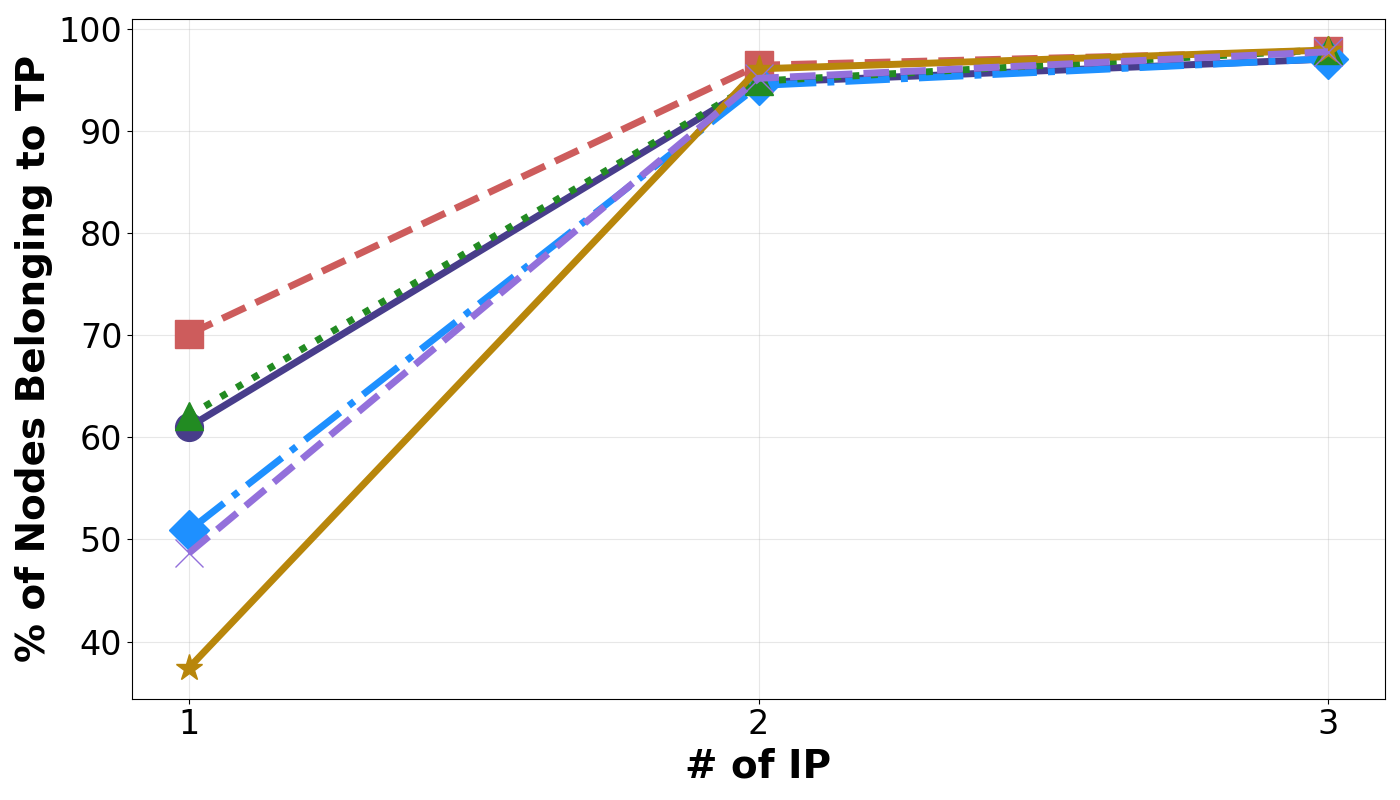}\label{fig:drl-drl-schemes-IP-sens-ds1}} 
  \subfigure[Varying \% of NO in Email Net.]{
    \includegraphics[width=0.3\textwidth, height=0.2\textwidth]{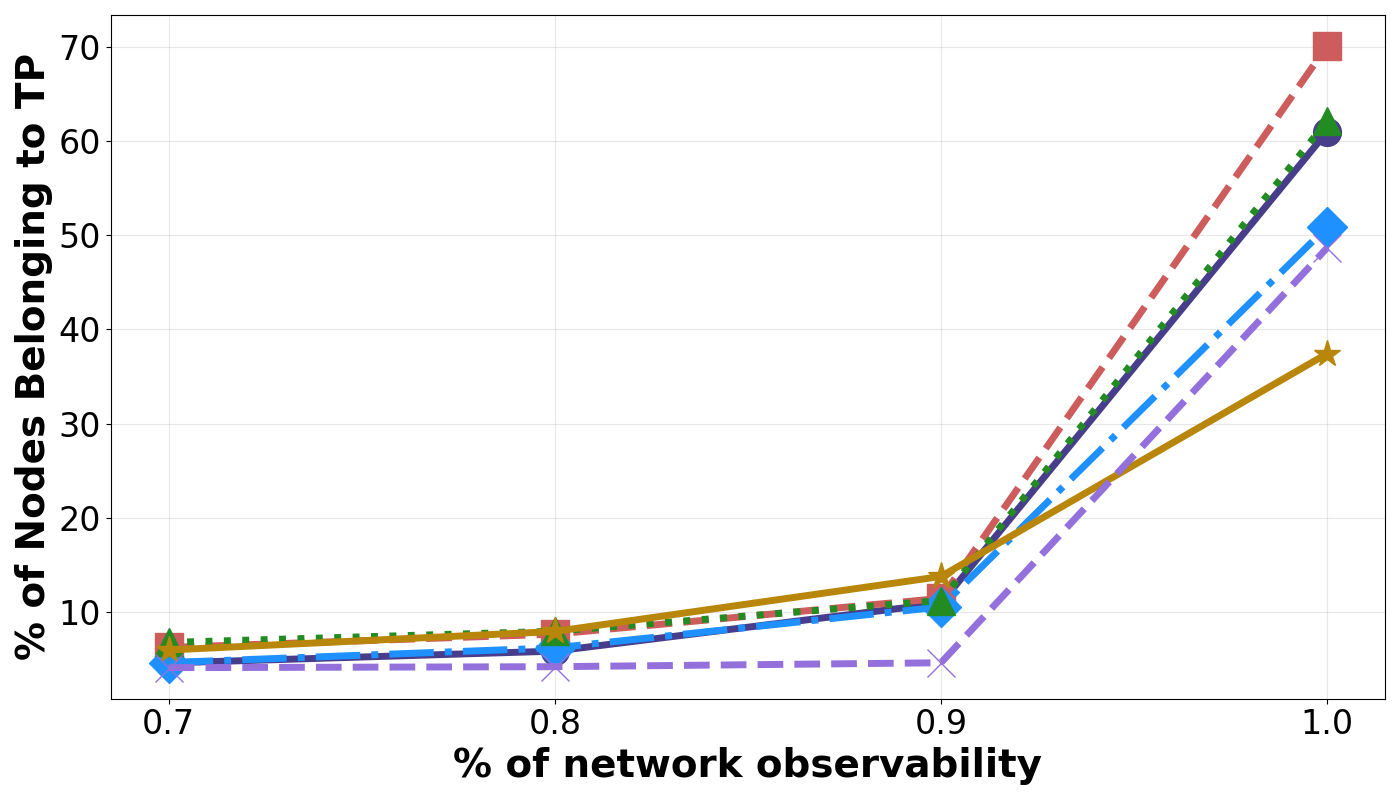} \label{fig:drl-drl-schemes-PON-sens-ds1}}
  \subfigure[Varying users' PB in Email Net.]{
    \includegraphics[width=0.3\textwidth, height=0.2\textwidth]{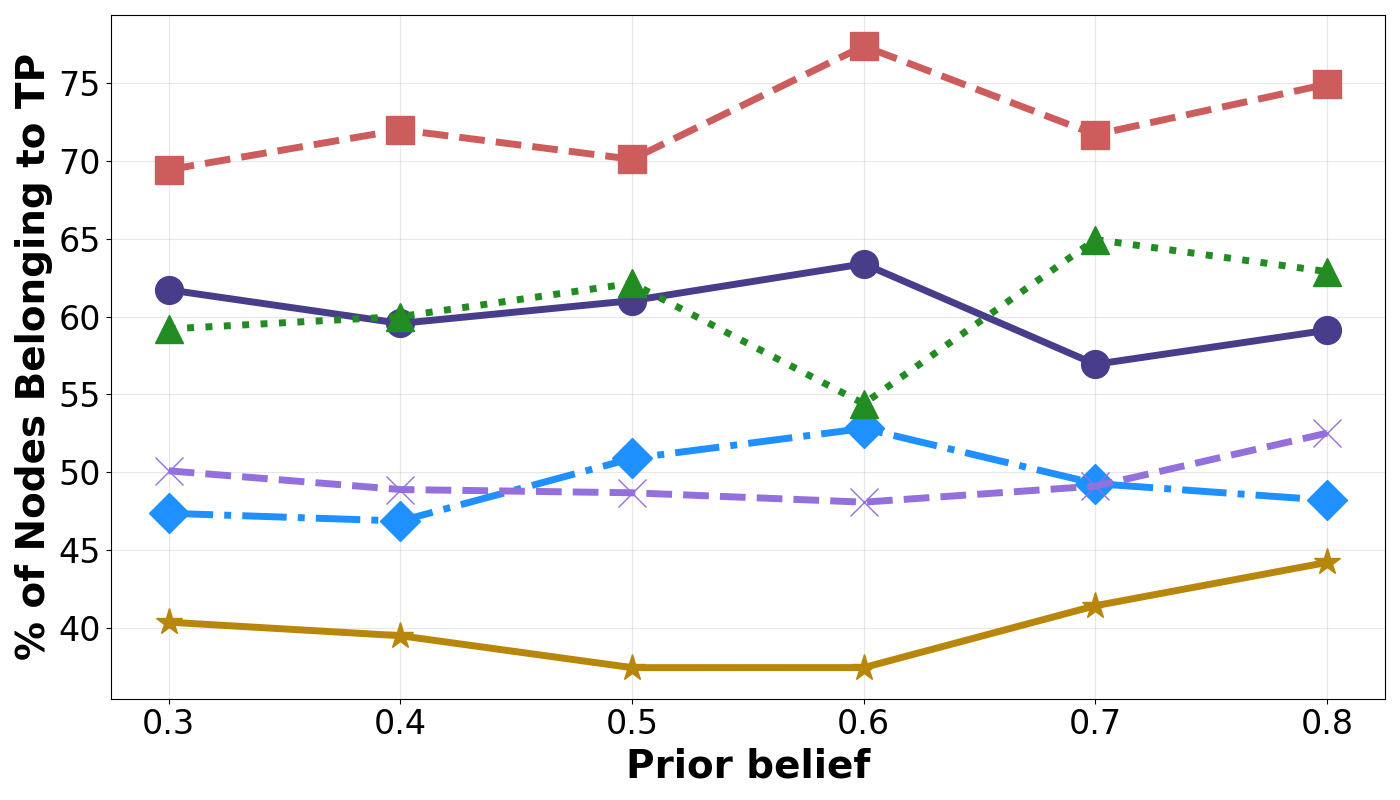} \label{fig:drl-drl-schemes-base-rate-sens-ds1}}
    \vspace{-2mm}
    
    \caption{TP's influence across CIM algorithms when TP and FP use DRL for seed selection in the URV Email network. \#IP denotes the number of information propagation by TP, `NO' represents network observability \%, and PB indicates users' prior belief in true information (i.e., $a$). In all cases except (c), we set $a=0.5$. 
    }
\label{fig:drl-sens-cim}
\end{figure*}

{\bf Sensitivity Analyses Under CIM Schemes:} This section examines how the number of true information propagations, network observability, and users' prior beliefs impact CIM algorithm performance. Both TP and FP use identical DRL-based agents for seed selection, enabling an assessment of TP's effectiveness against ``smart" opponents. Due to space constraints, Fig.~\ref{fig:drl-sens-cim} presents results from the URV Email dataset, while similar findings for the other two datasets are provided in Appendix B.2 of the supplement.

Fig.~\ref{fig:drl-drl-schemes-IP-sens-ds1} examines the impact of increasing TP's information propagations (IPs) from 1 to 3 while FP remains at a single propagation per seed. As expected, all schemes improve with more IPs, demonstrating that additional propagation enhances influence spread. Notably, the four DRIM-based schemes perform best with a single propagation per round, aligning with Fig.~\ref{fig:T-drl-F-schemes-vary-strategy_dataset1}, effectively countering false information with minimal resources. However, beyond two IPs, influence gains plateau as TP nears its maximum reach.

In Fig.~\ref{fig:drl-drl-schemes-PON-sens-ds1}, a clear trend shows improved performance across all schemes as network observability increases from 0.7 to 1. This indicates that greater visibility of the network structure significantly enhances performance. Under limited observability, all schemes exhibit reduced influence, highlighting the DRL agent's dependence on network state knowledge. The sharp performance increase between 0.9 and 1 suggests a high sensitivity to network observability.

Fig.~\ref{fig:drl-drl-schemes-base-rate-sens-ds1} examines the impact of varying prior beliefs across different schemes. Ideally, as prior belief in TP rises (from 0.3 to 0.8), influence maximization performance should improve, as users are more inclined to accept information from TP. However, this trend is not always evident. According to Eq.~\eqref{eq:expected-opinion}, prior belief primarily influences outcomes when users have nonzero vacuity $u$, allowing initial biases to shape their acceptance of information. In high-activity settings, frequent information exchange leads to greater certainty over time, reducing $u$ and diminishing the effect of initial beliefs. Thus, despite varying prior beliefs, differences in influence maximization remain subtle. Nevertheless, DRIM-based schemes, particularly DRIM-A-VDEE, consistently achieve the highest influence, even when initial user beliefs do not favor true information.

These results suggest that DRIM-based schemes exhibit robust and stable performance across varying network conditions, showing particular effectiveness in scenarios with limited information propagation and diverse prior beliefs.

\subsection{Performance Analyses of EE Strategies}

\begin{figure*}[htb]
  \centering
  \subfigure{
    \includegraphics[width=0.75\textwidth, height=0.025\textwidth]{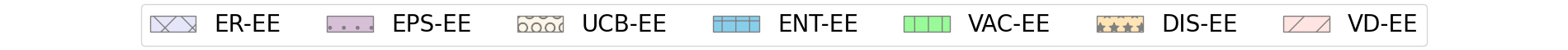}}
    \setcounter{subfigure}{0}
    \vspace{-3mm}    
    
  \subfigure[URV Email network]{
    \includegraphics[width=0.3\textwidth, height=0.2\textwidth]{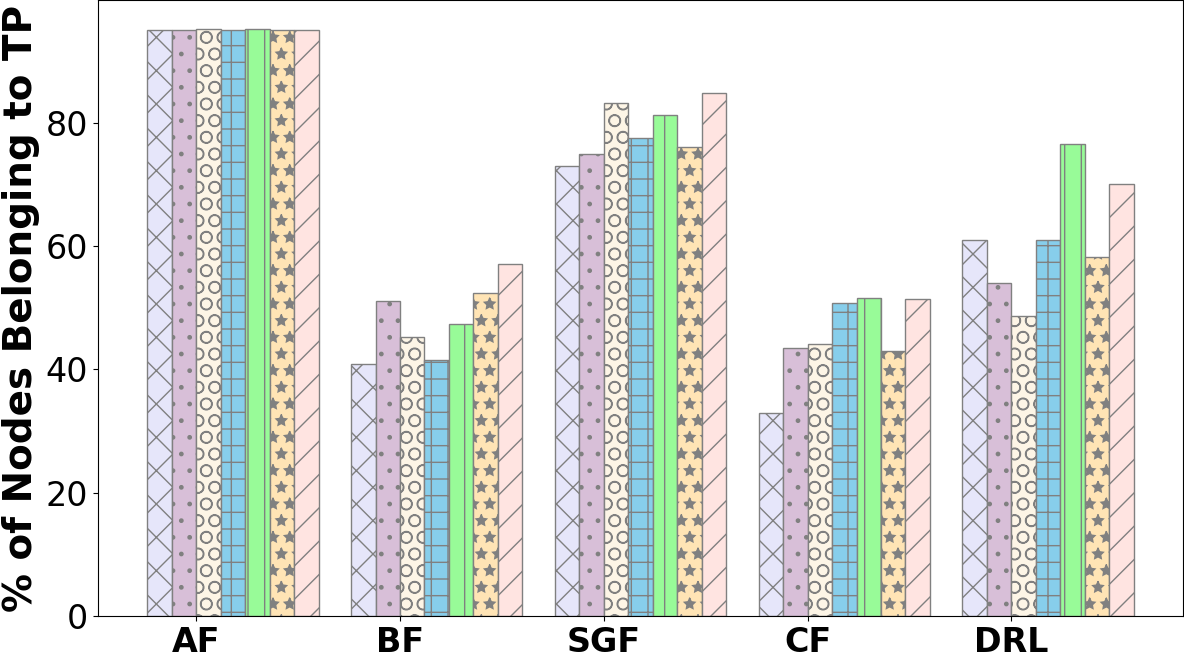}\label{fig:T-drl-F-ee-vary-strategy_dataset1}}
  \subfigure[Facebook network (FBN)]{
    \includegraphics[width=0.3\textwidth, height=0.2\textwidth]{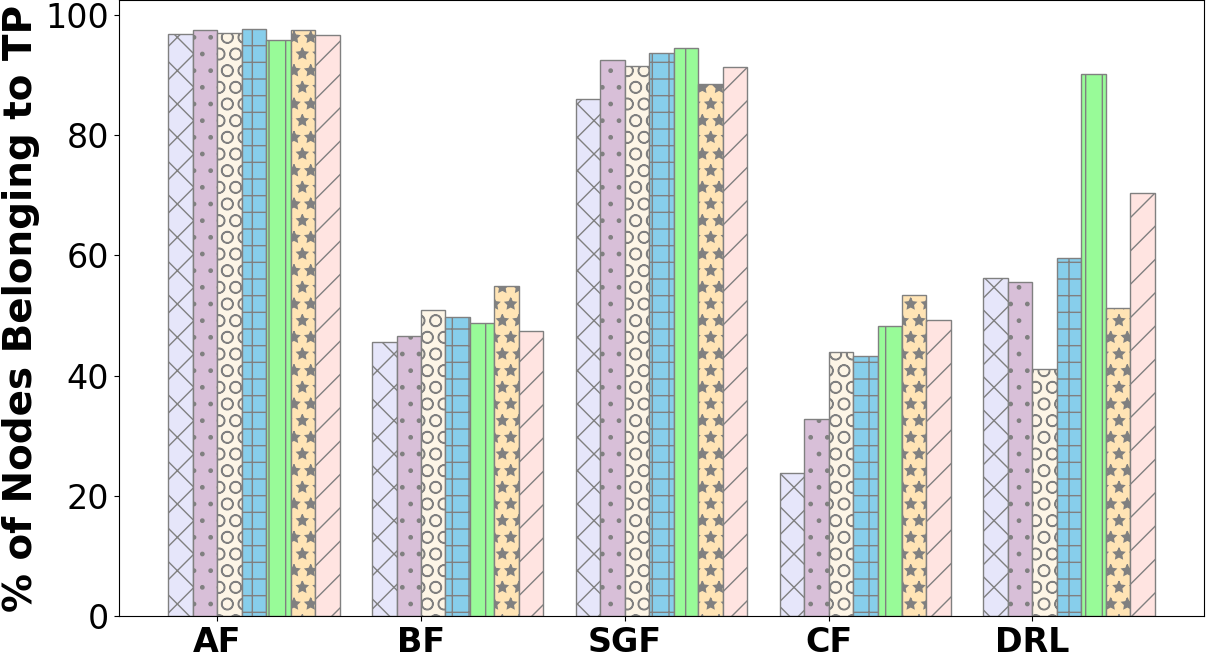} \label{fig:T-drl-F-ee-vary-strategy_dataset2}}
  \subfigure[Facebook page network (FBPN)]{
    \includegraphics[width=0.3\textwidth, height=0.2\textwidth]{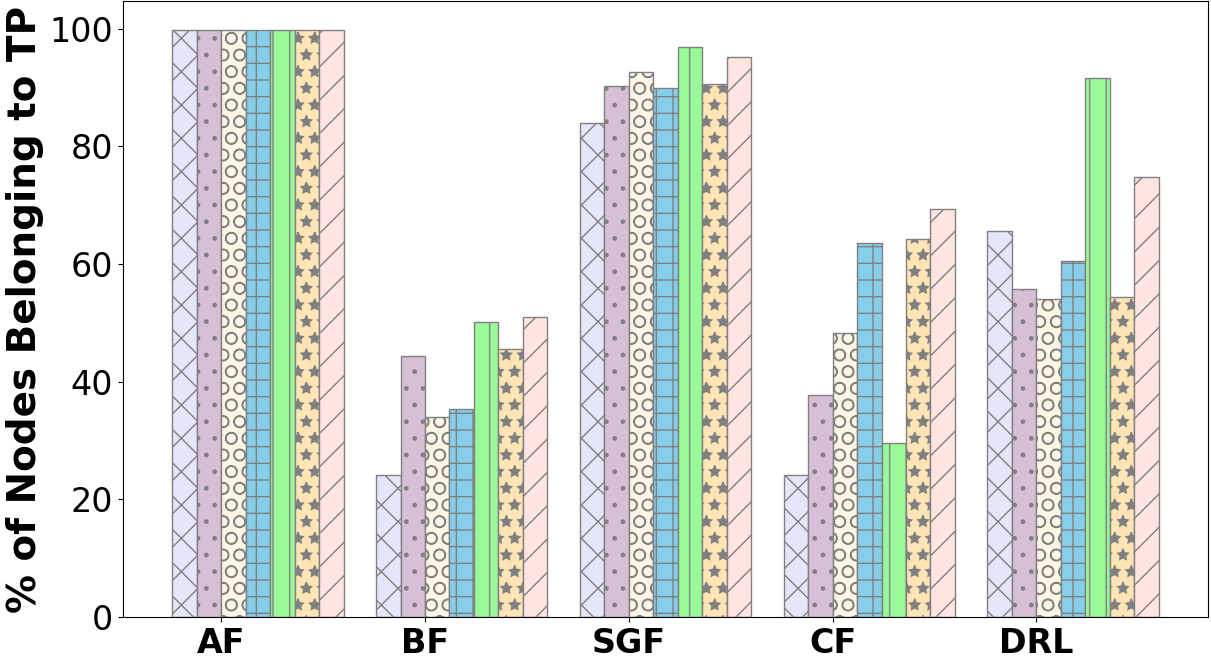} \label{fig:T-drl-F-ee-vary-strategy_dataset3}}
    \vspace{-2mm}
    
    \caption{TP's influence under various EE strategies under the three networks.
    }
\label{fig:T-drl-F-ee-vary-strategy-performance}
\end{figure*}

{\bf Comparative Analysis of EE Strategies Across Three Datasets:}  Fig.~\ref{fig:T-drl-F-ee-vary-strategy-performance} compares the performance of various EE strategies within the DRIM-A framework across three datasets.  In Fig.~\ref{fig:T-drl-F-ee-vary-strategy-performance}, we observe that TP dominates when FP employs AF, SGF, or DRL strategies while maintaining strong performance against BF and CF. This underscores the impact of FP's strategic choices on DRIM-A's influence. VD-EE consistently performs well across datasets and FP strategies. Unlike ER-EE, it benefits from incorporating uncertainty by considering both vacuity and dissonance. By accounting for missing (vacuity) and conflicting (dissonance) information, VD-EE optimizes exploitation timing. In contrast, traditional uncertainty-based strategies such as EPS-EE, UCB-EE, and ENT-EE show inconsistent results due to limited uncertainty modeling.

VAC-EE, relying solely on vacuity, excels under DRL-based node selection, even surpassing VD-EE despite lacking dissonance modeling. Since vacuity directly signals information gaps, it enables more targeted exploration in networks with uneven information distribution. Conversely, VD-EE’s dual consideration of vacuity and dissonance offers deeper insights but increases decision complexity. In rapidly changing environments requiring quick adaptation, this complexity may reduce its effectiveness in balancing exploration and exploitation.

\begin{figure*}[htb]
  \centering
  \subfigure{
    \includegraphics[width=0.75\textwidth, height=0.025\textwidth]{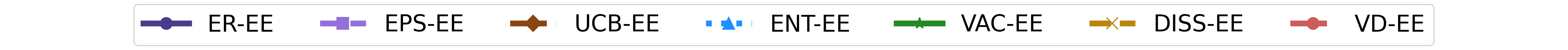}}
    \setcounter{subfigure}{0}
    \vspace{-3mm}    
    
  \subfigure[Varying \# of IP]{
    \includegraphics[width=0.3\textwidth, height=0.2\textwidth]{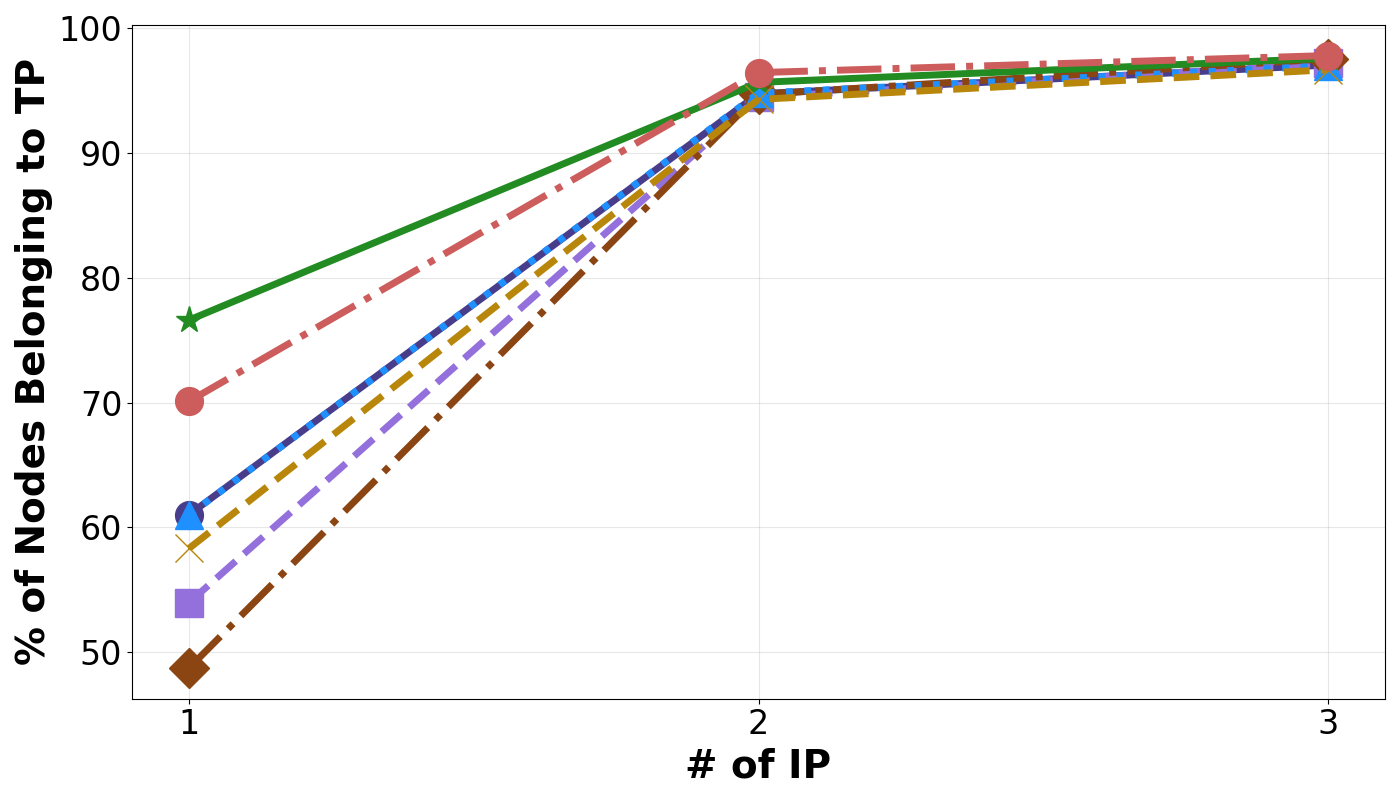}\label{fig:drl-drl-ee-IP-sens-ds1}} 
  \subfigure[Varying \% of NO]{
    \includegraphics[width=0.3\textwidth, height=0.2\textwidth]{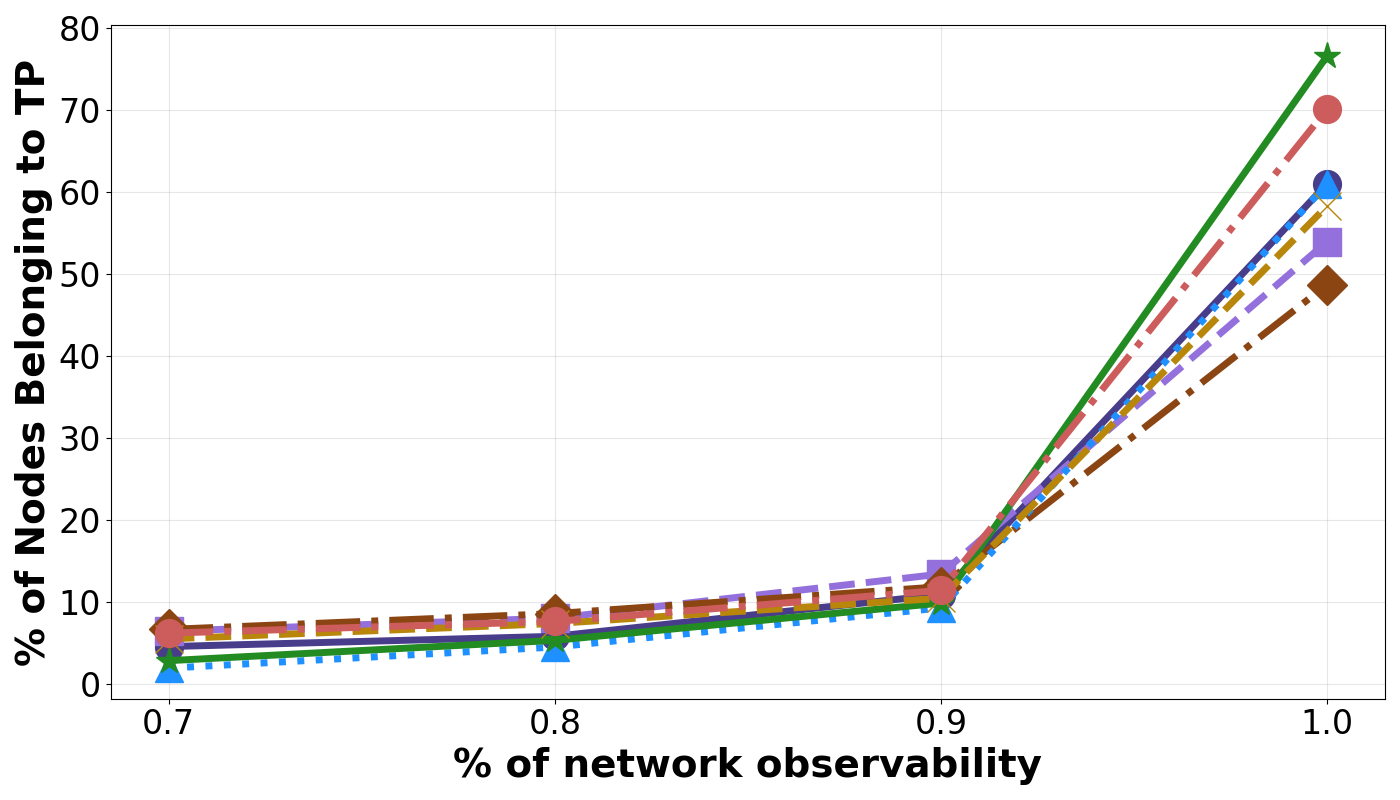} \label{fig:drl-drl-PON-sens-ds1}}
  \subfigure[Varying users' PB]{
    \includegraphics[width=0.3\textwidth, height=0.2\textwidth]{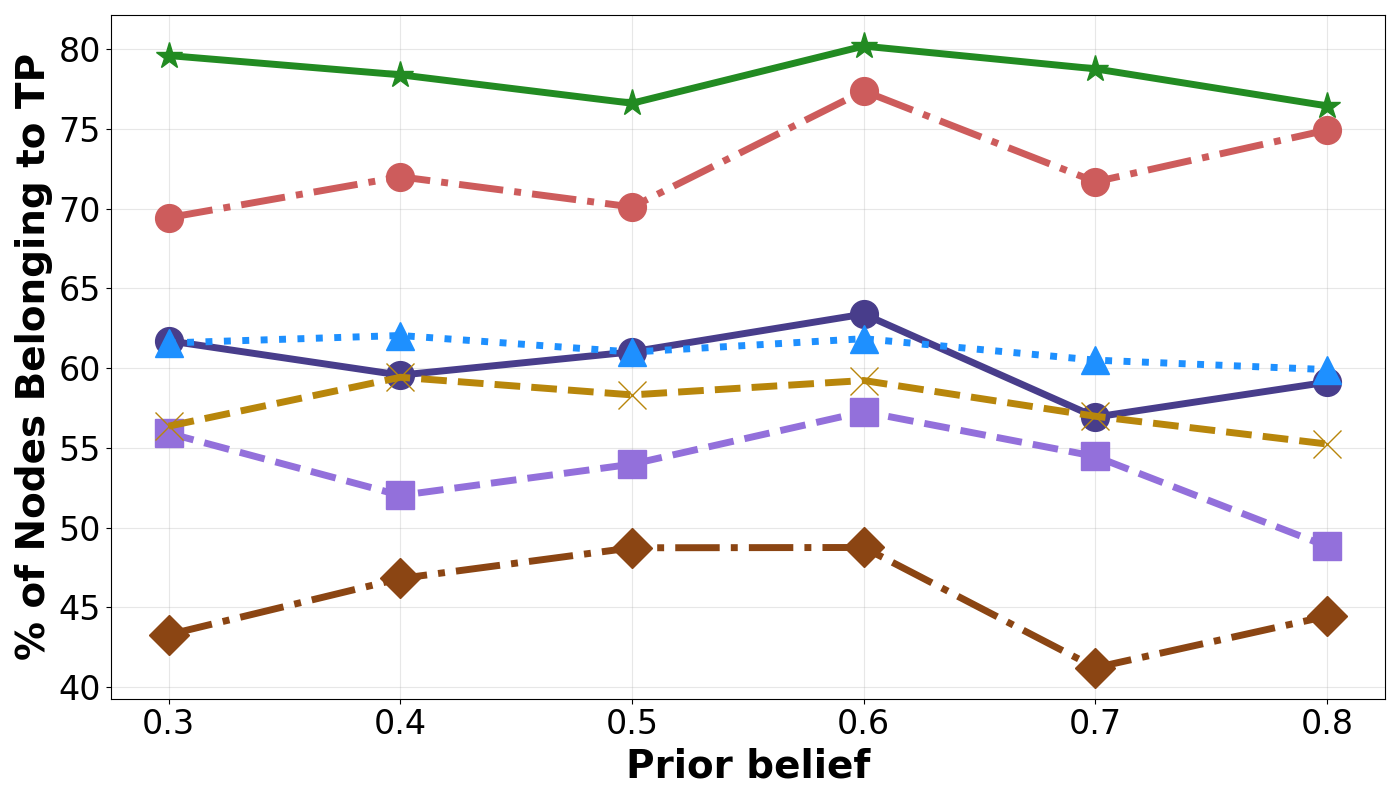} \label{fig:drl-drl-base-rate-sens-ds1}}
    \vspace{-2mm}
    
    \caption{TP's influence across EE strategies when both TP and FP use DRL for seed selection in the URV Email Network. \#IP is TP's information propagation count, `NO' is network observability (\%), and PB is users' prior belief in true information ($a$). Except (c), we set $a=0.5$.}
\label{fig:drl-sens-ee}
\vspace{-6mm}
\end{figure*}

{\bf Sensitivity Analyses Under Various EE Strategies:}  
Fig.~\ref{fig:drl-sens-ee} examines the effects of true information propagation, network observability, and users' prior beliefs on EE strategies. Both TP and FP employ DRL-based agents for seed selection, allowing evaluation against ``smart" opponents. Due to space constraints, results for the URV Email dataset are shown, while analyses for FB networks appear in Appendix B.3 of the supplement.

Fig.~\ref{fig:drl-drl-ee-IP-sens-ds1} shows that all EE strategies gain influence as information propagations (IP) increase from 1 to 3, confirming that additional propagation enhances effectiveness. Fig.~\ref{fig:drl-drl-PON-sens-ds1} reveals a significant boost for all strategies as network observability rises from 0.7 to 1, underscoring their reliance on visibility. VAC-EE and VD-EE improve steeply, excelling in well-observed networks. Fig.~\ref{fig:drl-drl-base-rate-sens-ds1} explores users' prior beliefs. As noted in Section~\ref{subsec:cim-PA}, high user activity mitigates belief influence. Still, VAC-EE and VD-EE outperform others, benefiting from true information acceptance and low uncertainty. DIS-EE and ENT-EE remain stable, suggesting greater dependence on network structure than user biases.

\subsection{Uncertainty Analyses of DRL-based Seed Node Selection Decision-Making}
This section evaluates agent performance by analyzing vacuity, dissonance, and entropy estimates at each decision point. For brevity, we present average values across all steps using the URV Email dataset, with analyses of the Facebook and Facebook Page Network datasets in Appendices B.4 and B.5 of the supplement.

\begin{figure*}[htb]
  \centering
  \subfigure{
    \includegraphics[width=0.75\textwidth, height=0.025\textwidth]{./Figs/legend-bar-schemes.png}}
    \setcounter{subfigure}{0}
    \vspace{-3mm}    
    
  \subfigure[Vacuity]{
    \includegraphics[width=0.3\textwidth, height=0.2\textwidth]{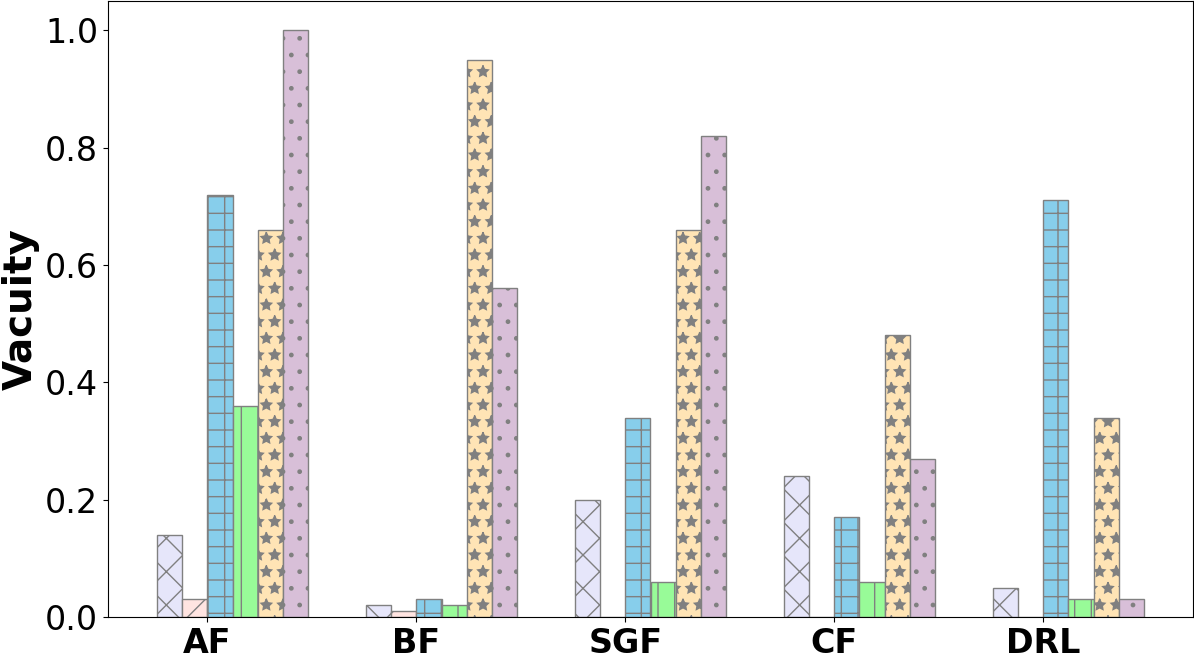}\label{fig:T-drl-F-schemes-vary-strategy_vac}}
  \subfigure[Dissonance]{
    \includegraphics[width=0.3\textwidth, height=0.2\textwidth]{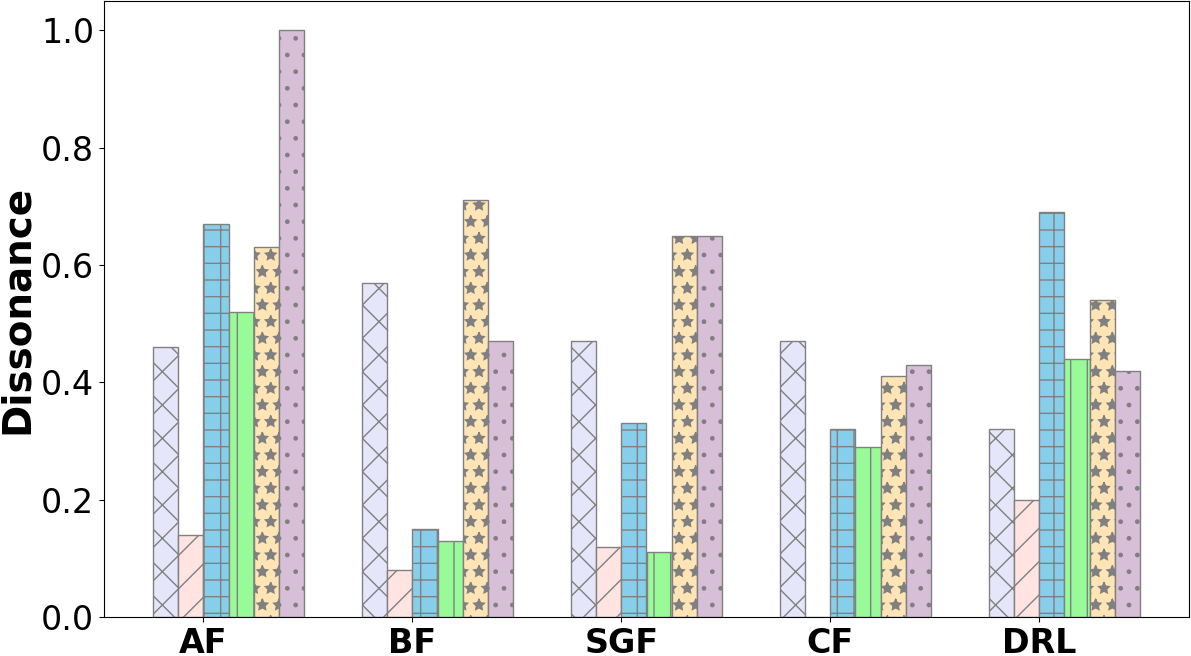} \label{fig:T-drl-F-schemes-vary-strategy_diss}}
  \subfigure[Entropy]{
    \includegraphics[width=0.3\textwidth, height=0.2\textwidth]{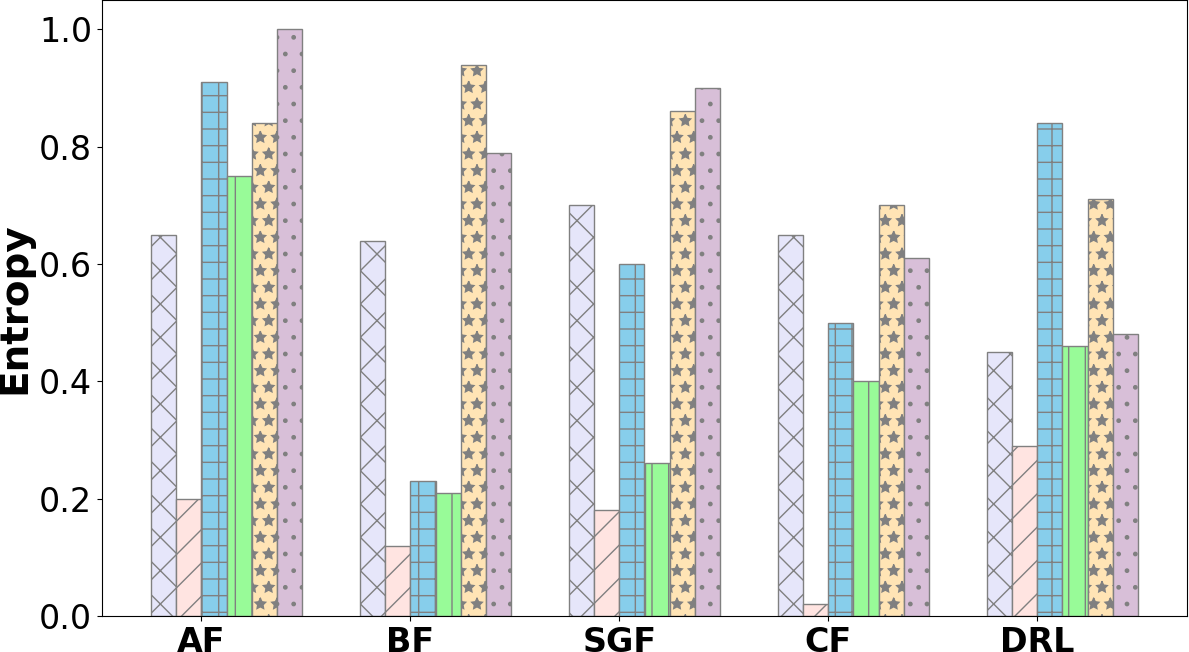} \label{fig:T-drl-F-schemes-vary-strategy_ent}}
    \vspace{-2mm}
    
    \caption{TP agent's uncertainty measurement under various CIM algorithms in the URV email network. 
    }
\label{fig:uncertainty-T-drl-F-vary-schemes-strategy}
\end{figure*}

{\bf Uncertainty Analyses of DRL-based Seed Node Selection Decision-Making Under Various CIM Schemes:}
First, we analyze vacuity, dissonance, and entropy across CIM schemes, with results in Fig.~\ref{fig:uncertainty-T-drl-F-vary-schemes-strategy}. When FP employs AF or BF, no significant correlation emerges between CIM scheme performance and uncertainty measures. Under AF, all schemes reach nearly 100\% influence regardless of uncertainty. With BF, despite higher vacuity, dissonance, and entropy, STORM and C-STORM outperform others, indicating that confidence in decisions does not guarantee better performance. Effectiveness depends on the opponent’s strategy. AF is ineffective in highly active OSNs, while DRIM-based schemes struggle against BF, unlike STORM and C-STORM.  However, under SGF, CF, or DRL, a clear relationship appears: lower performance corresponds to higher vacuity. For SGF and CF, STORM and C-STORM exhibit higher vacuity and lower performance, indicating uncertainty in node selection, which hinders effectiveness.

Comparing DRIM-A-ER and DRIM-A-VDEE under DRL, the higher-performing scheme consistently shows lower vacuity and dissonance. A similar trend holds between DRIM-NA-ER and DRIM-NA-VDEE, as well as STORM and C-STORM. Within the same framework, performance improves with greater decision confidence, reinforcing the impact of uncertainty-aware EE algorithms.  DRIM-A-VDEE maintains the lowest vacuity, dissonance, and entropy across all FP strategies while achieving superior performance, indicating an optimized balance between exploration and exploitation.

\begin{figure*}[htb]
  \centering
  \subfigure{
    \includegraphics[width=0.75\textwidth, height=0.025\textwidth]{./Figs/legend-bar-ee.png}}
    \setcounter{subfigure}{0}
    
  \subfigure[Vacuity]{
    \includegraphics[width=0.3\textwidth, height=0.2\textwidth]{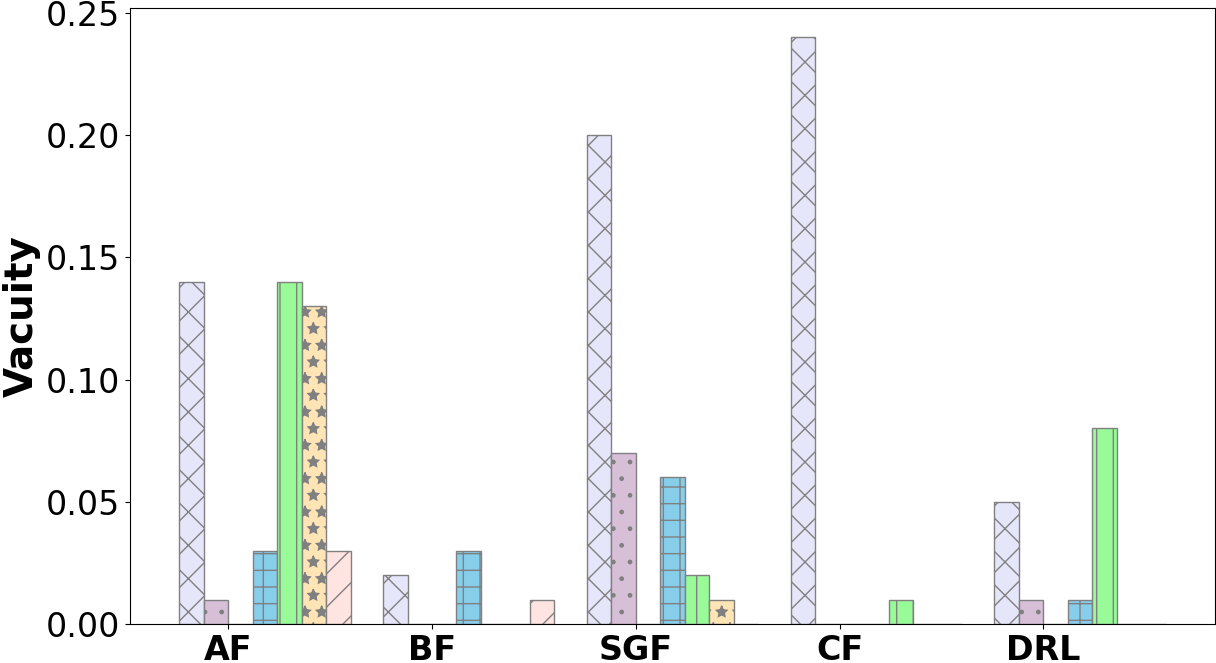}\label{fig:T-drl-F-ee-vary-strategy_vac}}
  \subfigure[Dissonance]{
    \includegraphics[width=0.3\textwidth, height=0.2\textwidth]{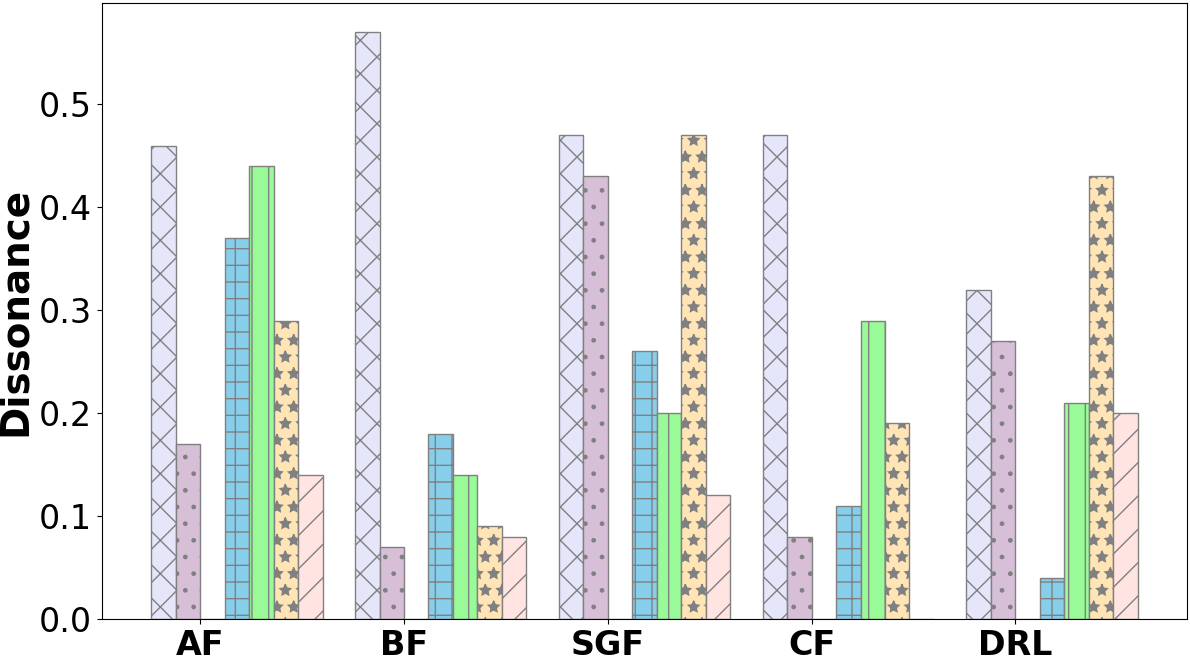} \label{fig:T-drl-F-ee-vary-strategy_diss}}
\subfigure[Entropy]{
    \includegraphics[width=0.3\textwidth, height=0.2\textwidth]{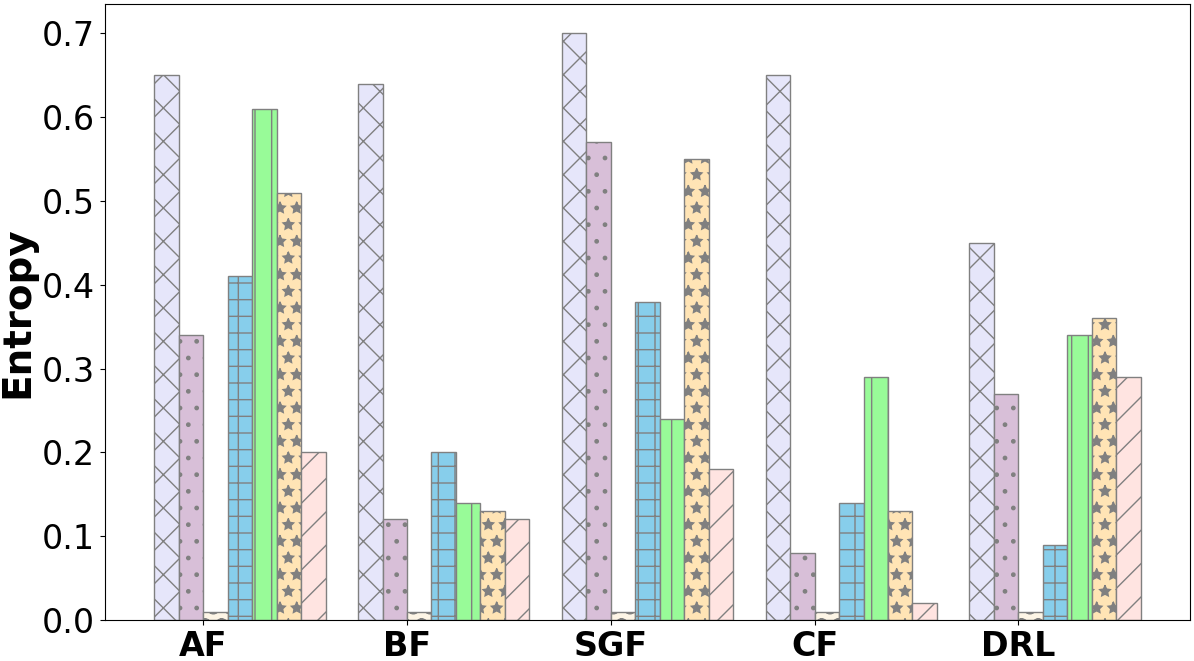} \label{fig:T-drl-F-ee-vary-strategy_ent}}
    \vspace{-2mm}
    
    \caption{TP agent's uncertainty measurement under various EE strategies in URV email dataset. 
    }
\label{fig:uncertainty-T-drl-F-vary-EE-strategy}
\end{figure*}

{\bf Uncertainty Analyses of DRL-Based Seed Node Selection Under Various EE Strategies:}  
Fig.~\ref{fig:uncertainty-T-drl-F-vary-EE-strategy} evaluates uncertainty estimates in DRL-based seed selection across EE strategies. Strategies with lower vacuity generally perform better, as more information enables wiser node selection. For instance, ER-EE exhibits high vacuity and lower influence when FP employs SGF or CF. However, since all EE strategies already maintain low vacuity, a moderate level is not necessarily detrimental. VAC-EE, despite having the highest vacuity, achieves the best performance when FP uses DRL, suggesting that beyond a certain confidence threshold, reducing uncertainty alone does not guarantee better performance.  Higher dissonance and entropy often correlate with varied performance. Strategies with high dissonance, such as ER-EE and DIS-EE, face conflicting information, complicating decision-making and influence spread, leading to fluctuating performance.

\subsection{Running Time Analyses of CIM Algorithms} 
\label{subsec:complexity-analysis}
Table~\ref{tab:running_time} demonstrates that DRIM-based schemes outperform C-STORM and STORM in terms of running time. We measured the time required to select a node and complete one round of information propagation, averaging these times over 5,000 cycles of node selection and propagation. While larger datasets naturally take longer, the trend within the same dataset remains clear. Specifically, in the URV Email network, DRIM-based strategy, especially DRIM-A-VDEE operates 6.8\% faster than C-STORM and 44.5\% faster than STORM. This gap widens to 47.8\% and 77.1\% respectively in the denser Facebook network (FBN), and 15.1\% and 61.6\% in the larger Facebook Page network (FBPN). Consequently, our DRIM-based approach proves to be significantly quicker, particularly in denser networks. Moreover, when comparing DRIM-A-VDEE with DRIM-A-ER, the incorporation of an uncertainty-based exploration-exploitation trade-off slightly increases running times, indicating that the efficiency impact of incorporating uncertainty measurement is acceptable.

\begin{table}[t]
\centering
\caption{\sc \centering Simulation Running Time (in sec.) of the CIM Algorithms} \label{tab:running_time}
\vspace{-3mm}
\scriptsize 
\begin{tabular}{|P{0.9cm}|P{1.1cm}|P{1cm}|P{1cm}|P{1.2cm}|P{0.9cm}|}
\hline
Dataset & DRIM-A-VDEE & DRIM-A-ER & DRIM-NA-ER & C-STORM & STORM \\
\hline
URV~\cite{rossi2015network_dataset_c} & 0.352 & 0.354 & 0.308 & 0.380 & 0.638\\
\hline
FBN~\cite{leskovec2012facebook} & 11.248 & 10.912 & 10.19 & 20.890 & 47.624 \\
\hline
FBPN~\cite{rozemberczki2021facebookpage} & 28.400 & 24.612 & 19.546 & 29.006 & 64.092\\

\hline
\end{tabular}
\vspace{-5mm}
\end{table}

\section{Conclusions \& Future Work}

We highlight the {\bf key contributions} of our approach as follows.  This paper comprehensively analyzes opinion models, CIM schemes, and uncertainty-aware EE strategies for combating misinformation. We introduce DRIM-based CIM schemes that integrate uncertainty-aware decision-making to enhance influence maximization. A comparative evaluation against state-of-the-art methods demonstrates the superiority of DRIM-based schemes in adaptability, efficiency, and robustness across various network conditions.

Our study reveals the following {\bf key findings}:First, our study highlights the {effectiveness of opinion models}, with UOM achieving the highest influence, emphasizing the importance of timely and accurate information. High vacuity (\( u \) values) increased receptiveness, while HOM and NOM underscored the challenge of countering early false information exposure. Second, {DRIM-based schemes demonstrated superiority}, consistently outperforming STORM and C-STORM, particularly against SGF, CF, and DRL strategies, showcasing their adaptability and efficiency. Third, we observed the {impact of key parameters}, where increased propagations improved influence spread, though DRIM-based schemes remained effective even with fewer propagations. Higher network observability significantly boosted performance, while prior beliefs had minimal impact due to frequent information exchanges. Fourth, {uncertainty measures correlated with performance}, as lower uncertainty consistently led to better outcomes. DRIM-A-VDEE exhibited the lowest uncertainty and highest influence, proving its robustness. Finally, in terms of {computational efficiency}, DRIM-based schemes ran significantly faster than STORM and C-STORM, particularly in dense networks. The exploration-exploitation trade-off in DRIM-A-VDEE had a negligible impact on runtime, ensuring efficiency.

As {future research}, we aim to enhance false information mitigation through adaptive countermeasures that adjust to evolving network conditions. To this end, first, scaling DRIM-based frameworks to larger real-world networks will validate their effectiveness beyond simulations. Second, incorporating heterogeneous factors like user behavior, trust, and temporal dynamics will improve real-world applicability. Third, strengthening robustness against adversarial misinformation will enhance model reliability. Lastly, integrating multi-agent and reinforcement learning will improve adaptability in dynamic environments, making influence maximization more effective.

\section*{Acknowledgment}
This work is partly supported by NSF under grants 2107449, 2107450, and 2107451.

\section*{Appendix}
\appendices
\section{Summary of Experimental Settings}

Table~\ref{tab:param-meaning-default} lists the key design parameters, their definitions, and default settings used in the experiments, as detailed in Section~6.

\begin{table}[h]
\footnotesize
    \caption{\sc \centering Key Parameters and Default Values}
    \label{tab:param-meaning-default}
    \centering
    \begin{tabular}{|P{0.7cm}|P{5cm}|P{1.1cm}|}\hline
    Param. &  Meaning & Def. Val.\\
    \hline
    $T$ & Number of information propagation rounds & 50 \\
    \hline
    $T_v$ & Vacuity threshold in UOM  & 0.01\\
    \hline
    $T_d$ & Dissonance threshold in UOM & 0.6\\
    \hline
    $lr_a$ & Learning rate in the actor-network & 0.0003 \\
    \hline
    $lr_c$ & Learning rate in the critic network & 0.001 \\
    \hline
    $K_{epo}$ & The number of epochs used in one PPO update & 80 \\
    \hline
    $\epsilon$ & Clipping parameter value used in PPO & 0.2 \\
    \hline
    $\gamma$ & Discount factor in DRL's reward function & 0.95 \\
    \hline
    $p^{TP}$ & Number of true information propagation by the true party's seed nodes & 1 \\
    \hline
    $p^{FP}$ & Number of false information propagation by the false party's seed nodes  & 1 \\
    \hline
    $d$ & Value used in the Subgreedy strategy & 2 \\
    \hline
    $a$ & Prior belief of legitimate user & 0.5 \\
    \hline
    \end{tabular}
\end{table}
\vspace{-5mm}
\section{Additional Experimental Results} \label{sec:additional-results}

\subsection{Comparative Analyses of TP's Uncertainty Across CIM Algorithms}

Table~\ref{tab:uncertainty-cim-table} summarizes TP's uncertainty across CIM algorithms with different opinion models (OMs) in the Email Dataset, corresponding to Figs.~3(a) and 7 in the main paper for comparative analysis.  The observed key trends are: (1) DRIM-A-VD-EE consistently achieving the lowest vacuity, indicating stronger opinion certainty; (2) DRIM-A-ER maintaining low uncertainty but with higher entropy than DRIM-A-VD-EE; (3) C-STORM and STORM exhibiting significantly higher uncertainty, particularly in vacuity and entropy, suggesting weaker opinion stabilization; and (4) DRIM-based approaches outperforming STORM variants, reinforcing the benefits of an SL-based dynamic opinion model.

\begin{table*}
\centering
\caption{\sc \centering True Party (TP)'s Uncertainty Under Various CIM Algorithms with Different OM}
\label{tab:uncertainty-cim-table}
\begin{tabular}{|c|c|c|c|c|c|}
\hline
\textbf{Scheme / Uncertainty}  & \textbf{AF} & \textbf{BF} & \textbf{SGF} & \textbf{CF} & \textbf{DRL} \\ \hline
DRIM-A-ER/vacuity & 0.14 & 0.02 & 0.2 & 0.24 & 0.05  \\
DRIM-A-ER/dissonance & 0.46 & 0.57 & 0.47 & 0.47 & 0.32  \\ 
DRIM-A-ER/entropy & 0.65 & 0.64 & 0.7 & 0.65 & 0.45  \\
DRIM-A-ER/performance & 1077.74 & 463.79 & 827.05 & 374.01 & 691.37 \\  \hline
DRIM-A-VD-EE/vacuity & \textbf{0.03} & \textbf{0.01} & \textbf{0} & \textbf{0} & \textbf{0} \\
DRIM-A-VD-EE/dissonance & \textbf{0.14} & \textbf{0.08} & 0.12 & \textbf{0} & \textbf{0.2}  \\
DRIM-A-VD-EE/entropy & \textbf{0.2} & \textbf{0.12} & \textbf{0.18} & \textbf{0.02} & \textbf{0.29}  \\ 
DRIM-A-VD-EE/performance & 1077.64 & 646.94 & 960.66 & \textbf{581.9} & \textbf{794.33}  \\\hline
DRIM-NA-ER/vacuity & 0.72 & 0.03 & 0.34 & 0.17 & 0.71 \\ 
DRIM-NA-ER/dissonance & 0.67 & 0.15 & 0.33 & 0.32 & 0.69  \\  
DRIM-NA-ER/entropy & 0.91 & 0.23 & 0.6 & 0.5 & 0.84  \\
DRIM-NA-ER/performance & \textbf{1078.26} & 529.53 & 959.63 & 525.7 & 576.47 \\ \hline
DRIM-NA-VD-EE/vacuity & 0.36 & 0.02 & 0.06 & 0.06 & 0.03 \\
DRIM-NA-VD-EE/dissonance & 0.52 & 0.13 & \textbf{0.11} & 0.29 & 0.44  \\
DRIM-NA-VD-EE/entropy & 0.75 & 0.21 & 0.26 & 0.4 & 0.46 \\ 
DRIM-NA-VD-EE/performance & 1077 & 547.16 & \textbf{961.48} & 439.94 & 704.01  \\\hline
C-STORM/vacuity & 1 & 0.56 & 0.82 & 0.27 & 0.03 \\ 
C-STORM/dissonance & 1 & 0.47 & 0.65 & 0.43 & 0.42 \\  
C-STORM/entropy & 1 & 0.79 & 0.9 & 0.61 & 0.48  \\ 
C-STORM/performance & 1070.748 & \textbf{1082.53} & 524.39 & 229.15 & 551.44  \\
\hline
STORM/vacuity & 0.66 & 0.95 & 0.66 & 0.48 & 0.34 \\ 
STORM/dissonance & 0.63 & 0.71 & 0.65 & 0.41 & 0.54  \\  
STORM/entropy & 0.84 & 0.94 & 0.86 & 0.7 & 0.71  \\
STORM/performance & 1072.45 & 1058.80 & 461.19 & 63.61 & 424.09  \\
\hline
\end{tabular}
\end{table*}

\subsection{Sensitivity Analyses Under Various CIM Schemes}
\label{sec:sen-cim-2-datasets}

This section analyzes how true information propagations, network observability, and users' prior beliefs impact CIM algorithm performance on Facebook and Facebook Page datasets. Both TP and FP use DRL-based agents for seed selection, allowing evaluation against ``smart" opponents.

\begin{figure*}[htb]
  \centering
  \subfigure{
    \includegraphics[width=0.75\textwidth, height=0.025\textwidth]{./Figs/legend-sens-schemes.png}}
    \setcounter{subfigure}{0}
    \vspace{-3mm}    
    
  \subfigure[Varying \# of IP in FBN]{
    \includegraphics[width=0.3\textwidth, height=0.2\textwidth]{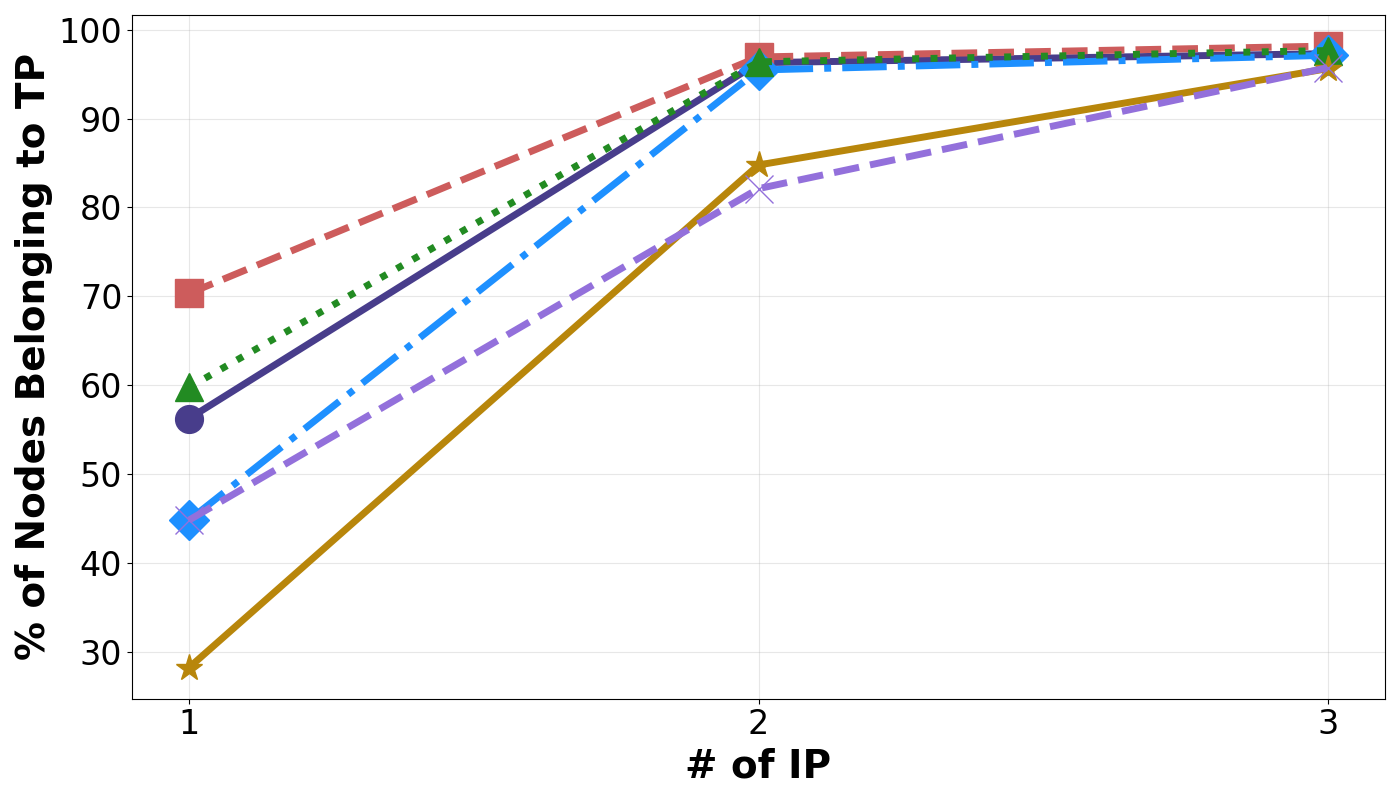}\label{fig:drl-drl-IP-schemes-sens-ds2}} 
  \subfigure[Varying \% of NO in FBN]{
    \includegraphics[width=0.3\textwidth, height=0.2\textwidth]{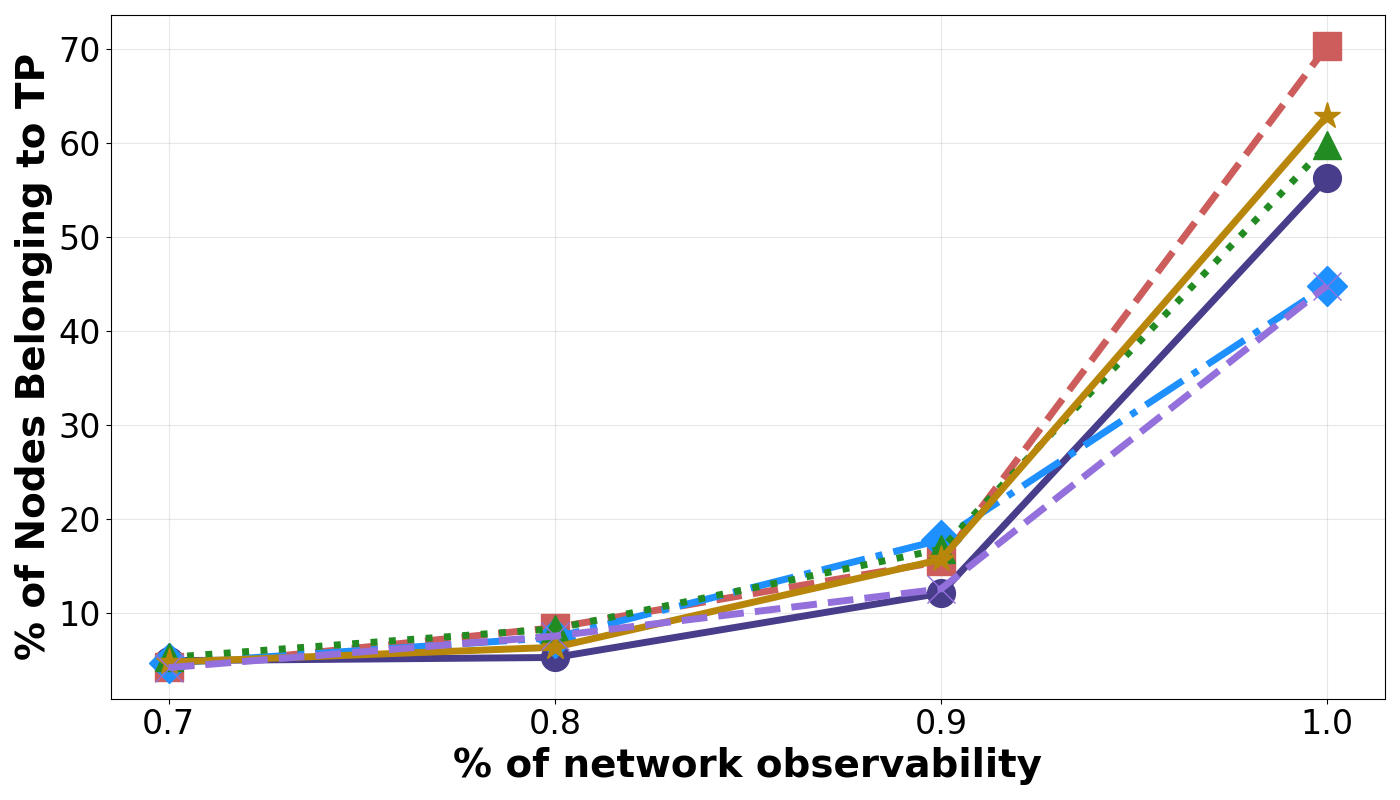} \label{fig:drl-drl-PON-schemes-sens-ds2}}
  \subfigure[Varying users' PB in FBN]{
    \includegraphics[width=0.3\textwidth, height=0.2\textwidth]{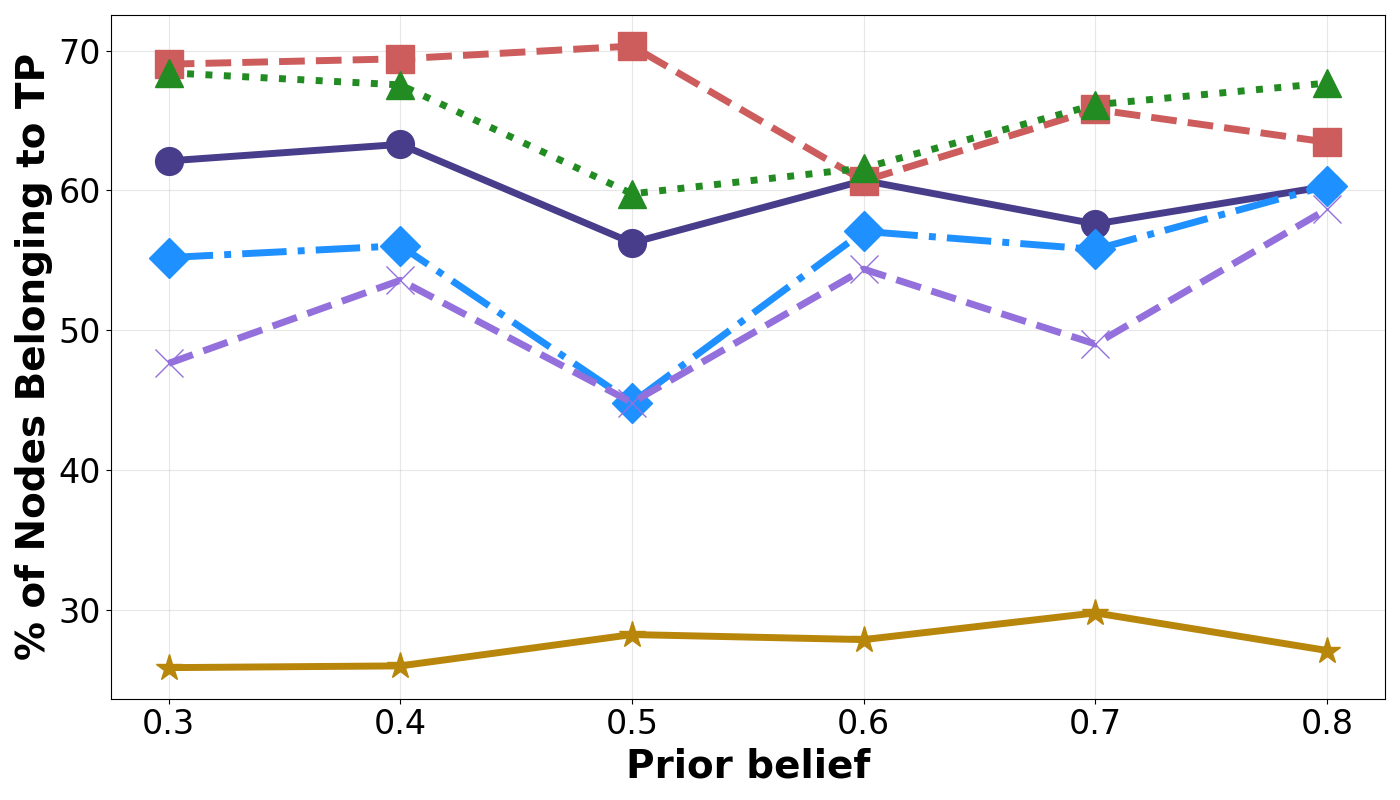} \label{fig:drl-drl-base-rate-schemes-sens-ds2}}
    \vspace{-2mm}

 \subfigure[Varying \# of IP in FBPN]{
    \includegraphics[width=0.3\textwidth, height=0.2\textwidth]{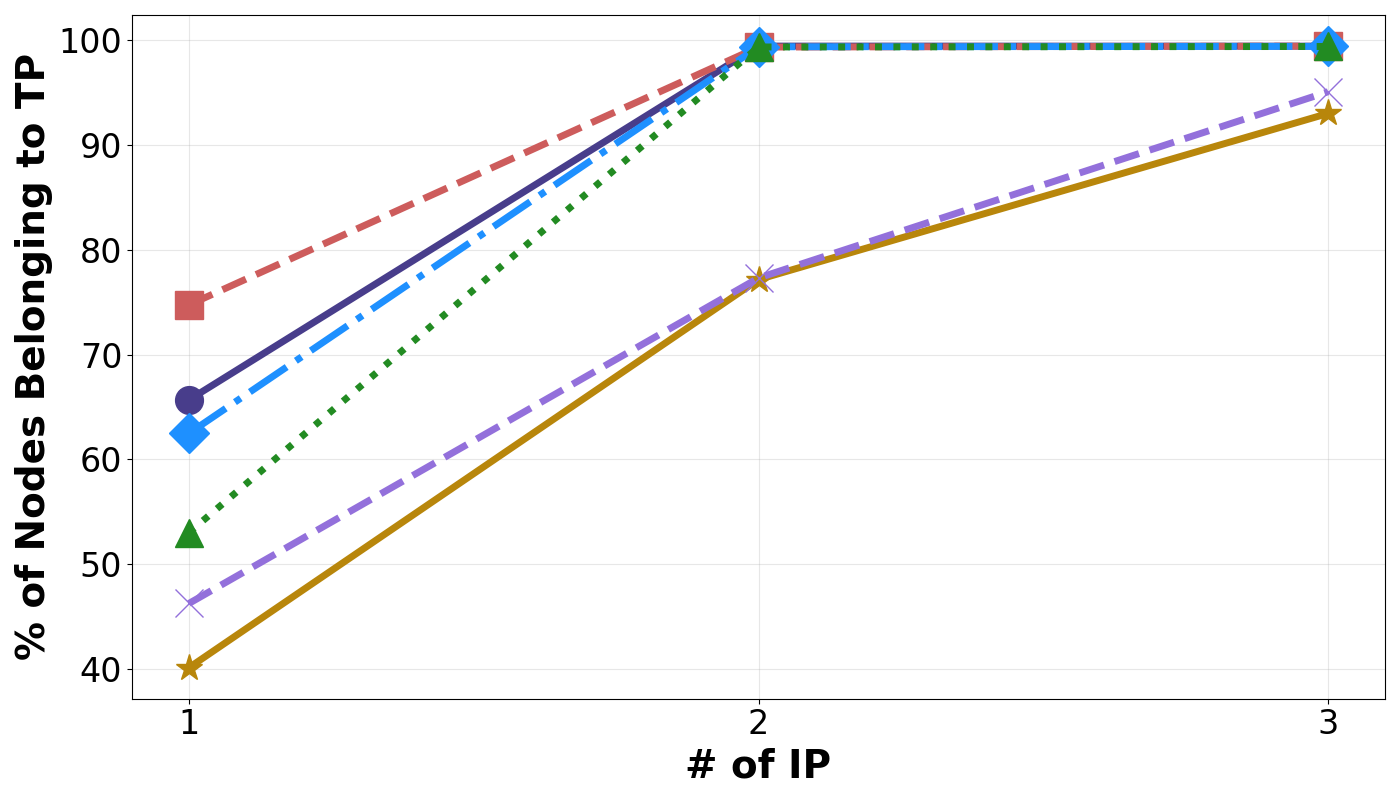}\label{fig:drl-drl-IP-schemes-sens-ds3}} 
  \subfigure[Varying \% of NO in FBPN]{
    \includegraphics[width=0.3\textwidth, height=0.2\textwidth]{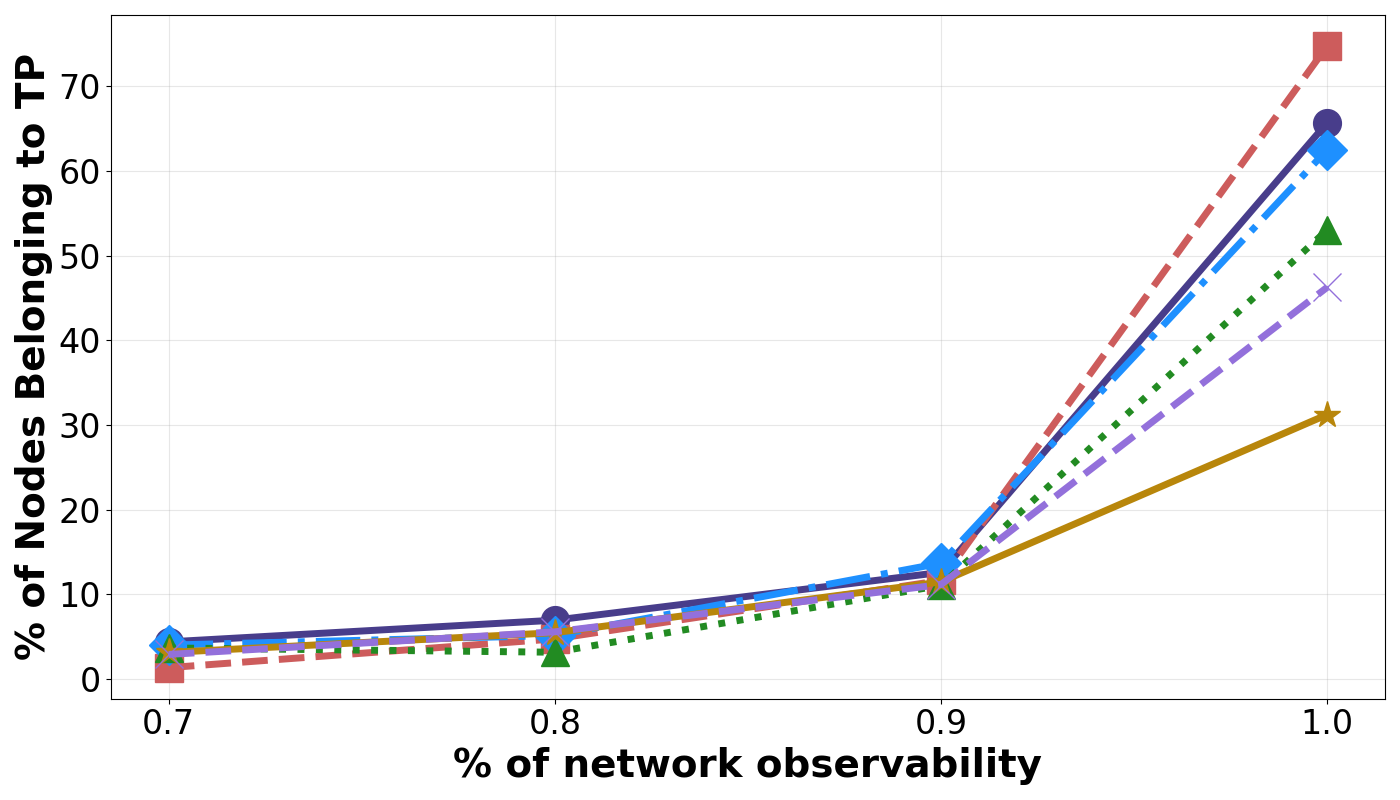} \label{fig:drl-drl-PON-schemes-sens-ds3}}
  \subfigure[Varying users' PB in FBPN]{
    \includegraphics[width=0.3\textwidth, height=0.2\textwidth]{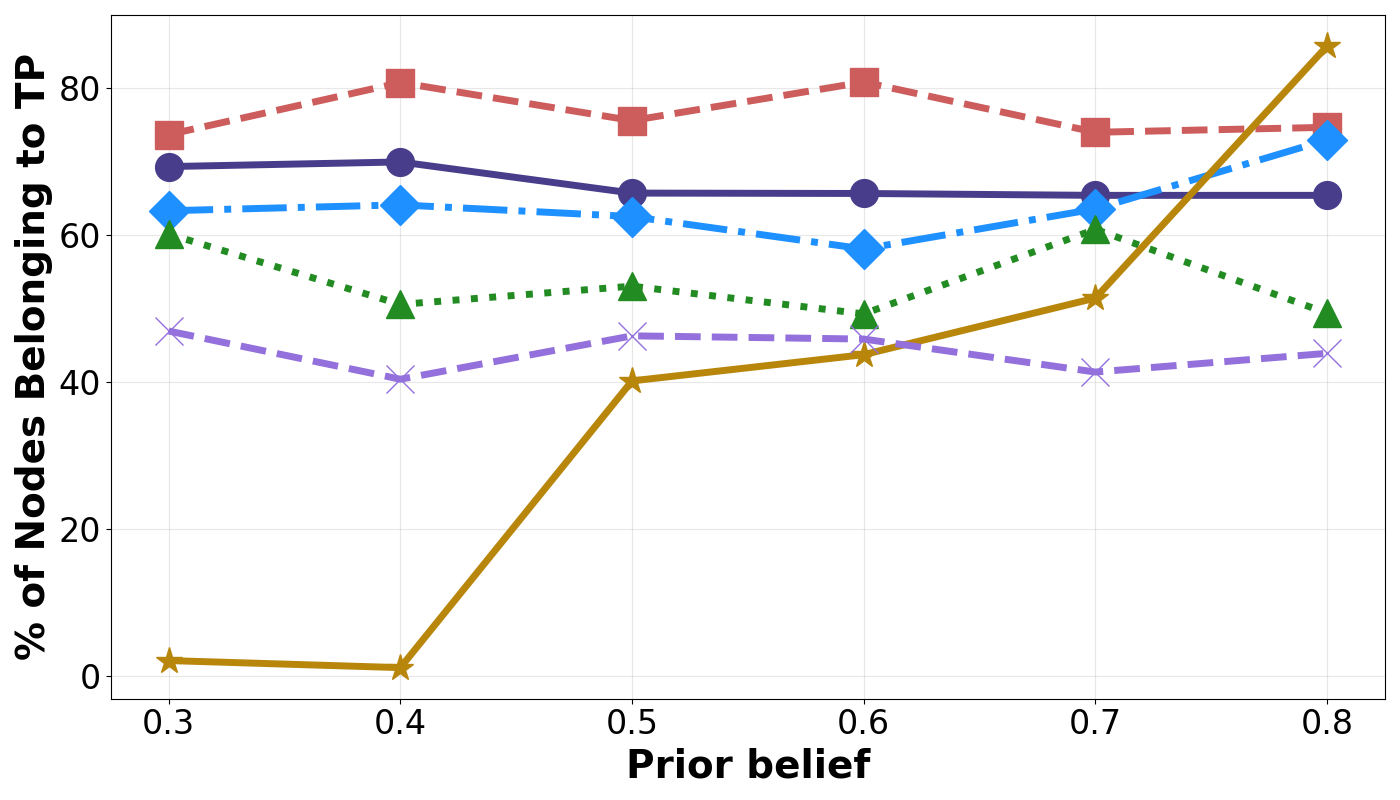} \label{fig:drl-drl-base-rate-schemes-sens-ds3}}
    \vspace{-2mm}
    
    \caption{TP's influence across CIM algorithms when both TP and FP use DRL for seed selection in the Facebook (FBN) and Facebook Page (FBPN) networks, as discussed in Section~6.3 (Network Datasets) of the main paper. \#IP denotes TP's number of information propagations, `NO' represents network observability (\%), and PB indicates users' prior belief in true information ($a$). Unless stated otherwise in (c) and (f), we set $a=0.5$. }
\label{fig:drl-sens-cim-2}
\end{figure*}

Figs.~\ref{fig:drl-drl-IP-schemes-sens-ds2} and \ref{fig:drl-drl-IP-schemes-sens-ds3} examine the effect of increasing TP's information propagations (IPs) from 1 to 3 while FP remains at a single propagation per seed. As expected, all schemes gain more influence with additional IPs, reinforcing that more iterations enhance spread. DRIM-based schemes perform best with a single propagation per round, aligning with Fig.~2(a) in the main paper, effectively countering false information with minimal resources. Beyond two IPs, influence plateaus as TP reaches maximum spread. Unlike results from the URV Email dataset, STORM and C-STORM fail to match DRIM-based schemes at two propagations, indicating their performance is network-dependent, whereas DRIM-based schemes maintain stability across structural variations.

Figs.~\ref{fig:drl-drl-PON-schemes-sens-ds2} and \ref{fig:drl-drl-PON-schemes-sens-ds3} show a clear trend of improved performance across all schemes as network observability increases from 0.7 to 1. Similar to the URV Email dataset, greater visibility significantly enhances performance. When observability is low, all schemes exhibit reduced influence, highlighting the DRL agent’s dependence on network state knowledge. The steep rise from 0.9 to 1 suggests these strategies are highly sensitive to changes in network visibility.

Figs.~\ref{fig:drl-drl-base-rate-schemes-sens-ds2} and \ref{fig:drl-drl-base-rate-schemes-sens-ds3} examine the impact of varying prior beliefs. Ideally, as prior belief in the True party increases (0.3 to 0.8), influence maximization performance should improve. However, this trend is not strongly observed across most schemes. According to Eq.~(3) in the main paper, prior belief primarily affects outcomes when users have nonzero vacuity $u$, allowing biases to shape information acceptance. In highly active networks, users gather substantial evidence throughout propagation, reducing vacuity and diminishing prior belief effects. Consequently, even with varying prior beliefs, differences in influence maximization remain minimal. However, DRIM-based schemes, particularly DRIM-A-VDEE, consistently maintain the highest influence across all prior belief levels.

An exception occurs with STORM in the Facebook Page network. As seen in Fig.~\ref{fig:drl-drl-base-rate-schemes-sens-ds3}, as prior belief increases, the number of nodes aligned with the True party also rises, surpassing all other schemes at prior belief 0.8. However, this result is not necessarily positive. When initial bias strongly affects final decisions, vacuity remains high, indicating insufficient information spread. This exposes a critical weakness of STORM: its inability to facilitate robust information dissemination. While it may exploit initial biases, it lacks reliability in scenarios requiring strong and genuine information exchange.

Overall, DRIM-based schemes exhibit stable performance across varying network conditions, particularly excelling in scenarios with limited information propagation and diverse prior beliefs.

\begin{figure*}[htb]
  \centering
  \subfigure{
    \includegraphics[width=0.75\textwidth, height=0.025\textwidth]{./Figs/legend-sens-ee.png}}
    \setcounter{subfigure}{0}
    \vspace{-3mm}    
    
  \subfigure[Varying \# of IP in FBN]{
    \includegraphics[width=0.3\textwidth, height=0.2\textwidth]{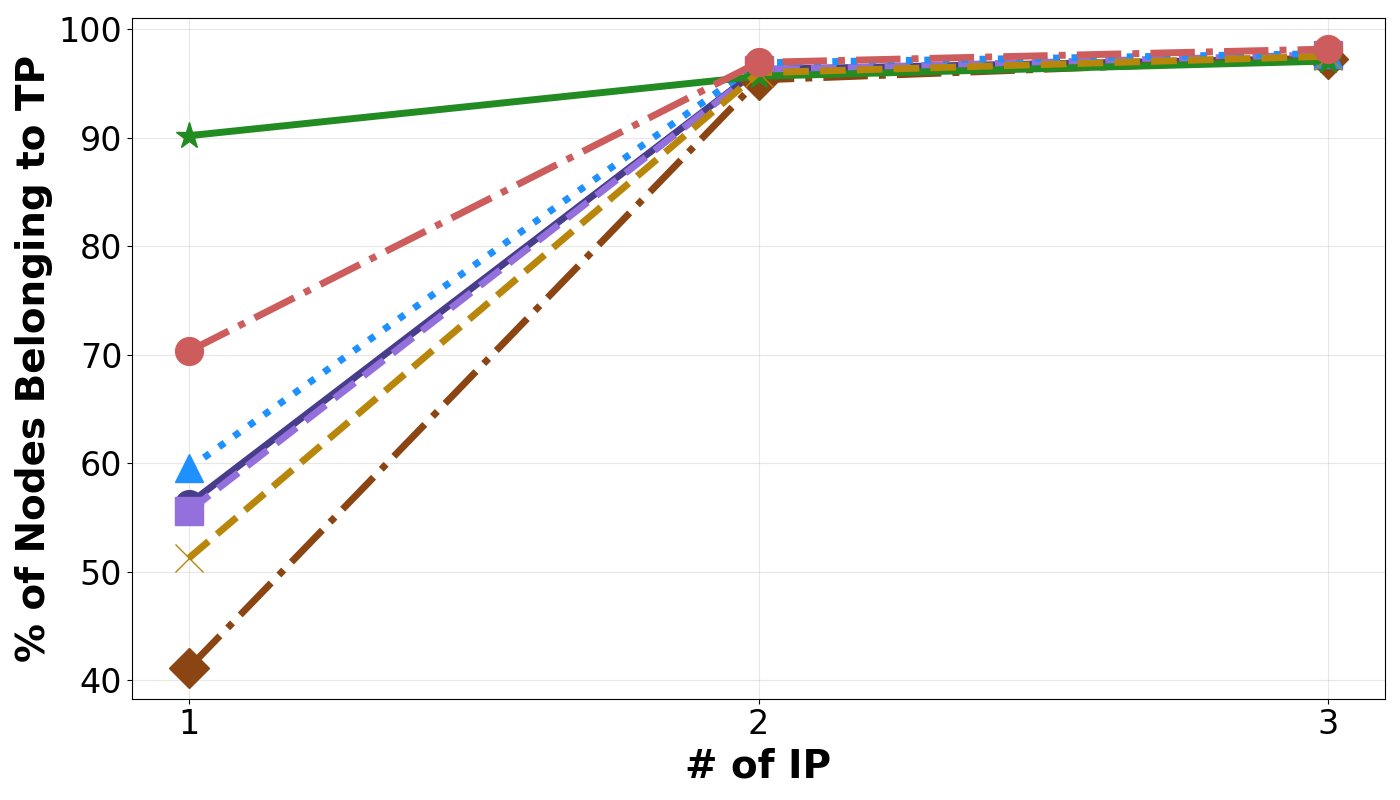}\label{fig:drl-drl-IP-ee-sens-ds2}} 
  \subfigure[Varying \% of NO in FBN]{
    \includegraphics[width=0.3\textwidth, height=0.2\textwidth]{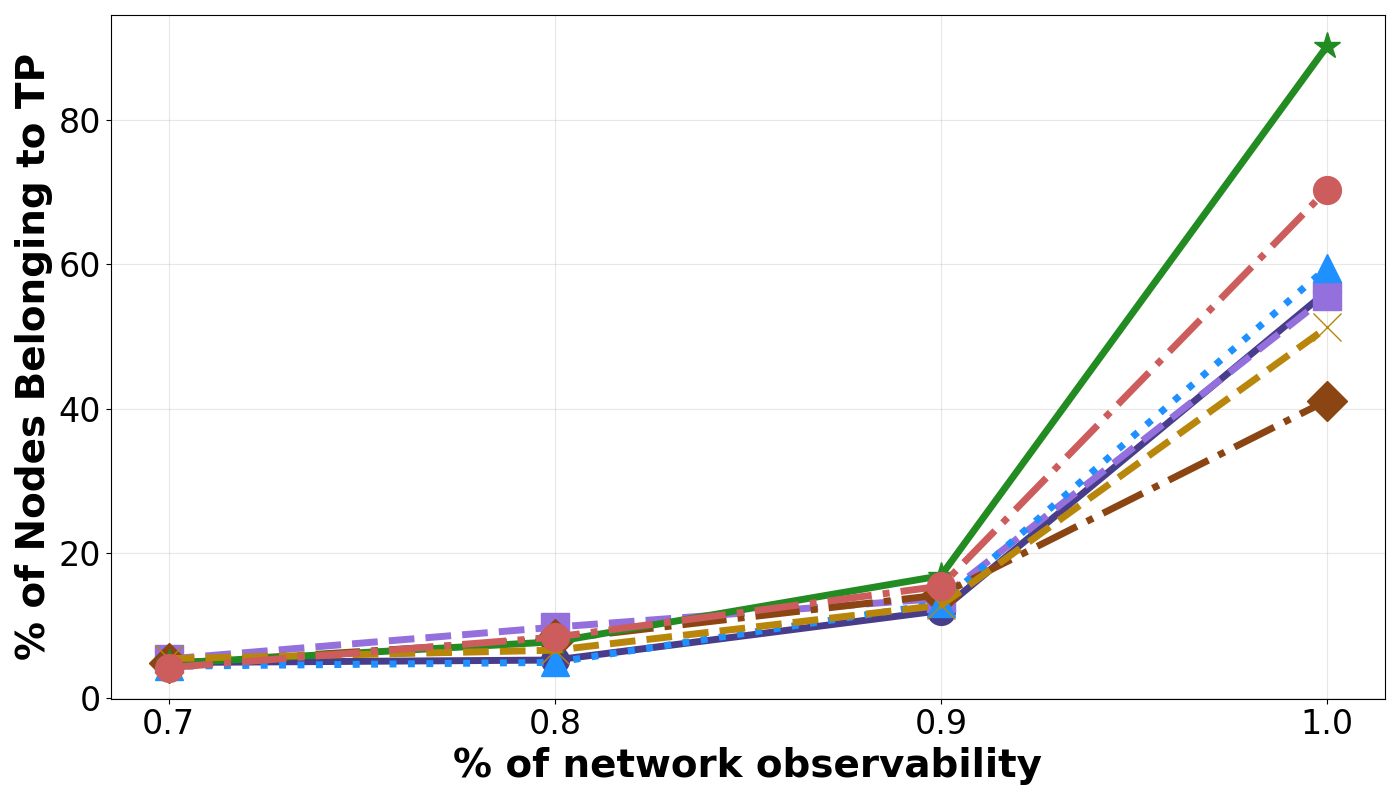} \label{fig:drl-drl-PON-ee-sens-ds2}}
  \subfigure[Varying users' PB in FBN]{
    \includegraphics[width=0.3\textwidth, height=0.2\textwidth]{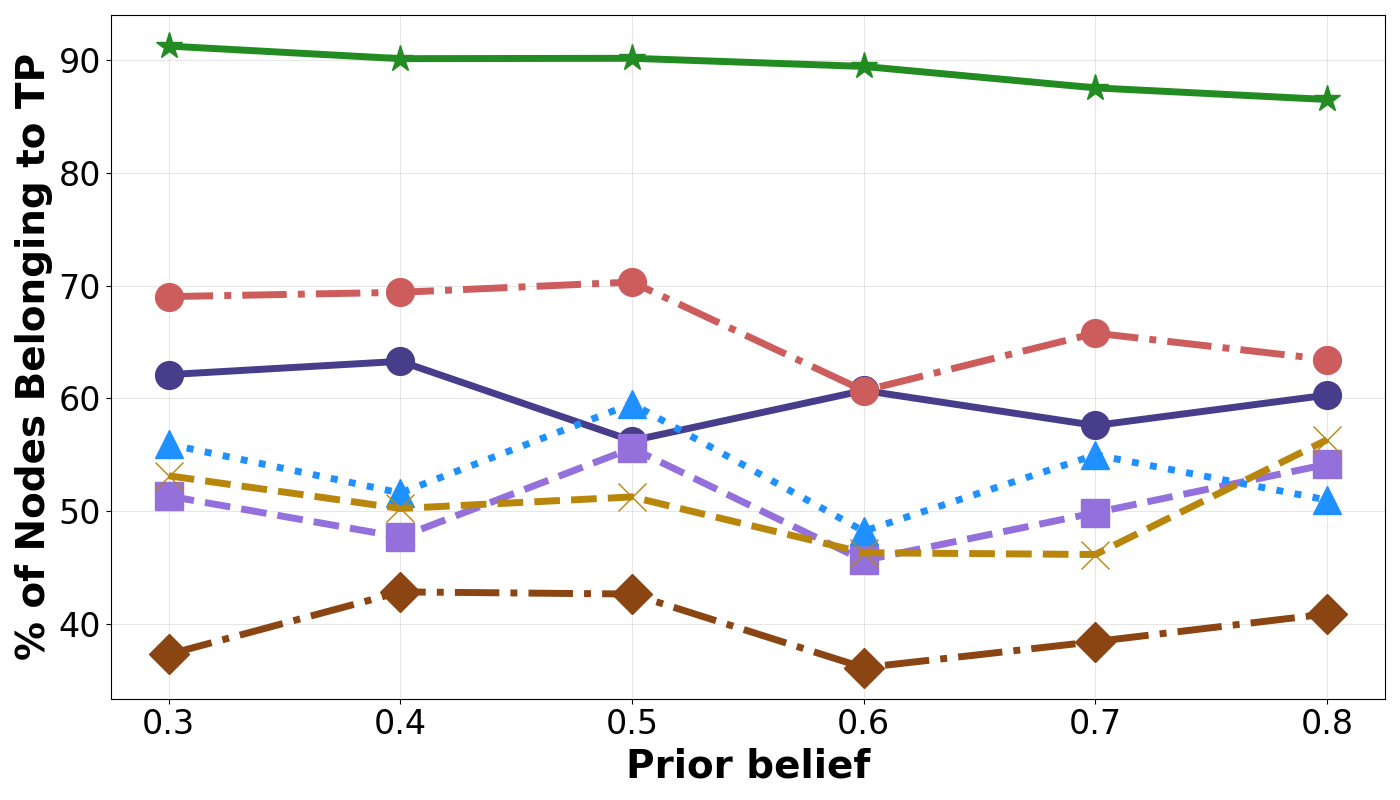} \label{fig:drl-drl-base-rate-ee-sens-ds2}}
    \vspace{-2mm}

   \subfigure[Varying \# of IP in FBPN]{
    \includegraphics[width=0.3\textwidth, height=0.2\textwidth]{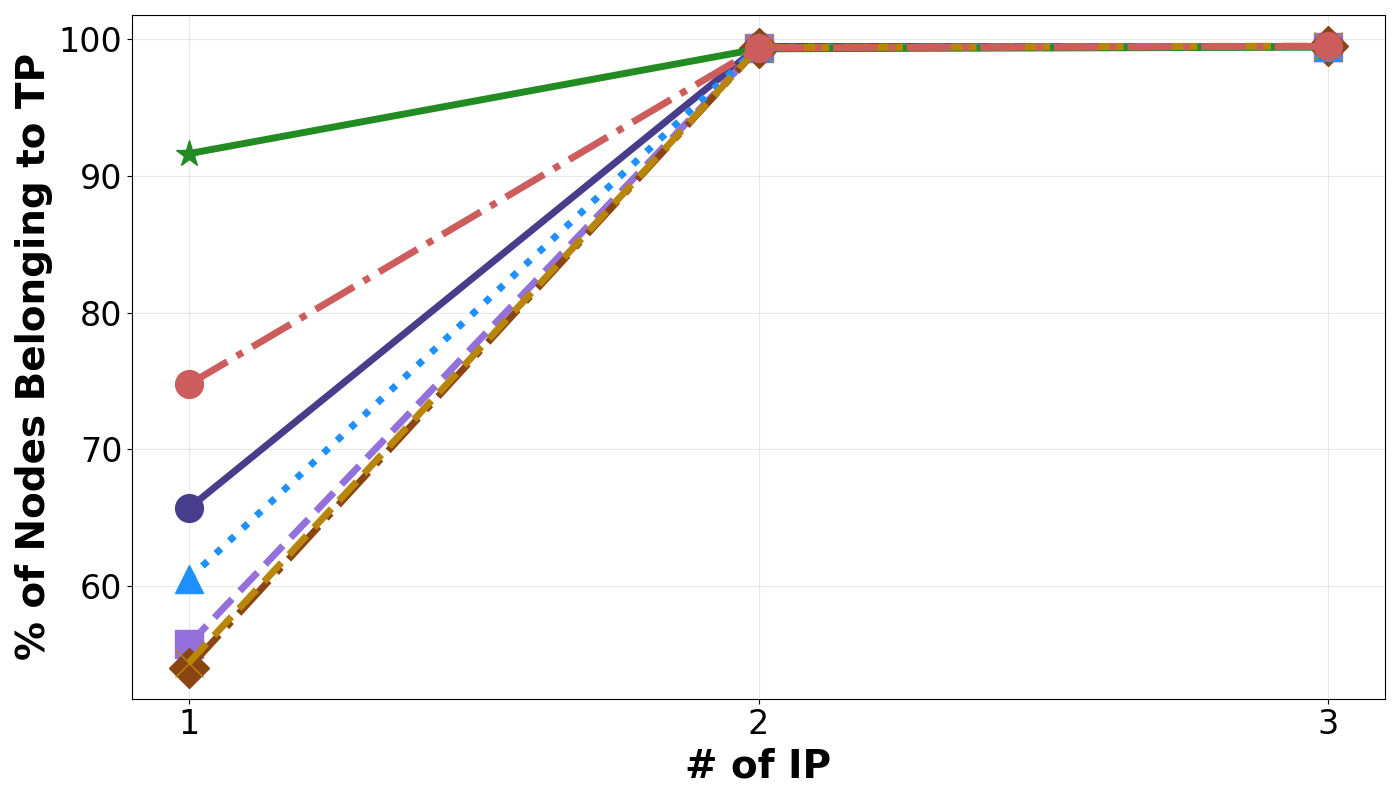}\label{fig:drl-drl-IP-ee-sens-ds3}} 
  \subfigure[Varying \% of NO in FBPN]{
    \includegraphics[width=0.3\textwidth, height=0.2\textwidth]{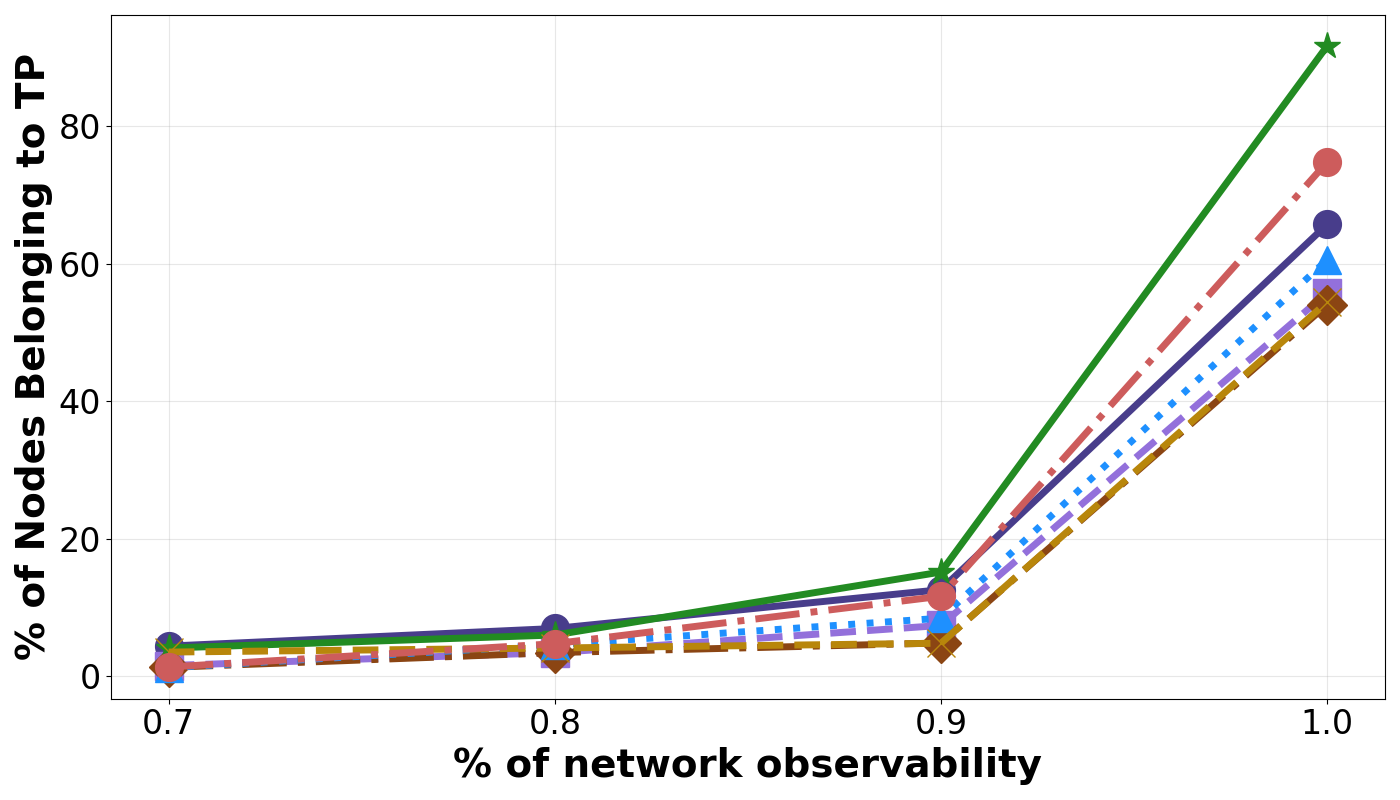} \label{fig:drl-drl-PON-ee-sens-ds3}}
  \subfigure[Varying users' PB in FBPN]{
    \includegraphics[width=0.3\textwidth, height=0.2\textwidth]{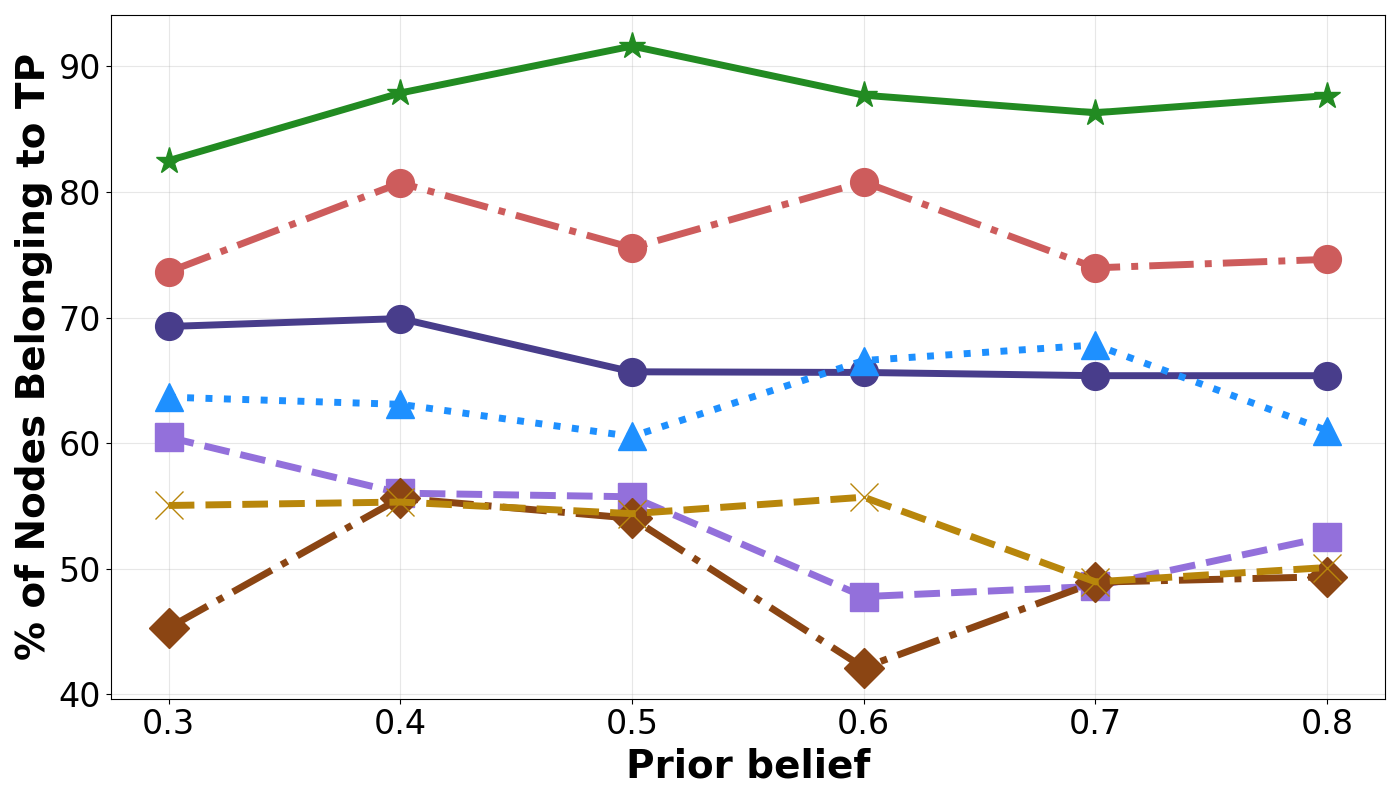} \label{fig:drl-drl-base-rate-ee-sens-ds3}}
    \vspace{-2mm}
    
    \caption{TP's influence across EE strategies when both TP and FP use DRL for seed selection in the Facebook (FBN) and Facebook Page (FBPN) networks, as discussed in Section~6.3 (Network Datasets) of the main paper. \#IP denotes TP's number of information propagations, `NO' represents network observability (\%), and PB indicates users' prior belief in true information ($a$). Unless stated otherwise in (c) and (f), we set $a=0.5$.}
\label{fig:drl-sens-ee-2}
\end{figure*}
\subsection{Sensitivity Analyses of Various EE Strategies Under Facebook and Facebook Page-to-Page Datasets}

We analyze how true information propagations, network observability, and users' prior beliefs impact EE strategy performance. Both TP and FP use DRL-based agents for seed selection, enabling evaluation against \textit{smart} opponents.

Figs.~\ref{fig:drl-drl-IP-ee-sens-ds2} and \ref{fig:drl-drl-IP-ee-sens-ds3} show that all strategies gain influence as the number of information propagation (IP) events increases from 1 to 3, reinforcing that additional rounds enhance effectiveness across different network densities and sizes.  Figs.~\ref{fig:drl-drl-PON-ee-sens-ds2} and \ref{fig:drl-drl-PON-ee-sens-ds3} reveal significant performance improvements as network observability rises from 0.7 to 1, highlighting these strategies' reliance on network visibility. VAC-EE and VD-EE show particularly steep gains, demonstrating their effectiveness in well-observed networks.  Figs.~\ref{fig:drl-drl-base-rate-ee-sens-ds2} and \ref{fig:drl-drl-base-rate-ee-sens-ds3} examine prior belief effects. As observed in Section~7.3, high user activity reduces prior belief influence. However, VAC-EE and VD-EE consistently outperform others across varying beliefs, while DIS-EE and ENT-EE remain stable, relying more on network structure and dynamics than user biases.

\subsection{Uncertainty Analyses in DRL-based Seed Node Selection Decision-Making Under Various CIM Schemes}

\begin{figure*}[htb]
  \centering
  \subfigure{
    \includegraphics[width=0.75\textwidth, height=0.025\textwidth]{./Figs/legend-bar-schemes.png}}
    \setcounter{subfigure}{0}
    
  \subfigure[Vacuity in FBN]{
    \includegraphics[width=0.3\textwidth, height=0.2\textwidth]{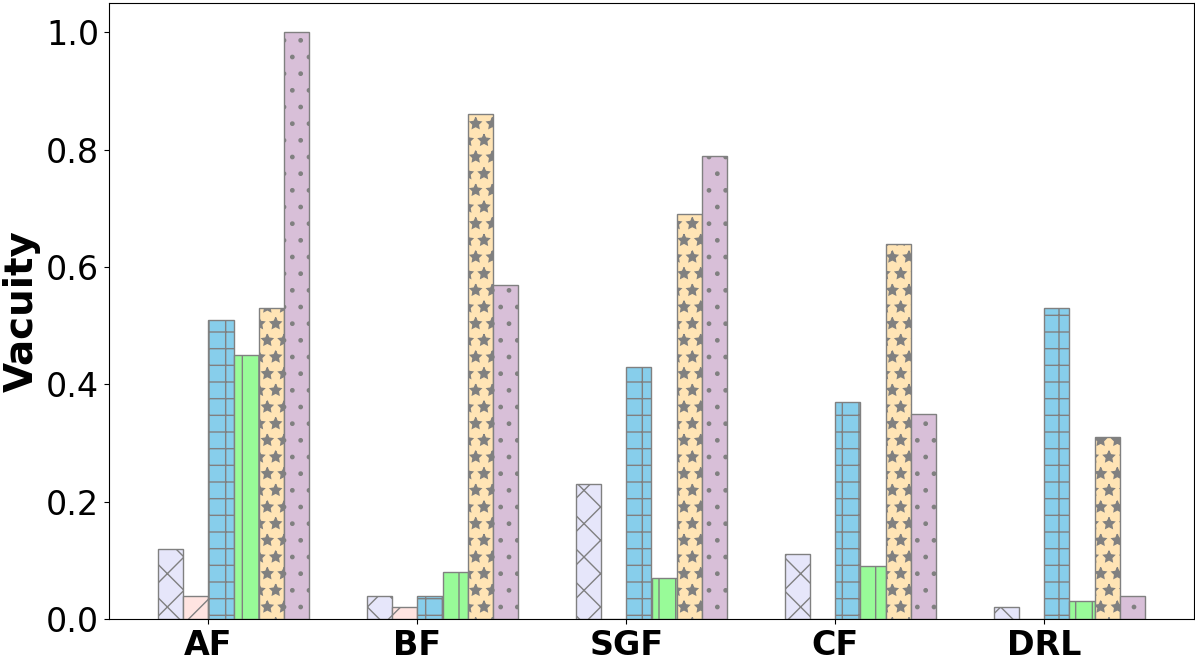}\label{fig:T-drl-F-schemes-vary-strategy_vac_ds2}}
  \subfigure[Dissonance in FBN]{
    \includegraphics[width=0.3\textwidth, height=0.2\textwidth]{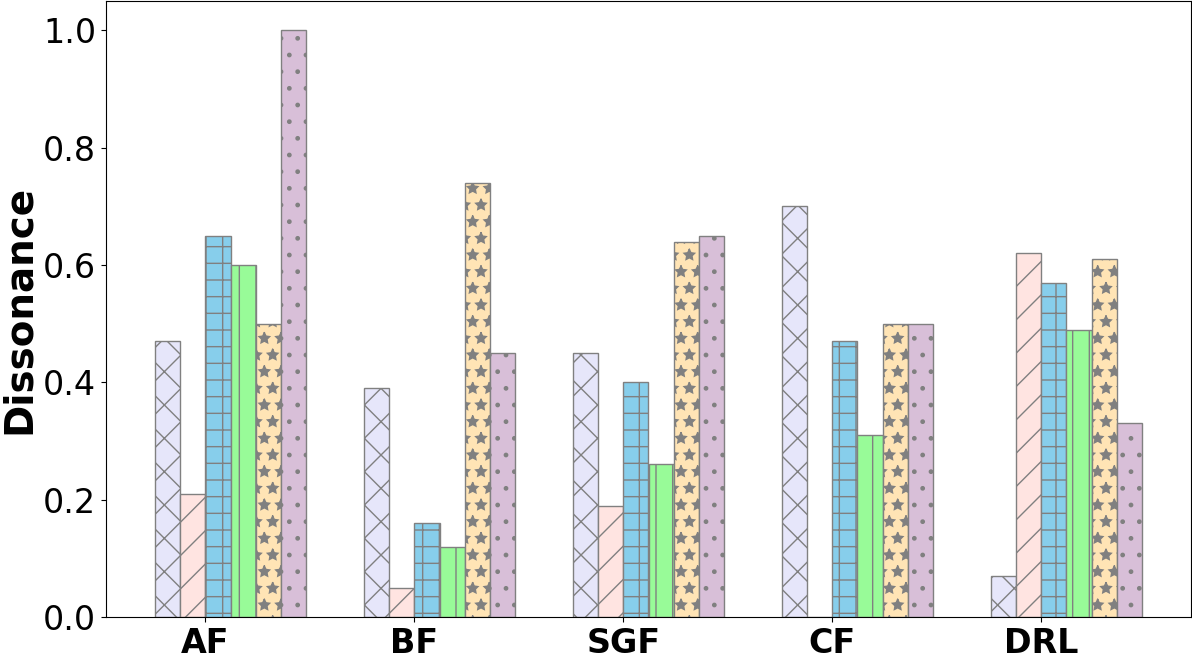} \label{fig:T-drl-F-schemes-vary-strategy_diss_ds2}}
  \subfigure[Entropy in FBN]{
    \includegraphics[width=0.3\textwidth, height=0.2\textwidth]{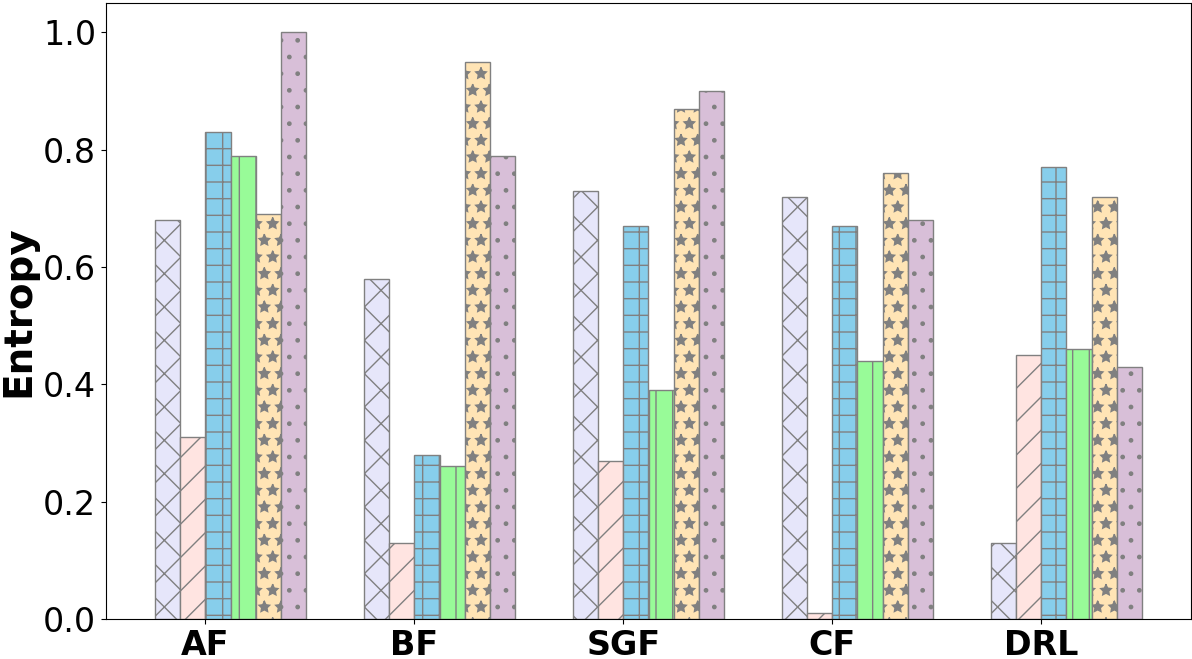} \label{fig:T-drl-F-schemes-vary-strategy_ent_ds2}}
    \vspace{-2mm}

    \subfigure[Vacuity in FBNP]{
    \includegraphics[width=0.3\textwidth, height=0.2\textwidth]{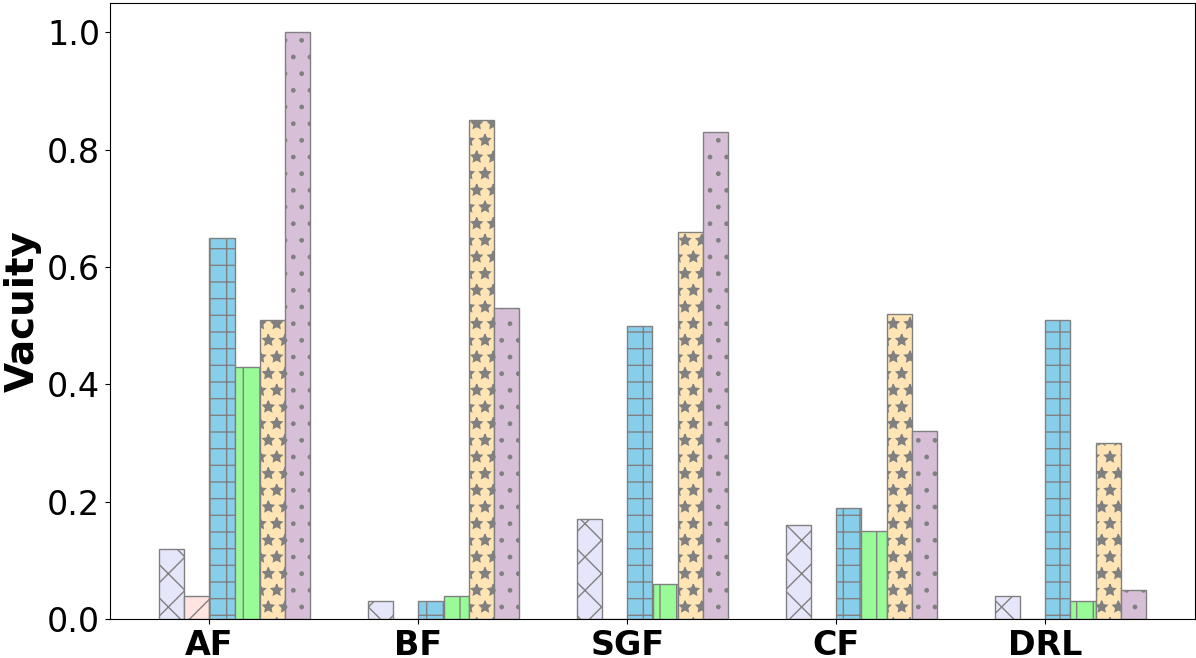}\label{fig:T-drl-F-schemes-vary-strategy_vac_ds3}}
  \subfigure[Dissonance in FBNP]{
    \includegraphics[width=0.3\textwidth, height=0.2\textwidth]{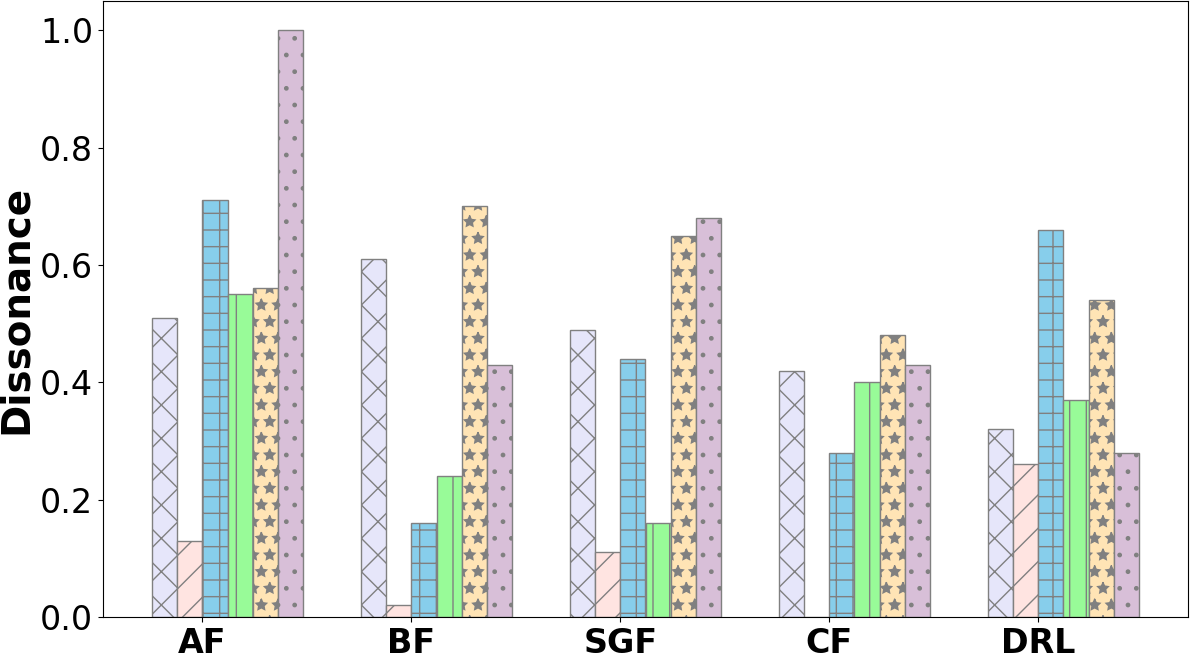} \label{fig:T-drl-F-schemes-vary-strategy_diss_ds3}}
  \subfigure[Entropy in FBNP]{
    \includegraphics[width=0.3\textwidth, height=0.2\textwidth]{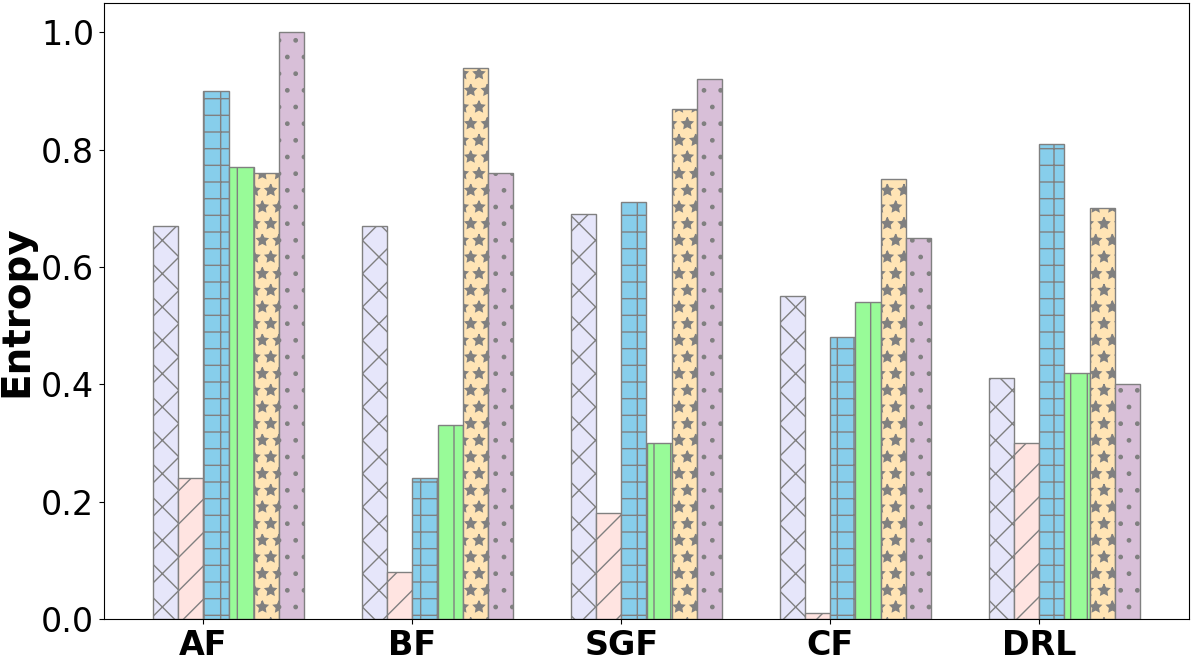} \label{fig:T-drl-F-schemes-vary-strategy_ent_ds3}}
    \vspace{-2mm}
    
    \caption{TP agent's uncertainty across CIM algorithms in FBN and FBPN, as described in Section 6.3 (Network Datasets).}
\label{fig:uncertainty-T-drl-F-vary-schemes-strategy-2}
\end{figure*}

Fig.~\ref{fig:uncertainty-T-drl-F-vary-schemes-strategy-2} examines user opinion uncertainty, including vacuity, dissonance, and entropy, across CIM schemes. As observed in Section~7.4 for the URV Email network, FP using AF or BF shows no clear correlation between CIM scheme performance and uncertainty measures. Under AF, all schemes achieve nearly 100\% influence regardless of uncertainty. With BF, despite higher vacuity, dissonance, and entropy, STORM and C-STORM still outperform others.

When FP employs SGF, CF, or DRL, a clear relationship emerges between performance and uncertainty in DRIM-A-VDEE and DRIM-NA-VDEE. These schemes achieve higher performance with the lowest vacuity, indicating that their confidence stems from sufficient exploration and effective node selection. However, dissonance does not directly correlate with performance. Notably, in DRIM-A-VDEE, dissonance follows the same trend as performance, suggesting a non-linear relationship. While lower dissonance is often linked to better outcomes, our data indicates that when dissonance drops below a critical threshold, performance declines. This implies that a certain level of dissonance is beneficial, as overly low dissonance may reduce decision-making flexibility.

DRIM-A-VDEE consistently maintains the lowest vacuity and entropy across all FP strategies, achieving superior performance. This highlights its well-optimized strategy, effectively balancing exploration and exploitation for optimal results.

\subsection{Uncertainty Analyses in DRL-based Seed Node Selection Decision-Making Under Various EE Strategies}
\label{subsubsec:uncertainty-analysis-ee}
\begin{figure*}[htb]
  \centering
  \subfigure{
    \includegraphics[width=0.75\textwidth, height=0.025\textwidth]{./Figs/legend-bar-ee.png}}
    \setcounter{subfigure}{0}
    
  \subfigure[Vacuity in FBN]{
    \includegraphics[width=0.3\textwidth, height=0.2\textwidth]{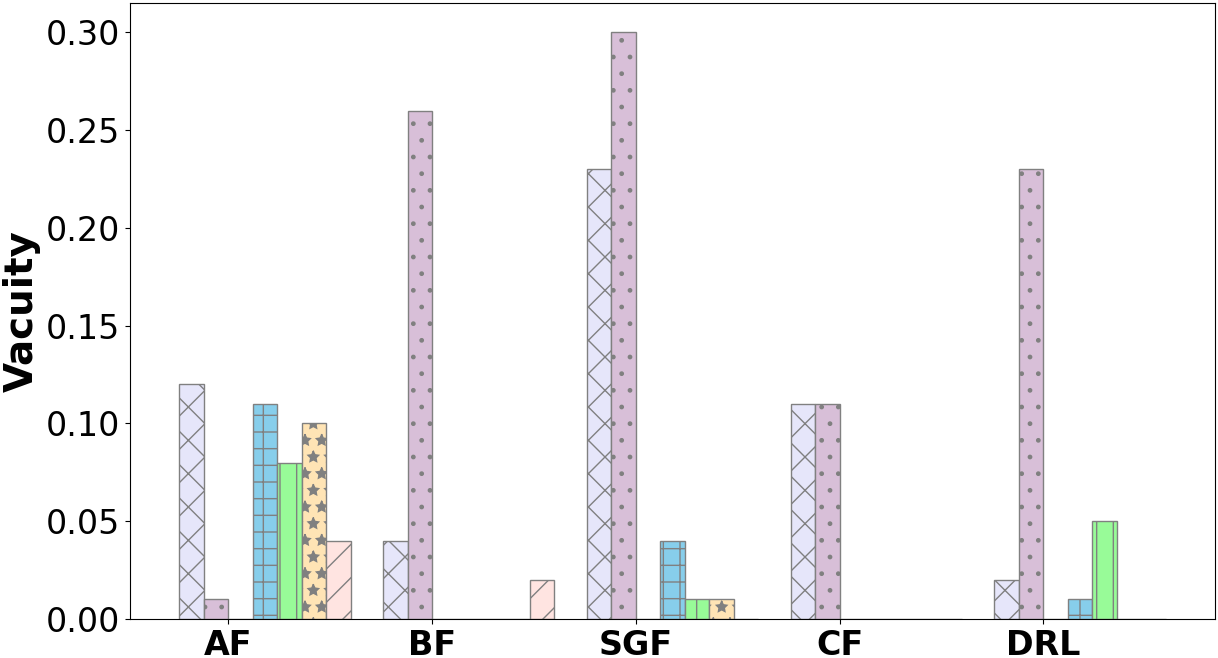}\label{fig:T-drl-F-ee-vary-strategy_vac_ds2}}
  \subfigure[Dissonance in FBN]{
    \includegraphics[width=0.3\textwidth, height=0.2\textwidth]{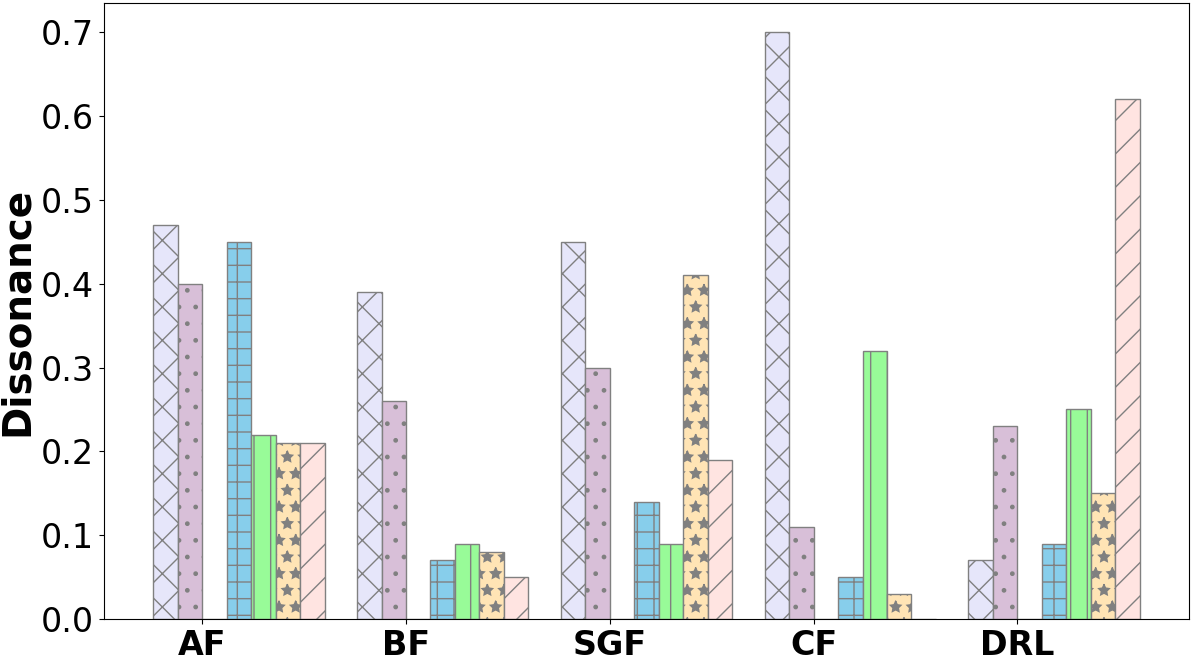} \label{fig:T-drl-F-ee-vary-strategy_diss_ds2}}
  \subfigure[Entropy in FBN]{
    \includegraphics[width=0.3\textwidth, height=0.2\textwidth]{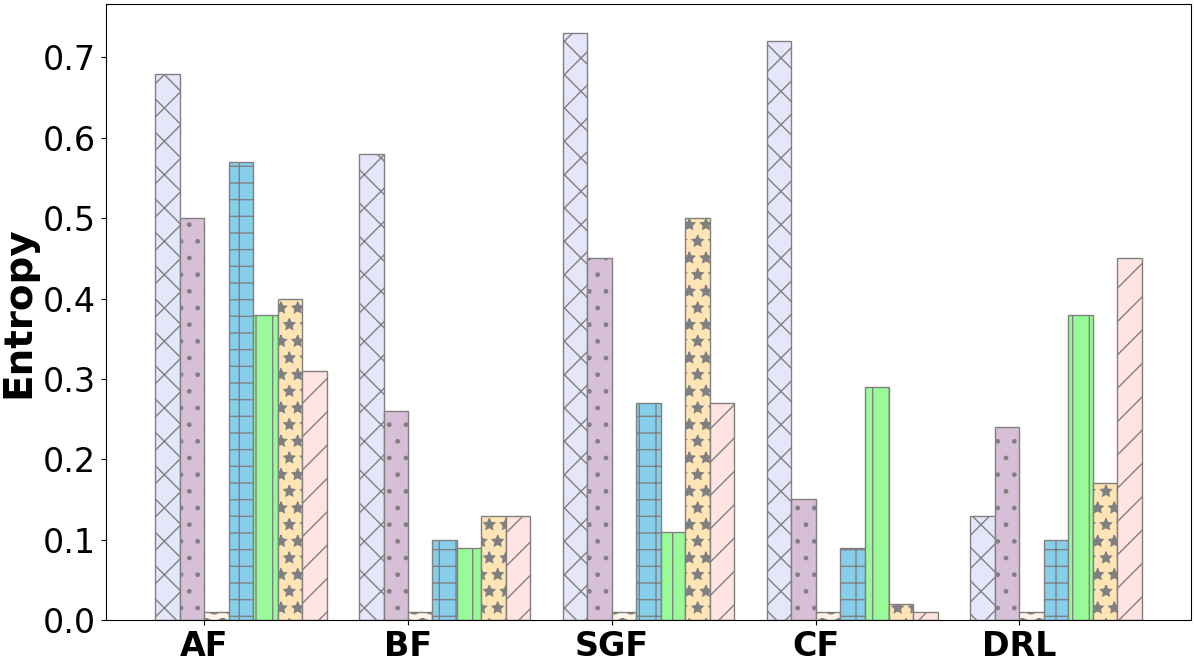} \label{fig:T-drl-F-ee-vary-strategy_ent_ds2}}
    \vspace{-2mm}

    \subfigure[Vacuity in FBNP]{
    \includegraphics[width=0.3\textwidth, height=0.2\textwidth]{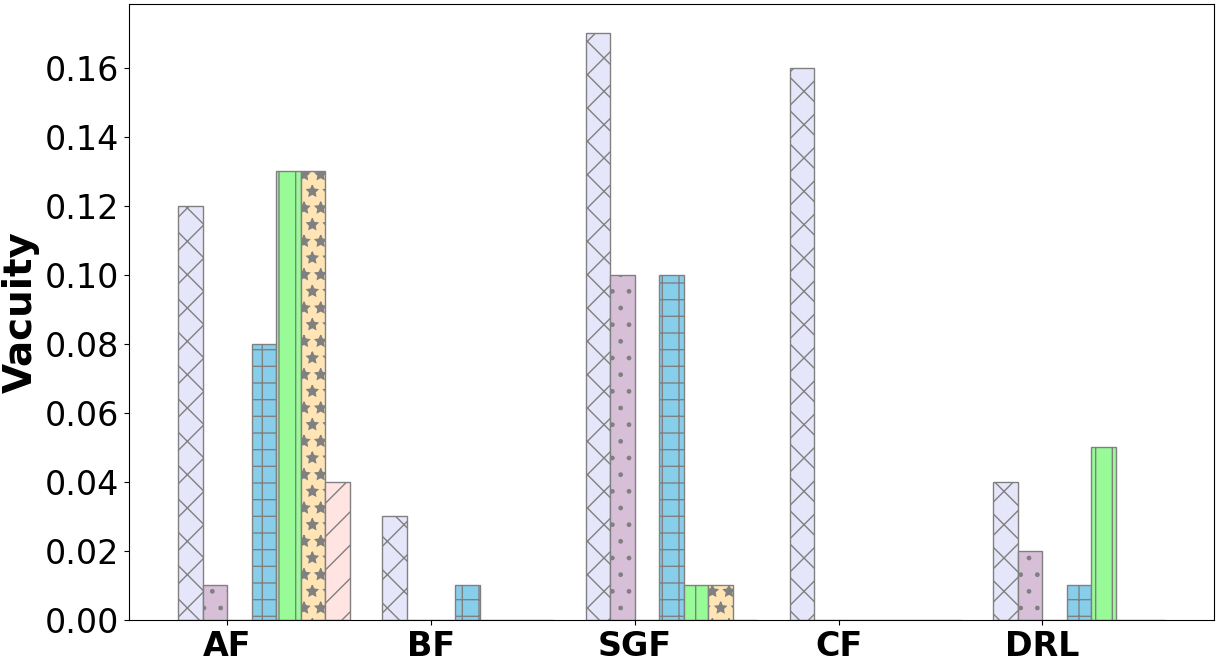}\label{fig:T-drl-F-ee-vary-strategy_vac_ds3}}
  \subfigure[Dissonance in FBNP]{
    \includegraphics[width=0.3\textwidth, height=0.2\textwidth]{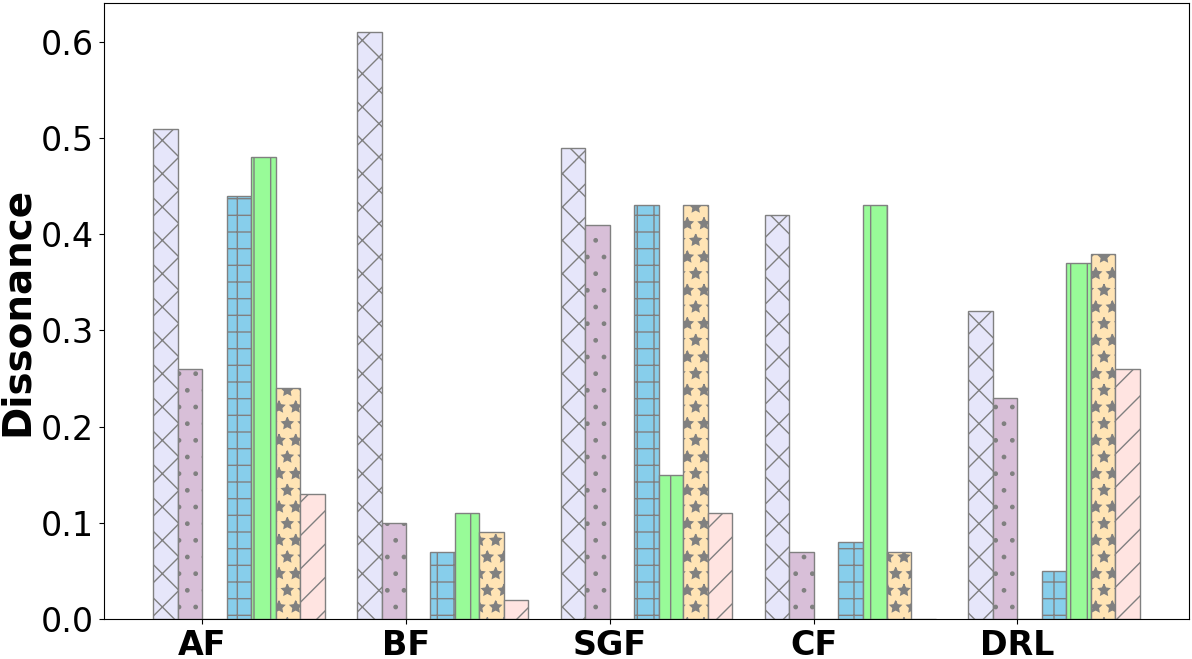} \label{fig:T-drl-F-ee-vary-strategy_diss_ds3}}
  \subfigure[Entropy in FBNP]{
    \includegraphics[width=0.3\textwidth, height=0.2\textwidth]{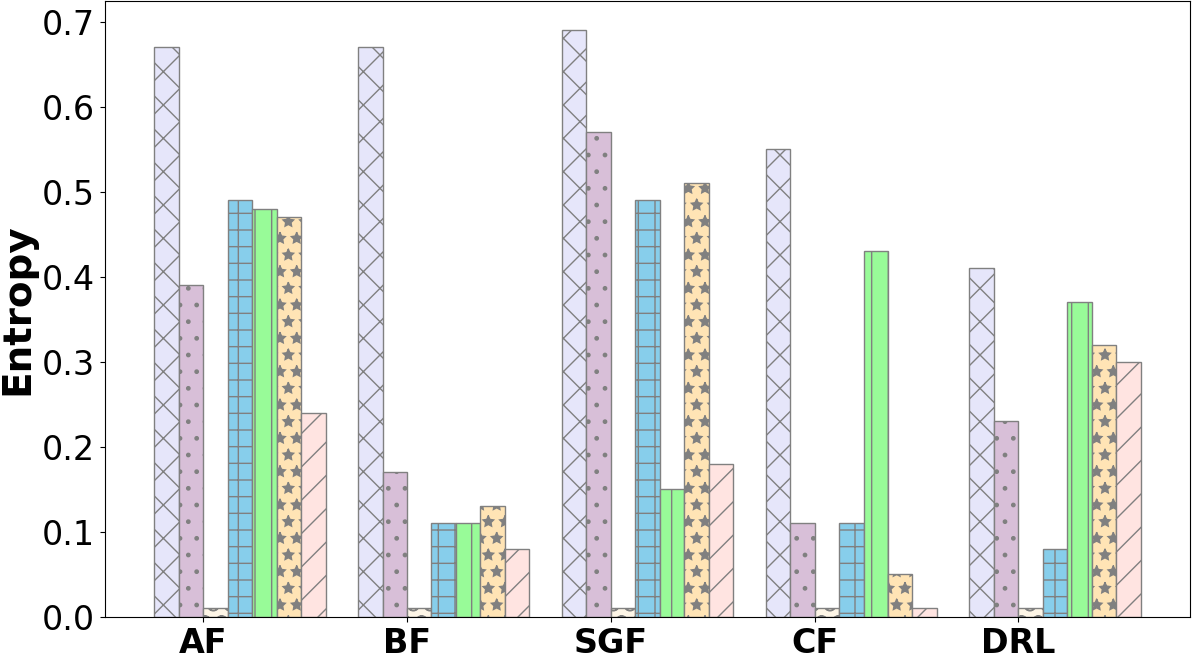} \label{fig:T-drl-F-ee-vary-strategy_ent_ds3}}
    \vspace{-2mm}
    
    \caption{TP agent's uncertainty across various EE strategies in FBN and FBPN, as described in Section 6.3 (Network Datasets).}
\label{fig:uncertainty-T-drl-F-vary-EE-strategy-2}
\end{figure*}

Fig.~\ref{fig:uncertainty-T-drl-F-vary-EE-strategy-2} examines the impact of uncertainty on performance across EE strategies. At first glance, no strong correlation appears, as performance is primarily driven by strategy choices and FP's node selection, with EE strategies having a minor influence.  VD-EE generally outperforms ER-EE while exhibiting lower uncertainty, except when FP employs AF in the Facebook Page dataset. However, all EE strategies maintain low uncertainty, suggesting that moderate uncertainty is not necessarily detrimental. Notably, VAC-EE achieves the highest performance under DRL, despite having higher uncertainty than VD-EE.  Higher dissonance and entropy often correlate with poorer performance, as these strategies struggle with conflicting information. Across both datasets, ER-EE shows higher dissonance and lower performance when FP uses non-DRL strategies. However, under DRL, VAC-EE performs best despite not having the lowest dissonance and entropy. Detailed values indicate that when dissonance and entropy remain below $0.4$, VAC-EE and VD-EE perform well, while ER-EE's performance does not directly correlate with these factors.

\section{Discussion of Highlighting the Contribution of Our Proposed Approach Compared to the Existing Related Research}

\begin{table*}[t]
\centering
\caption{\sc \new{Comparison of Opinion and Information Propagation Models with Our Work}}
\label{tab:comparison_opinion}
\begin{tabular}{|p{3cm}|p{7cm}|p{6cm}|}
\hline
\multicolumn{1}{|c|}{\textbf{Work}} & \multicolumn{1}{c|}{\textbf{Contribution}} & \multicolumn{1}{c|}{\textbf{Difference from Our Work}} \\ 
\hline
\cite{liggett2013stochastic, degroot1974reaching, friedkin1999social} (Voter, DeGroot, FJ) & Models opinion dynamics via neighbor interactions; FJ accounts for stubborn users, leading to polarization. & Focuses on consensus formation rather than competitive influence. \\ 
\hline
\cite{deffuant2000mixing, hegselmann2002opinion} (DW, HK) & Uses bounded confidence for clustering, limiting interactions to similar opinions. & Does not model strategic influence in competitive settings. \\ 
\hline
\cite{galam2008sociophysics} (Galam) & Uses majority rule in random groups to model binary opinions. & Oversimplifies influence propagation, lacking opinion evolution. \\ 
\hline
\cite{sznajd2000opinion} (Sznajd) & Relies on local interactions to drive consensus or polarization. & Lacks large-scale influence modeling, limiting applicability to CIM. \\ 
\hline
\cite{grauwin2012opinion, su2017noise} (Noisy Models) & Introduces randomness into existing models to simulate external influences. & Randomness complicates strategic decision-making in CIM. \\ 
\hline
\cite{kempe2003maximizing, li2015getreal, liang2023targeted, tong2022novel} (IC and Variants) & Models independent influence spread; widely used in CIM. & Assumes static parameters, not accounting for dynamic opinion shifts. \\ 
\hline
\cite{Pham19-CIM, bagheri2024community} (LT and Variants) & Activates nodes based on cumulative influence thresholds. & Static threshold assumptions limit adaptability to misinformation. \\ 
\hline
\textbf{Our Work} & Integrates dynamic opinion evolution and misinformation resilience into CIM models. & Moves beyond static models by incorporating evolving user beliefs and adversarial influence. \\ 
\hline
\end{tabular}
\end{table*}

\begin{table*}[t]
\centering
\caption{\sc \new{Comparison of Exploitation-Exploration (EE) Strategies in DRL with Our Work}}
\label{tab:comparison_ee}
\begin{tabular}{|p{3cm}|p{6cm}|p{7cm}|}
\hline
\multicolumn{1}{|c|}{\textbf{Work}} & \multicolumn{1}{c|}{\textbf{Contribution}} & \multicolumn{1}{c|}{\textbf{Difference from Our Work}} \\ 
\hline
\cite{sutton2018reinforcement} (Epsilon-greedy) & Uses a decaying $\epsilon$ to balance exploration and exploitation. & Simplistic approach, inefficient in complex environments. \\ 
\hline
\cite{devidze2022exploration} (Reward Shaping) & Introduces auxiliary rewards to improve learning in sparse environments. & Requires manual reward engineering, limiting generalizability. \\ 
\hline
\cite{ziebart2010modeling} (Entropy Regularization, SAC) & Encourages diverse actions by maximizing policy entropy. & Does not explicitly model uncertainty or strategic exploration. \\ 
\hline
\cite{burda2018exploration} (RND) & Uses prediction errors of a random network as intrinsic rewards. & Focuses on novelty but does not account for task-specific uncertainties. \\ 
\hline
\cite{bellemare2016unifying, badia2020never} (Pseudo-counts, NGU) & Rewards visits to less frequent states; NGU detects novelty via episodic memory. & Encourages exploration but lacks explicit uncertainty quantification. \\ 
\hline
\cite{ecoffet2019goexplore, florensa2017reverse} (Goal-based Exploration) & Guides exploration using goal selection and reverse curriculum learning. & Requires goal specification, limiting adaptability to dynamic environments. \\ 
\hline
\cite{Lai85UCB} (UCB) & Selects actions using an optimism-based confidence bound. & Assumes well-defined uncertainty estimates, which may not generalize. \\ 
\hline
\cite{osband2016boostrappedDQN, kalweit2017boostrappedDDPG} (Bootstrapped DQN) & Uses multiple Q-networks for uncertainty estimation. & Computationally expensive and prone to underfitting due to data inefficiency. \\ 
\hline
\cite{houthooft2016vime} (VIME) & Uses Bayesian neural networks to maximize information gain. & Focuses on intrinsic rewards but lacks fine-grained uncertainty modeling. \\ 
\hline
\textbf{Our Work} & Incorporates uncertainty-aware EE strategies tailored for adversarial influence settings. & Moves beyond traditional EE by leveraging evidential deep learning for adaptive decision-making. \\ 
\hline
\end{tabular}
\end{table*}

\begin{table*}[t]
\centering
\caption{\sc \new{Comparison of Evidential Deep Learning (EDL) Approaches with Our Work}}
\label{tab:comparison_edl}
\begin{tabular}{|p{4cm}|p{6cm}|p{6cm}|}
\hline
\multicolumn{1}{|c|}{\textbf{Work}} & \multicolumn{1}{c|}{\textbf{Contribution}} & \multicolumn{1}{c|}{\textbf{Difference from Our Work}} \\ 
\hline
\cite{sensoy2018EDL} (EDL) & Refined DST-based EDL, introducing a loss function to balance accuracy and uncertainty. & Improves uncertainty quantification but lacks adversarial robustness in competitive environments. \\ 
\hline
\cite{hu2021muENN, wang2022stereoMatchingUncertainty} (WGAN, NIG) & Integrated Wasserstein GANs (WGANs) and Normal Inverse Gamma (NIG) for better OOD detection. & Focuses on stereo matching uncertainty rather than strategic decision-making. \\ 
\hline
\cite{deng2023FisherInfoEDL} (IEDL) & Introduced Fisher Information Matrix (FIM)-based loss reweighting for high-uncertainty samples. & Enhances estimation but does not address CIM-specific uncertainty. \\ 
\hline
\cite{shao2024dual} (DDEF) & Developed Dual-level Deep Evidential Fusion (DDEF) for multimodal learning. & Focuses on evidence fusion, not adversarial influence modeling. \\ 
\hline
\cite{ancha2024deep} (Dirichlet-based Framework) & Proposed pixel-wise uncertainty estimation for semantic segmentation in robotics. & Optimized for vision tasks, lacking influence propagation modeling. \\ 
\hline
\cite{duan2024evidential} (Variance-based EDL) & Extended uncertainty quantification using total covariance for richer class-level insights. & Provides finer uncertainty estimates but does not model opinion dynamics. \\ 
\hline
\cite{li2023region} (Medical Imaging) & Applied EDL to brain tumor segmentation for reliable uncertainty mapping. & Application-specific, does not generalize to CIM scenarios. \\ 
\hline
\cite{schreck2024evidential} (Earth System Science) & Demonstrated EDL’s effectiveness in weather and climate modeling. & Focuses on scientific prediction rather than competitive influence modeling. \\ 
\hline
\textbf{Our Work} & Integrates EDL with CIM, leveraging uncertainty-aware opinion dynamics. & Extends EDL beyond classification to adversarial influence and misinformation resilience. \\ 
\hline
\end{tabular}
\end{table*}

\bibliographystyle{IEEEtranN}
\bibliography{ref}
\end{document}